\documentclass{article}

\usepackage{fullpage}
\usepackage{amsfonts}
\usepackage{graphicx}

\def\ms#1{\null\ifmmode\mathord{\mathcode`-="702D\it #1\mathcode`\-="2200}%
	\else$\mathord{\mathcode`-="702D\it #1\mathcode`\-="2200}$\fi}

\newcommand{\cws}[2]
	{\\ \centerline{$#2$} \\[-#1pt]}

\newlength{\spacelen}
\newcommand{\tabspace}[1]
        {\settowidth{\spacelen}{$#1$}
         \hspace*{\spacelen}}

\newcommand{\lsp}
	{[ \! [}
\newcommand{\rsp}
	{] \! ]}

\newcommand{\cala}
        {{\cal A}}

\newcommand{\calb}
        {{\cal B}}

\newcommand{\calc}
        {{\cal C}}

\newcommand{\calcu}
        {\ms{\cal CU}}

\newcommand{\cale}
        {{\cal E}}

\newcommand{\calf}
        {{\cal F}}

\newcommand{\calli}
        {\ms{\cal LI}}

\newcommand{\caloali}
        {\ms{\cal OALI}}

\newcommand{\calp}
        {{\cal P}}

\newcommand{\calr}
        {{\cal R}}

\newcommand{\calsf}
        {{\cal SF}}

\newcommand{\cals}
        {{\cal S}}

\newcommand{\calv}
        {{\cal V}}

\newcommand{\calx}
        {{\cal X}}

\newcommand{\caly}
        {{\cal Y}}

\newcommand{\natns}
	{\mathbb{N}}

\newcommand{\arrow}[2]
        {\, {\auxarrow\limits^{#1}}_{#2} \,}
\newcommand{\auxarrow}
	{\mathop{- \!\! - \!\!\!\! \longrightarrow}}

\newcommand{\warrow}[2]
        {\, {\wauxarrow\limits^{#1}}_{#2} \,}
\newcommand{\wauxarrow}
	{\mathop{= \!\! = \!\! = \!\!\!\! \Longrightarrow}}

\newcommand{\larrow}[2]
        {\, {\auxlarrow\limits^{#1}}_{#2} \,}
\newcommand{\auxlarrow}
        {\mathop{- \hspace{-0.2cm} - \hspace{-0.2cm} - \hspace{-0.2cm}
	- \hspace{-0.2cm} - \hspace{-0.2cm} - \hspace{-0.2cm}
	- \hspace{-0.2cm} - \hspace{-0.3cm} \longrightarrow}}

\newcommand{\nil}
	{\underline 0}

\newcommand{\eqdef}
	{\buildrel \Delta \over =}

\newcommand{\wbis}[1]
        {\approx_{#1}}

\newcommand{\obis}[1]
        {\simeq_{#1}}

\newcommand{\pco}[1]
	{\mathop{\Vert_{#1}}}

\newcommand{\infr}[2]
	{\renewcommand{\arraystretch}{1.5}
	\begin{array}{c}
	#1\\
	\hline
	#2
	\end{array}}

\newcommand{\fullbox}
	{{\mbox{ }\nolinebreak\hfill{$\rule{2mm}{2mm}$}}}

\newcommand{\aemilia}
	{\textrm{{\AE}milia}}

\newtheorem{new_theorem}
	{Theorem}[section]

\newtheorem{new_definition}
	[new_theorem]{Definition}

\newtheorem{new_remark}
	[new_theorem]{Remark}

\newtheorem{new_example}
	[new_theorem]{Example}

\newtheorem{new_lemma}
	[new_theorem]{Lemma}

\newtheorem{new_proposition}
	[new_theorem]{Proposition}

\newtheorem{new_corollary}
	[new_theorem]{Corollary}

\newenvironment{definition}
	{\begin{new_definition}\rm}
	{\end{new_definition}}

\newenvironment{example}
	{\begin{new_example}\rm}
	{\end{new_example}}

\newenvironment{theorem}
	{\begin{new_theorem}\rm}
	{\end{new_theorem}}

\newenvironment{proof}
	{\medskip\noindent{\bf Proof}$\ $}
	{}

\begin{document}

\title{Process Algebraic Architectural Description Languages: \\[-0.1cm]
       Generalizing Component-Oriented Mismatch Detection \\[-0.1cm]
       in the Presence of Nonsynchronous Communications \\[0.3cm]
       {\small Addendum to: \\[-0.1cm]
	       ``Handling Communications in Process Algebraic 
	       Architectural Description Languages: \\[-0.1cm]
	       Modeling, Verification, and Implementation'' \\[-0.5cm]
	       Journal of Systems and Software~83:1404--1429, August 2010}}

\author{Marco Bernardo \quad Edoardo Bont\`a \quad Alessandro Aldini}

\date{\small Universit\`a di Urbino ``Carlo Bo'' -- Italy}

\maketitle


\begin{abstract}
In the original paper, we showed how to enhance the expressiveness of a typical process algebraic
architectural description language by including the capability of representing nonsynchronous
communications. In particular, we extended the language by means of additional qualifiers enabling the
designer to distinguish among synchronous, semi-synchronous, and asynchronous ports. Moreover, we showed how
to modify techniques for detecting coordination mismatches such as the compatibility check for star
topologies and the interoperability check for cycle topologies, in such a way that those two checks are
applicable also in the presence of nonsynchronous communications. In this addendum, we generalize those
results by showing that it is possible to verify in a component-oriented way an arbitrary property of a
certain class (not only deadlock) over an entire architectural type having an arbitrary topology (not only
stars and cycles) by considering also behavioral variations, exogenous variations, endogenous variations,
and multiplicity variations, so to deal with the possible presence of nonsynchronous communications. The
proofs are at the basis of some results mentioned in the book ``A Process Algebraic Approach to Software
Architecture Design'' by Alessandro Aldini, Marco Bernardo, and Flavio Corradini, published by Springer in
2010.
\end{abstract}

%
%
\section*{Introduction}
\label{sec:intro}
%
%

In~\cite{BBA1}, we showed how to enhance the expressiveness of a typical process algebraic architectural
description language by including the capability of representing nonsynchronous communications. We focused
on PADL~\cite{BCD1,AB1} and extended it by means of additional qualifiers enabling the designer to
distinguish among synchronous, semi-synchronous, and asynchronous ports.

Semi-synchronous ports are not blocking. A semi-synchronous port of a component succeeds if there is another
component ready to communicate with it, otherwise it raises an exception so as not to block the component to
which it belongs. For example, a semi-synchronous input port can be used to model accesses to a tuple space
via input or read probes. A semi-synchronous output port can instead be used to model the interplay between
a graphical user interface and an underlying application, as the former must not block whenever the latter
cannot do certain tasks requested by the user.

Likewise, asynchronous ports are not blocking because the beginning and the end of the communications in
which these ports are involved are completely decoupled. For instance, an asynchronous output port can be
used to model output operations on a tuple space. An asynchronous input port can instead be used to model
the periodical check for the presence of information received from an event notification service. The
semantic treatment of asynchronous ports requires the addition of implicit repository-like components that
implement the decoupling.

As far as verification is concerned, in~\cite{BBA1} we showed how to modify techniques for detecting
coordination mismatches. In particular, we addressed the compatibility check for star topologies and the
interoperability check for cycle topologies, both introduced in~\cite{BCD1}, in such a way that those two
checks are applicable also in the presence of nonsynchronous communications.

This is accomplished by viewing certain activities carried out through semi-synchronous and asynchronous
ports as internal activities when performing the above mentioned checks. The reason is that each such
activity has a specific outcome and takes place at a specific time instant when considered from the point of
view of the individual component executing that activity. However, in the overall architecture, the same
activity can raise an exception (if the port is semi-synchronous and the other ports are not ready to
communicate with it) or can be delayed (if the port is asynchronous and the communication is buffered).
Thus, if we do not regard exceptions and all the activities carried out through asynchronous ports as
internal activities at verification time, the compatibility or interoperability check may fail even in the
absence of a real coordination mismatch.

In Thms.~4.7 and~4.9 of~\cite{BBA1}, we showed results based on adaptations of the compatibility check for
star topologies and the interoperability check for cycle topologies, which correspond to Thms.~4.5 and~5.2
of~\cite{BCD1}. However, in Thms.~4.14, 4.15, 4.23, 4.25, and~4.26 of~\cite{AB1}, the results of~\cite{BCD1}
were generalized by showing that it is possible to verify in a component-oriented way an arbitrary property
of a certain class (not only deadlock) over an entire architectural type having an arbitrary topology (not
only stars and cycles) by considering also behavioral variations, exogenous variations, endogenous
variations, and multiplicity variations.

In this addendum to~\cite{BBA1}, we extend the general results of~\cite{AB1} to deal with the possible
presence of nonsynchronous communications. Observing that Prop.~5.1 of~\cite{ABC1} is a slight extension of
Thm.~4.7 of~\cite{BBA1} and that Props.~5.2 and~5.3 of~\cite{ABC1} are slight extensions of Thm.~4.9
of~\cite{BBA1}, this work gives rise to Thm.~5.1 and Cors.~5.1, 5.2, 5.3, and 5.4 of~\cite{ABC1}, which
respectively extend Thms.~4.14, 4.15, 4.23, 4.25, and~4.26 of~\cite{AB1}.

%
%
\section*{Arbitrary Topologies and General Properties}
\label{sec:arb_top_gen_prop}
%
%

As in~\cite{AB1,ABC1}, we start with an architectural description $\cala$ that has an arbitrary topology and
a property~$\calp$ that belongs to the class $\Psi$ of properties each of which (i)~is expressed only in
terms of the possibility/necessity of executing certain local interactions in a certain order through a
logic that does not allow negation to be freely used and (ii)~comes equipped with a weak behavioral
equivalence $\wbis{\calp}$ coarser than weak bisimilarity that preserves $\calp$ and is a congruence with
respect to static process algebraic operators. Following the notation used in~\cite{ABC1}, we denote by
$\calf_{C_{1}, \dots, C_{n}}$ the frontier of a set of AEIs $\{ C_{1}, \dots, C_{n} \}$, by $\calcu_{C}$ the
cyclic union of an AEI $C$, and by $\calcu(\kappa)$ the set of cyclic unions generated by a cycle covering
algorithm~$\kappa$; moreover, subscript bbm, standing for before behavioral modifications, replaces bbv,
standing for before behavioral variations.


\medskip\noindent\textbf{Theorem 5.1 of~\cite{ABC1}}
Let $\cala$ be an architectural description and $\calp \in \Psi$ be a property for which the following two
conditions hold:

		\begin{enumerate}

\item For each $K \in \cala$ belonging to an acyclic portion or to the intersection of some cycle with
acyclic portions of the abstract enriched flow graph of $\cala$, $K$ is $\calp$-compatible with every $C \in
\calb_{K} - \calcu_{K}$.

\item If $\cala$ is cyclic, then there exists a total cycle covering algorithm $\kappa$ such that for each
cyclic union $\{ C_{1}, \dots, C_{n} \} \in \calcu(\kappa)$:

			\begin{enumerate}

\item If $\calf_{C_{1}, \dots, C_{n}} = \emptyset$, then there exists $C_{j} \in \{ C_{1}, \dots, C_{n} \}$
that $\calp$-interoperates with the other AEIs in the cyclic union.

\item If $\calf_{C_{1}, \dots, C_{n}} \neq \emptyset$, then every $C_{j} \in \calf_{C_{1}, \dots, C_{n}}$
$\calp$-interoperates with the other AEIs in the cyclic union.

\item If no $C_{j} \in \calf_{C_{1}, \dots, C_{n}}$ is such that $\lsp C_{j} \rsp^{\rm pc; wob}_{\cala}$
satisfies $\calp$ and there exists $C_{g} \in \{ C_{1}, \dots, C_{n} \} - \calf_{C_{1}, \dots, C_{n}}$ such
that $\lsp C_{g} \rsp^{\rm pc; wob}_{\cala}$ satisfies $\calp$, then at least one such $C_{g}$
$\calp$-interoperates with the other AEIs in the cyclic union.

			\end{enumerate}

		\end{enumerate}

\noindent
Then $\lsp \cala \rsp^{{\rm pc}; \#\cala}_{\rm bbm}$ satisfies $\calp$ iff so does $\lsp C \rsp^{\rm pc;
wob}_{\cala}$ for some $C \in \cala$.

\pagebreak

		\begin{proof}
We proceed by induction on the number $m \in \natns$ of cycles in the abstract enriched flow graph of
$\cala$:

			\begin{itemize}

\item If $m = 0$, then the abstract enriched flow graph of $\cala$ is acyclic. We prove the result by
induction on the number $s \in \natns_{\ge 1}$ of stars in the abstract enriched flow graph of $\cala$:

				\begin{itemize}

\item If $s = 1$, then there is only one star in the abstract enriched flow graph of $\cala$, which we
assume to be composed of the AEIs $K, C_{1}, \dots, C_{n}$ and centered on $K$. In order to avoid trivial
cases, let us assume $n > 0$. We distinguish among the following three cases:

					\begin{itemize}

\item Case i: $\lsp K \rsp^{\rm pc;wob}_{\cala}$ satisfies $\calp$. 
By virtue of condition~1, since $K$ is $\calp$-compatible with all the AEIs in $\calb_{K}$, from Prop.~5.1
of~\cite{ABC1} we derive that also $\lsp K, \calb_{K} \rsp^{{\rm tc};\#K,\calb_{K};K}_{K, \calb_{K}} / \,
\mathop{\cup}\limits_{l = 1}^{n} (H_{K, C_{l}} \cup E_{K, C_{l}})$ satisfies~$\calp$. Since $\lsp K,
\calb_{K} \rsp^{{\rm tc};\#K,\calb_{K};K}_{K, \calb_{K}} = \lsp \cala \rsp^{{\rm tc};\#\cala;K}_{\rm bbm}$,
it holds that $\lsp \cala \rsp^{{\rm pc}; \#\cala}_{\rm bbm}$ satisfies $\calp$ too, because $\calp$ does
not contain any free use of negation.

\item Case ii: $\lsp K \rsp^{\rm pc;wob}_{\cala}$ does not satisfy $\calp$, but there exists $C_{j} \in
\calb_{K}$ such that $\lsp C_{j} \rsp^{\rm pc;wob}_{\cala}$ satisfies $\calp$. \linebreak
By virtue of condition~1, $C_{j}$ is $\calp$-compatible with $K$:
\cws{0}{\hspace*{-2.4cm} (\lsp C_{j} \rsp^{{\rm pc}; \#K}_{\cala} \pco{\cals(C_{j}, K; \cala)} \, \lsp K
\rsp^{{\rm tc};\#C_{j}}_{C_{j}, \calb_{C_{j}}}) \, / \, (H_{C_{j}, K} \cup E_{C_{j}, K}) \: \wbis{\calp} \:
\lsp C_{j} \rsp^{\rm pc;wob}_{\cala}}
where we observe that:
\cws{0}{\hspace*{-2.4cm}\begin{array}{rcl}
\lsp K \rsp^{{\rm tc};\#C_{j}}_{C_{j}, \calb_{C_{j}}}
& \!\!\! = \!\!\! &
\lsp K \rsp^{\rm pc;\#C_{j}}_{C_{j}, \calb_{C_{j}}} \, / \, \varphi_{K, {\rm async}}(\caloali_{K}) \\
& \!\!\! \wbis{\calp} \!\!\! &
\lsp K \rsp^{\rm pc;\#C_{j}}_{\cala} \, / \, (\ms{Name} - \calv_{K; C_{j}}) \, / \,
\varphi_{K, {\rm async}}(\caloali_{K}) \\
\end{array}}
By virtue of condition~1, since $K$ is $\calp$-compatible with all the AEIs in $\calb_{K}$, from Prop.~5.1
of~\cite{ABC1} we derive in particular that:
\cws{0}{\hspace*{-2.4cm}
\lsp K, \calb_{K} - \{C_{j}\} \rsp^{{\rm tc};\#K,\calb_{K};K}_{K, \calb_{K}} \, / \, 
\mathop{\cup}\limits_{l = 1, l \neq j}^{n} (H_{K, C_{l}} \cup E_{K, C_{l}}) \: \wbis{\calp} \: 
\lsp K \rsp^{\rm pc;wob}_{\cala}}
Since $\lsp K \rsp^{\rm pc;\#C_{j}}_{\cala}$ -- occurring in the former of the last two equalities -- is
given by $\lsp K \rsp^{\rm pc;wob}_{\cala}$ -- occurring in the latter of the last two equalities -- in
parallel with the buffers associated with the originally asynchronous local interactions of~$K$ attached to
$C_{j}$, from the last equality \linebreak we derive that:
\cws{0}{\hspace*{-2.4cm}\begin{array}{c}
\lsp K \rsp^{\rm pc;\#C_{j}}_{\cala} \, / \, (\ms{Name} - \calv_{K; C_{j}}) \, / \,
\varphi_{K, {\rm async}}(\caloali_{K})
\\ \wbis{\calp} \\
\lsp K, \calb_{K} - \{C_{j}\} \rsp^{{\rm tc};\#K,\calb_{K};K}_{K, \calb_{K}} \, / \, 
\mathop{\cup}\limits_{l = 1, l \neq j}^{n} (H_{K, C_{l}} \cup E_{K, C_{l}}) \, / \,
(\ms{Name} - \calv_{K; C_{j}}) \, / \, \varphi_{K, {\rm async}}(\caloali_{K})
\end{array}}
Thanks to the last two hidings, all the actions but those in $\varphi_{K; C_{j}}(\calli_{K;C_{j}})$ are
hidden, hence the first hiding in the right-hand term above is redundant and we obtain that:
\cws{0}{\hspace*{-2.4cm}\begin{array}{c}
\lsp K, \calb_{K} - \{C_{j}\} \rsp^{{\rm tc};\#K,\calb_{K};K}_{K, \calb_{K}} 
\, / \, \mathop{\cup}\limits_{l = 1, l \not= j}^{n} (H_{K, C_{l}} \cup E_{K, C_{l}}) 
\, / \, (\ms{Name} - \calv_{K; C_{j}})
\, / \, \varphi_{K, {\rm async}}(\caloali_{K})
\\ \wbis{\calp} \\
\lsp K, \calb_{K} - \{C_{j}\} \rsp^{{\rm tc};\#K,\calb_{K};K}_{K, \calb_{K}} 
\, / \, (\ms{Name} - \calv_{K; C_{j}})
\, / \, \varphi_{K, {\rm async}}(\caloali_{K})
\end{array}}
By definition of totally closed semantics, we then derive that:
\cws{0}{\hspace*{-2.4cm}\begin{array}{c}
\lsp K, \calb_{K} - \{C_{j}\} \rsp^{{\rm tc};\#K,\calb_{K};K}_{K, \calb_{K}} 
\, / \, (\ms{Name} - \calv_{K; C_{j}})
\, / \, \varphi_{K, {\rm async}}(\caloali_{K})
\\ \wbis{\calp} \\
\lsp K, \calb_{K} - \{C_{j}\} \rsp^{{\rm tc};\#K,\calb_{K};K}_{K, \calb_{K}}
\, / \, (\ms{Name} - \calv_{K; C_{j}})
\end{array}}
Hence, summarizing, we have proved that:
\cws{0}{\hspace*{-2.4cm} \lsp K \rsp^{{\rm tc};\#C_{j}}_{C_{j}, \calb_{C_{j}}} \: \wbis{\calp} \: \lsp K,
\calb_{K} - \{C_{j}\} \rsp^{{\rm tc};\#K,\calb_{K};K}_{K, \calb_{K}} \, / \, (\ms{Name} - \calv_{K; C_{j}})}
From the first equality at the beginning of this case and the congruence property of $\wbis{\calp}$ with
respect to static process algebraic operators, we obtain that:
\cws{0}{\hspace*{-2.4cm}\begin{array}{c}
(\lsp C_{j} \rsp^{{\rm pc};\#K}_{\cala} \pco{\cals(C_{j}, K; \cala)} \, (\lsp K, \calb_{K} - \{C_{j}\}
\rsp^{{\rm tc};\#K,\calb_{K};K}_{K, \calb_{K}} \, / \, (\ms{Name} - \calv_{K; C_{j}}))) \, / \,
(H_{C_{j}, K} \cup E_{C_{j}, K}) 
\\ \wbis{\calp} \\
\lsp C_{j} \rsp^{\rm pc;wob}_{\cala}
\end{array}} 
Since $\wbis{\calp}$ preserves $\calp$, the left-hand term of the previous equality satisfies $\calp$. From
the fact that $\calp$ does not contain any free use of negation, we derive that $\lsp C_{j}, \! K, \!
\calb_{K} - \{C_{j}\} \rsp^{{\rm tc};\#K,\calb_{K};C_{j},K}_{K,\calb_{K}}$ satisfies $\calp$. Since $\lsp
C_{j}, K, \calb_{K} - \{C_{j}\} \rsp^{{\rm tc};\#K,\calb_{K};C_{j},K}_{K,\calb_{K}} = \lsp \cala \rsp^{{\rm
tc};\#\cala;C_{j},K}_{\rm bbm}$, it holds that $\lsp \cala \rsp^{{\rm pc};\#\cala}_{\rm bbm}$ satisfies
$\calp$ too, because $\calp$ does not contain any free use of negation.

\item Case iii: no AEI in the star satisfies $\calp$.
By following the same arguments as case~i, we reduce the star to the AEI $K$, which does not satisfy
$\calp$, from which it immediately follows that not even $\lsp \cala \rsp^{{\rm pc}; \#\cala}_{\rm bbm}$
satisfies $\calp$.

					\end{itemize}

\item Let the result hold for a certain $s \ge 1$ and suppose that the abstract enriched flow graph of
$\cala$ is composed of $s + 1$ stars. Due to the acyclicity of the abstract enriched flow graph of $\cala$,
there must be a star -- say composed of the AEIs $K, C_{1}, \dots, C_{n}$ and centered on $K$ -- that is
attached only to one other star in the abstract enriched flow graph of $\cala$ -- say with $C_{i}$. Then, we
distinguish among the following four cases:

					\begin{itemize}

\item Case I: $\lsp C_{i} \rsp^{\rm pc;wob}_{\cala}$ does not satisfy $\calp$, but there exists $C_{j} \in
\calb_{K} - \{C_{i}\}$ such that $\lsp C_{j} \rsp^{\rm pc;wob}_{\cala}$ satisfies $\calp$.
By considering $\calb_{K} - \{C_{i}\}$ in place of $\calb_{K}$ and following the same arguments as case~ii,
it is straightforward to obtain that $\lsp C_{j}, K, \calb_{K} - \{C_{j}, C_{i}\}\rsp^{{\rm
tc};\#K,\calb_{K};C_{j},K}_{K,\calb_{K}}$ satisfies $\calp$ too. Now, by virtue of condition~1, $K$ is
$\calp$-compatible with $C_{i}$:
\cws{0}{\hspace*{-2.4cm} (\lsp K \rsp^{{\rm pc};\#C_{i}}_{\cala} \pco{\cals(K, C_{i}; \cala)} \, \lsp C_{i}
\rsp^{{\rm tc; \#K}}_{K, \calb_{K}}) \, / \, (H_{K, C_{i}} \cup E_{K, C_{i}}) \: \wbis{\calp} \: \lsp K
\rsp^{\rm pc;wob}_{\cala}}
Since $\wbis{\calp}$ is a congruence with respect to static process algebraic operators, we derive that:
\cws{0}{\hspace*{-2.4cm}\begin{array}{c}
(\lsp C_{j}, K, \calb_{K} - \{C_{j}, C_{i}\} \rsp^{{\rm tc};\#K,\calb_{K};C_{j},K}_{K,\calb_{K}}
\pco{\cals(K, C_{i}; \cala)}
\lsp C_{i} \rsp^{{\rm tc}; \#K}_{K, \calb_{K}}) \, / \, (H_{K, C_{i}} \cup E_{K, C_{i}})  
\\ \wbis{\calp} \\
\lsp C_{j}, K, \calb_{K} - \{C_{j}, C_{i}\} \rsp^{{\rm tc};\#K,\calb_{K};C_{j},K}_{K,\calb_{K}}
\end{array}} 
and, as a consequence, the left-hand term of this equality satisfies $\calp$. Note that such a term is
$\wbis{\calp}$-equivalent to $\lsp K, \calb_{K} \rsp^{{\rm tc};\#K,\calb_{K};C_{j},K}_{K,\calb_{K}} \, / \,
(H_{K, C_{i}} \cup E_{K, C_{i}})$. Since $\calp$ does not contain any free use of negation, we derive that
also $\lsp K, \calb_{K} \rsp^{{\rm tc};\#K,\calb_{K};C_{j},K}_{K, \calb_{K}}$ satisfies $\calp$ and, for the
same reason, so does $\lsp K, \calb_{K} \rsp^{{\rm tc}; \#K,\calb_{K};C_{j},K}_{\cala}$. Since $\calp$ is
expressed only in terms of local interactions, it holds that $\lsp K, \calb_{K} \rsp^{{\rm tc};
\#K,\calb_{K};C_{j},K}_{\cala} \, / \, E_{K, \calb_{K}}$ satisfies $\calp$ too, where $E_{K, \calb_{K}}$ is
the set of exceptions that may be raised by semi-synchronous interactions involved in attachments between
$K$ and the AEIs in $\calb_{K}$.

Now, consider the architectural description $\cala'$ obtained by replacing the AEIs $K, C_{1}, \dots, C_{n}$
with a new AEI $K'$ isomorphic to $\lsp K, \calb_{K} \rsp^{{\rm tc}; \#K,\calb_{K};C_{j},K}_{\cala} \, / \,
E_{K, \calb_{K}}$. It turns out that $\cala'$ has an acyclic topology with one fewer star with respect to
$\cala$, so the induction hypothesis is applicable to $\cala'$ if we show that all of its AEIs satisfy
condition~1. It will then follow that $\lsp \cala' \rsp^{{\rm pc};\#\cala'}_{\rm bbm}$ satisfies $\calp$
because so does $\lsp K' \rsp^{\rm pc;wob}_{\cala'}$ and hence, since $\calp$ does not contain any free use
of negation, we will derive that $\lsp \cala \rsp^{{\rm pc};\#\cala}_{\rm bbm}$ satisfies $\calp$ because so
does $\lsp C_{j} \rsp^{\rm pc;wob}_{\cala}$.

It is easy to see that $K'$ satisfies condition~1. If $C$ is an arbitrary AEI attached to $K'$ because it
was previously attached to $C_{i}$, by virtue of condition~1 in $\cala$ we have that:
\cws{0}{\hspace*{-2.4cm} (\lsp C_{i} \rsp^{{\rm pc}; \#C}_{\cala} \pco{\cals(C_{i}, C; \cala)} \, \lsp C
\rsp^{{\rm tc};\#C_{i}}_{C_{i}, \calb_{C_{i}}}) \, / \, (H_{C_{i},C} \cup E_{C_{i},C}) \: \wbis{\calp} \:
\lsp C_{i} \rsp^{\rm pc;wob}_{\cala}}
from which it follows that in $\cala'$:
\cws{0}{\hspace*{-2.4cm} (\lsp K' \rsp^{{\rm pc}; \#C}_{\cala'} \pco{\cals(K', C; \cala')} \, \lsp C
\rsp^{{\rm tc};\#K'}_{K', \calb_{K'}}) \, / \, (H_{K',C} \cup E_{K',C}) \: \wbis{\calp} \: \lsp K' \rsp^{\rm
pc;wob}_{\cala'}}
because $\wbis{\calp}$ is a congruence with respect to static process algebraic operators.

Also any such $C$ satisfies condition~1 in $\cala'$. Starting from the fact that by virtue of condition~1 in
$\cala$ we have that:
\cws{0}{\hspace*{-2.4cm} (\lsp C \rsp^{{\rm pc};\#C_{i}}_{\cala} \pco{\cals(C, C_{i}; \cala)} \, \lsp C_{i}
\rsp^{{\rm tc};\#C}_{C, \calb_{C}}) \, / \, (H_{C, C_{i}} \cup E_{C, C_{i}}) \: \wbis{\calp} \: \lsp C
\rsp^{\rm pc;wob}_{\cala}}
we have to prove that in $\cala'$:
\cws{0}{\hspace*{-2.4cm} (\lsp C \rsp^{{\rm pc};\#K'}_{\cala'} \pco{\cals(C, K'; \cala')} \, \lsp K'
\rsp^{{\rm tc};\#C}_{C, \calb_{C}}) \, / \, (H_{C, K'} \cup E_{C, K'}) \: \wbis{\calp} \: \lsp C \rsp^{\rm
pc;wob}_{\cala'}}
which can be accomplished by proving that: 
\cws{0}{\hspace*{-2.4cm} \lsp K' \rsp^{{\rm tc};\#C}_{C, \calb_{C}}  \: \wbis{\calp} \: \lsp C_{i}
\rsp^{{\rm tc};\#C}_{C, \calb_{C}}}
On the one hand, since $K'$ is attached to $C$ in $\cala'$ because $C_i$ is attached to $C$ in $\cala$, it
holds that:
\cws{0}{\hspace*{-2.4cm}\begin{array}{rcl}
\lsp K' \rsp^{{\rm tc};\#C}_{C, \calb_{C}} & \!\!\! \wbis{\calp} \!\!\! &
\lsp K, \calb_{K}\rsp^{{\rm tc}; \#C,K,\calb_{K};C_{j}}_{\cala} \, / \, E_{K, \calb_{K}}
\, / \, \varphi_{C_{j}, {\rm async}}(\caloali_{C_{j}}) \, / \, (\ms{Name} - \calv_{C_{i}; C}) \\
& \!\!\! \wbis{\calp} \!\!\! &
\lsp K, \calb_{K}\rsp^{{\rm tc}; \#C,K,\calb_{K}}_{\cala} \, / \, E_{K, \calb_{K}} \, / \, (\ms{Name}
- \calv_{C_{i}; C})
\end{array}}
On the other hand, it holds that:
\cws{0}{\hspace*{-2.4cm}\begin{array}{rcl}
\lsp C_{i} \rsp^{{\rm tc};\#C}_{C, \calb_{C}} & \!\!\! \wbis{\calp} \!\!\! &
\lsp C_{i} \rsp^{{\rm pc};\#C}_{C, \calb_{C}} \, / \, \varphi_{C_{i}, {\rm async}}(\caloali_{C_{i}}) \\
& \!\!\! \wbis{\calp} \!\!\! &
\lsp C_{i} \rsp^{{\rm pc};\#C}_{\cala} \, / \, \varphi_{C_{i}, {\rm async}}(\caloali_{C_{i}}) \, / \,
(\ms{Name} - \calv_{C_{i}; C})
\end{array}}
and by virtue of condition~1:
\cws{0}{\hspace*{-2.4cm}\begin{array}{rcl}
\lsp C_{i} \rsp^{\rm pc;wob}_{\cala} & \!\!\! \wbis{\calp} \!\!\! &
(\lsp C_{i} \rsp^{\rm pc;\#K}_{\cala} \pco{\cals(C_{i}, K; \cala)}
\lsp K \rsp^{{\rm tc};\#C_{i}}_{C_{i},\calb_{C_{i}}}) \, / \, (H_{C_{i},K} \cup E_{C_{i},K}) \\
& \!\!\! \wbis{\calp} \!\!\! &
\lsp C_{i},K \rsp^{{\rm tc};\#C_{i},K;C_{i}}_{C_{i},\calb_{C_{i}}} \, / \, (H_{C_{i},K} \cup E_{C_{i},K})
\end{array}}
Since $\lsp C_{i} \rsp^{{\rm pc};\#C}_{\cala}$ is given by $\lsp C_{i} \rsp^{\rm pc;wob}_{\cala}$ in
parallel with the buffers associated with the originally asynchronous local interactions of $C_{i}$ attached
to $C$, we derive that:
\cws{0}{\hspace*{-2.4cm}\begin{array}{c}
\lsp C_{i} \rsp^{{\rm pc};\#C}_{\cala}  
\, / \, \varphi_{C_{i}, {\rm async}}(\caloali_{C_{i}}) \, / \, (\ms{Name} - \calv_{C_{i}; C})
\\ \wbis{\calp} \\
\lsp C_{i},K \rsp^{{\rm tc};\#C,C_{i},K;C_{i}}_{C_{i},\calb_{C_{i}}} \, / \, (H_{C_{i},K} \cup E_{C_{i},K})
\, / \, \varphi_{C_{i}, {\rm async}}(\caloali_{C_{i}}) \, / \, (\ms{Name} - \calv_{C_{i}; C})
\\ \wbis{\calp} \\
\lsp C_{i},K \rsp^{{\rm tc};\#C,C_{i},K;C_{i}}_{C_{i},\calb_{C_{i}}}
\, / \, E_{C_{i},K}
\, / \, \varphi_{C_{i}, {\rm async}}(\caloali_{C_{i}}) \, / \, (\ms{Name} - \calv_{C_{i}; C})
\end{array}}
because $H_{C_{i},K} \subseteq (\ms{Name} - \calv_{C_{i}; C})$. Note that the term above includes $\lsp K
\rsp^{\rm pc;wob}_{\cala}$, which, by virtue of condition~1 and Prop.~5.1 of~\cite{ABC1}, satisfies:
\cws{0}{\hspace*{-2.4cm} \lsp K \rsp^{\rm pc;wob}_{\cala} \wbis{\calp}  
\lsp K, \calb_{K} - \{C_{i}\} \rsp^{{\rm tc};\#K,\calb_{K};K}_{K, \calb_{K}} 
\, / \, \mathop{\cup}\limits_{l = 1, l\not= i}^{n} (H_{K, C_{l}} \cup E_{K, C_{l}})}
Since $\wbis{\calp}$ is a congruence with respect to static process algebraic operators, from the equalities
above we derive that:
\cws{0}{\hspace*{-2.4cm}\begin{array}{c}
\lsp C_{i},K \rsp^{{\rm tc};\#C,C_{i},K;C_{i}}_{C_{i},\calb_{C_{i}}}
\, / \, E_{C_{i},K} 
\, / \, \varphi_{C_{i}, {\rm async}}(\caloali_{C_{i}}) \, / \, (\ms{Name} - \calv_{C_{i}; C}) 
\\ \wbis{\calp} \\
\lsp C_{i},K,\calb_{K} - \{C_{i}\}\rsp^{{\rm tc};\#C,K,\calb_{K};C_{i}}_{C_{i},\calb_{C_{i}}}
\, / \, \mathop{\cup}\limits_{l = 1, l\not= i}^{n} (H_{K, C_{l}} \cup E_{K, C_{l}})
\, / \, E_{C_{i},K} 
\, / \, \varphi_{C_{i}, {\rm async}}(\caloali_{C_{i}}) \, / \, (\ms{Name} - \calv_{C_{i}; C}) 
\end{array}}
Since $\mathop{\cup}\limits_{l = 1, l\not= i}^{n} H_{K, C_{l}} \subseteq (\ms{Name} - \calv_{C_{i}; C})$, by
definition of totally closed semantics the right-hand term of the last equality is $\wbis{\calp}$-equivalent
to:
\cws{0}{\hspace*{-2.4cm}\begin{array}{c}
\lsp K,\calb_{K}\rsp^{{\rm tc};\#C,K,\calb_{K}}_{C_{i},\calb_{C_{i}}} 
\, / \, \mathop{\cup}\limits_{l = 1}^{n} E_{K, C_{l}}
\, / \, (\ms{Name} - \calv_{C_{i}; C})
\\ \wbis{\calp} \\
\lsp K,\calb_{K}\rsp^{{\rm tc};\#C,K,\calb_{K}}_{C_{i},\calb_{C_{i}}} 
\, / \, E_{K, \calb_{K}} \, / \, (\ms{Name} - \calv_{C_{i}; C})
\end{array}}
Since the hiding operation hides all the actions but the interactions from $C_{i}$ attached to $C$, this
term is $\wbis{\calp}$-equivalent to $\lsp K,\calb_{K}\rsp^{{\rm tc};\#C,K,\calb_{K}}_{\cala} \, / \, E_{K,
\calb_{K}} \, / \, (\ms{Name} - \calv_{C_{i}; C})$. Therefore, we have shown that $\lsp K' \rsp^{{\rm
tc};\#C}_{C, \calb_{C}} \: \wbis{\calp} \: \lsp C_{i} \rsp^{{\rm tc};\#C}_{C, \calb_{C}}$.

\item Case II: $\lsp C_{i} \rsp^{\rm pc;wob}_{\cala}$ satisfies $\calp$.
The proof straightforwardly derives from case~I; in particular, $K'$ turns out to be isomorphic to $\lsp K,
\calb_{K}\rsp^{{\rm tc};\#K,\calb_{K};C_{i}}_{\cala}$.

\item Case III: $\lsp K \rsp^{\rm pc;wob}_{\cala}$ satisfies $\calp$.
The proof straightforwardly derives from case~i and case~I; \linebreak in particular, $K'$ turns out to be
isomorphic to $\lsp K, \calb_{K}\rsp^{{\rm tc};\#K,\calb_{K};K}_{\cala}$.

\item Case IV: no AEI in the star satisfies $\calp$.
It is sufficient to apply the same arguments as the previous case and then observe that $K'$ does not
satisfy $\calp$.

					\end{itemize}

				\end{itemize}

\item Let the result hold for a certain $m \ge 0$ and suppose that the abstract enriched flow graph of
$\cala$ has $m + 1$ cycles. Since the cycle covering algorithm $\kappa$ of condition~2 is total, let $\caly
= \{ C_{1}, \dots, C_{n}\}$ be a cyclic union in $\calcu(\kappa)$ that directly interacts with at most one
cyclic union in $\calcu(\kappa)$. In the following, we let $I = \{C_{g}\} \cup \calf_{C_1,\ldots,C_{n}}$ if
there exists $C_{g}$ satisfying condition~2.c, and $I = \calf_{C_1,\ldots,C_{n}}$ otherwise.

Now, we replace the AEIs $C_{1}, \dots, C_{n}$ with a new AEI $C$ whose behavior is isomorphic to: 
\cws{0}{\hspace*{-0.8cm} \lsp \caly \rsp^{{\rm tc};\#\caly;I}_{\cala} \, / \, (\ms{Name} -
\mathop{\cup}\limits_{C' \in I} \calv_{C'; \cala}) \, / \, \mathop{\cup}\limits_{C' \in I} (H_{C',\caly}
\cup E_{C', \caly})}
thus obtaining an architectural description $\cala'$ such that:

				\begin{enumerate}

\item $\lsp C \rsp^{\rm pc;wob}_{\cala'}$ satisfies $\calp$ iff so does at least one AEI in $\caly$. Indeed,
one such AEI exists in~$\caly$ iff, by virtue of conditions~2.b and~2.c, $I$ includes an AEI $C'$ that
$\calp$-interoperates with $\caly$ such that $\lsp C' \rsp^{\rm pc;wob}_{\cala}$ satisfies $\calp$, which
means that $\lsp \caly \rsp^{{\rm tc};\#\caly;C'}_{\cala} \, / \, (\ms{Name} - \calv_{C'; \cala}) \, / \,
(H_{C', \caly} \cup E_{C', \caly})$ satisfies~$\calp$ and hence so does $\lsp C \rsp^{\rm pc;wob}_{\cala'}$
because $\calp$ does not contain any free use of negation.

\item $C$ preserves condition~1. In fact, let $K$ be an arbitrary AEI attached to $C$ because it was
previously attached to an AEI $C_{j}$ of $\calf_{C_{1}, \dots, C_{n}}$. It holds that $C$ is
$\calp$-compatible with $K$ and vice versa. On $C$ side, we have that in $\cala$:
\cws{0}{\hspace*{-0.8cm} (\lsp C_{j} \rsp^{{\rm pc};\#K}_{\cala} \pco{\cals(C_{j}, K; \cala)} \, \lsp K 
\rsp^{{\rm tc};\#C_{j}}_{C_{j}, \calb_{C_{j}}}) \, / \, (H_{C_{j},K} \cup E_{C_{j},K}) \: \wbis{\calp} \: 
\lsp C_{j} \rsp^{\rm pc;wob}_{\cala}}
from which it follows that in $\cala'$:
\cws{0}{\hspace*{-0.8cm} (\lsp C \rsp^{{\rm pc};\#K}_{\cala'} \pco{\cals(C, K; \cala')} \, 
\lsp K \rsp^{{\rm tc};\#C}_{C, \calb_{C}}) \, / \, (H_{C,K} \cup E_{C,K}) \: \wbis{\calp} \: 
\lsp C \rsp^{\rm pc;wob}_{\cala'}}
because $\wbis{\calp}$ is a congruence with respect to static process algebraic operators.

On $K$ side, it can be similarly shown that from:
\cws{0}{\hspace*{-0.8cm} (\lsp K \rsp^{{\rm pc};\#C_{j}}_{\cala} \pco{\cals(K, C_{j}; \cala)} \, 
\lsp C_{j} \rsp^{{\rm tc};\#K}_{K, \calb_{K}}) \, / \, (H_{K, C_{j}} \cup E_{K, C_{j}}) \: \wbis{\calp} \: 
\lsp K \rsp^{\rm pc;wob}_{\cala}}
we derive that:
\cws{0}{\hspace*{-0.8cm} (\lsp K \rsp^{{\rm pc};\#C}_{\cala'} \pco{\cals(K, C; \cala')} \, 
\lsp C \rsp^{{\rm tc};\#K}_{K, \calb_{K}}) \, / \, (H_{K, C} \cup E_{K, C}) \: \wbis{\calp} \: 
\lsp K \rsp^{\rm pc, wob}_{\cala'}}
because $C_{j}$ $\calp$-interoperates with the other AEIs in $\caly$ due to condition~2.b.

\item If $\cala'$ is cyclic, then condition~2 is preserved. In fact, let $\calcu'(\kappa)$ be the set of
cyclic unions for $\cala'$ obtained from $\calcu(\kappa)$ by replacing in each original cyclic union every
occurrence of $C_{1}, \dots, C_{n}$ with $C$. Every cyclic union in $\calcu'(\kappa)$ that does not include
$C$ has a corresponding topologically equivalent cyclic union in $\calcu(\kappa)$. 

Now, consider a cyclic union $\calx' \in \calcu'(\kappa)$ formed by the AEIs $H_{1}, \dots, H_{m}, C$. Then,
$\calcu(\kappa)$ includes a cyclic union $\calx$ formed by the AEIs $H_{1}, \dots, H_{m}, C_{j}$, where
$C_{j} \in \calf_{C_{1}, \dots, C_{n}}$. By virtue of condition~2.b:
\cws{0}{\hspace*{-0.8cm} \lsp \calx \rsp^{{\rm tc};\#\calx;C_{j}}_{\cala} \, / \, (\ms{Name} - \calv_{C_{j};
\cala}) \, / \, (H_{C_{j}, \calx} \cup E_{C_{j}, \calx}) \: \wbis{\calp} \: \lsp C_{j} \rsp^{\rm
pc;wob}_{\cala}}
Since $\wbis{\calp}$ is a congruence with respect to static process algebraic operators:
\cws{0}{\hspace*{-0.8cm} \lsp \calx' \rsp^{{\rm tc};\#\calx';C}_{\cala'} \, / \, (\ms{Name} - \calv_{C;
\cala'}) \, / \, (H_{C, \calx'} \cup E_{C, \calx'}) \: \wbis{\calp} \: \lsp C \rsp^{\rm pc;wob}_{\cala'}}
Therefore, if $\calf_{H_{1}, \dots, H_{m}, C} = \emptyset$, then condition~2.a is preserved; otherwise, if
$C \in \calf_{H_{1}, \dots, H_{m}, C}$, then $C$ preserves condition~2.b and so does each $H_{l} \in
\calf_{H_{1}, \dots, H_{m}, C} - \{ C \}$ as from:
\cws{0}{\hspace*{-0.8cm} \lsp \calx \rsp^{{\rm tc};\#\calx;H_{l}}_{\cala} \, / \, (\ms{Name} - \calv_{H_{l};
\cala}) \, / \, (H_{H_{l}, \calx} \cup E_{H_{l}, \calx}) \: \wbis{\calp} \: \lsp H_{l} \rsp^{\rm
pc;wob}_{\cala}}
we derive that:
\cws{0}{\hspace*{-0.8cm} \lsp \calx' \rsp^{{\rm tc};\#\calx';H_{l}}_{\cala'} \, / \, (\ms{Name} -
\calv_{H_{l}; \cala'}) \, / \, (H_{H_{l}, \calx'} \cup E_{H_{l}, \calx'}) \: \wbis{\calp} \: \lsp H_{l}
\rsp^{\rm pc;wob}_{\cala'}}
because $C_{j}$ $\calp$-interoperates with its cyclic union.

Now, let us consider condition~2.c and assume that no AEI in the frontier of $\calx$ satisfies $\calp$. If,
by virtue of condition~2.c, there is $H_{g} \in \calx$ such that $\lsp H_{g}\rsp^{\rm pc;wob}_{\cala}$
satisfies $\calp$ and $H_{g}$ $\calp$-interoperates with $\calx$, then, by virtue of the same arguments used
for $H_{l}$, we immediately derive that $H_{g}$ $\calp$-interoperates with $\calx'$, thus preserving
condition~2.c. 

On the other hand, if $C_{j}$ is such that $\lsp C_{j}\rsp^{\rm pc;wob}_{\cala}$ satisfies $\calp$, then we
have shown that $\lsp C \rsp^{\rm pc;wob}_{\cala'}$ satisfies $\calp$ and $\calp$-interoperates with
$\calx'$. Hence, $C$ preserves condition~2.c in the case it does not belong to the frontier of $\calx'$.

\item The abstract enriched flow graph of $\cala'$ has at most $m$ cycles.

				\end{enumerate}

\noindent Then, by the induction hypothesis, the theorem holds for $\lsp \cala' \rsp^{{\rm
pc};\#\cala'}_{\rm bbm}$. Since $\calp$ does not contain any free use of negation, we immediately derive
that the theorem holds also for $\lsp \cala \rsp^{{\rm pc};\#\cala}_{\rm bbm}$. 
\fullbox

			\end{itemize}

		\end{proof}


%
%
\section*{Behavioral Variations}
\label{sec:behav_var}
%
%

We continue by extending the result to behavioral variations, i.e., to instances of an AT whose observable
behaviors conform to each other according to weak bisimilarity $\wbis{\rm B}$ as defined in~\cite{AB1,ABC1}.


\medskip\noindent\textbf{Corollary 5.1 of~\cite{ABC1}}
Let $\cala$ be an architectural description and $\calp \in \Psi$ be a property for which the two conditions
of Thm.~5.1 of~\cite{ABC1} hold. Whenever $\wbis{\rm B} \, \subseteq \, \wbis{\calp}$, then for each AT
instance $\cala'$ that strictly behaviorally conforms to $\cala$ it turns out that $\lsp \cala' \rsp^{{\rm
pc}; \#\cala'}_{\rm bbm}$ satisfies $\calp$ iff so does $\lsp C \rsp^{\rm pc; wob}_{\cala}$ for some $C \in
\cala$.

\pagebreak

		\begin{proof}
Due to behavioral conformity, $\lsp \cala' \rsp^{{\rm pc}; \#\cala'}_{\rm bbm} \wbis{\rm B} \lsp \cala
\rsp^{{\rm pc}; \#\cala}_{\rm bbm}$ up to an injective relabeling function that matches local interactions
occurring in $\cala'$, $\cala$, and $\calp$. Therefore, $\lsp \cala' \rsp^{{\rm pc}; \#\cala'}_{\rm bbm}
\wbis{\calp} \lsp \cala \rsp^{{\rm pc}; \#\cala}_{\rm bbm}$ up to the same relabeling function, because
$\wbis{\rm B} \, \subseteq \, \wbis{\calp}$, and hence $\lsp \cala' \rsp^{{\rm pc}; \#\cala'}_{\rm bbm}$
satisfies $\calp$ iff so does $\lsp \cala \rsp^{{\rm pc}; \#\cala}_{\rm bbm}$. The result then follows from
Thm.~5.1 of~\cite{ABC1}.
\fullbox

		\end{proof}


%
%
\section*{Exogenous Variations}
\label{sec:exo_var}
%
%

We now extend the result to topological variations of exogenous nature, which take place at the topological
frontier formed by architectural interactions as explained in~\cite{AB1,ABC1}. Following the notation used
in~\cite{ABC1}, we denote by $\calsf_{C_{1}, \dots, C_{n}}$ the semi-frontier of a set of AEIs $\{ C_{1},
\dots, C_{n} \}$; moreover, we consider partially/totally semi-closed interacting semantics, in which
architectural interactions are left visible, and the related $\calp$-semi-compatibility and
$\calp$-semi-interoperability checks, together with the notion of exo-coverability.


\medskip\noindent\textbf{Corollary 5.2 of~\cite{ABC1}}
Let $\cala$ be an architectural description and $\calp \in \Psi$ be a property for which the two conditions
of Thm.~5.1 of~\cite{ABC1} hold. Let $\cala'$ be an AT instance resulting from a strictly topologically
conformant exogenous variation of $\cala$ for which the following additional conditions hold:

		\begin{enumerate}

\item[3.] For each $K \in \cala$ belonging to an acyclic portion or to the intersection of some cycle with
acyclic portions of the abstract enriched flow graph of $\cala$, if $K$ is of the same type as an AEI having
architectural interactions at which the exogenous variation takes place, then $K$ is $\calp$-semi-compatible
with every $C \in \calb_{K} - \calcu^{\cala}_{K}$.

\item[4.] If $\cala'$ is cyclic, then $\cala'$ is exo-coverable by $\kappa$ and, for each $C_{j} \in
\calsf_{C_{1}, \dots, C_{n}}$ with $\{ C_{1}, \dots, C_{n} \} \in \calcu^{\cala}(\kappa)$, if $C_{j}$ is of
the same type as an AEI having architectural interactions at which the exogenous variation takes place, then
$C_{j}$ $\calp$-semi-interoperates with the other AEIs in $\{ C_{1}, \dots, C_{n} \}$.

		\end{enumerate}

\noindent
Then $\lsp \cala' \rsp^{{\rm pc}; \#\cala'}_{\rm bbm}$ satisfies $\calp$ iff so does $\lsp C \rsp^{\rm pc;
wob}_{\cala}$ for some $C \in \cala$.

		\begin{proof}
We show that $\cala'$ satisfies the two conditions of Thm.~5.1 of~\cite{ABC1}, from which the result will
immediately follow:

			\begin{itemize}

\item $\cala'$ satisfies condition~1. 
Consider an AEI $K \in \cala'$ belonging to an acyclic portion or to the intersection of some cycle with
acyclic portions of the abstract enriched flow graph of $\cala'$, and an AEI $C \in \calb_{K} -
\calcu^{\cala'}_{K}$. We distinguish among the following four cases:

				\begin{itemize}

\item Both AEIs are in $\cala$. On the one hand, if $K$ is not an AEI having architectural interactions at
which the exogenous extension takes place, then $K$ is $\calp$-compatible with $C$ by virtue of condition~1
applied to $\cala$. On the other hand, if $K$ is an AEI having architectural interactions at which the
exogenous extension takes place, then by virtue of condition~3 it holds that $K$ is $\calp$-semi-compatible
with $C$ in $\cala$, from which we derive that $K$ is $\calp$-compatible with $C$ in $\cala'$.

\item $K \in \cala$ and $C$ is an additional AEI. By hypothesis, in $\cala$ there is an attachment between
an AEI $K'$ and $\ms{corr}(C)$, such that $K'$ is of the same type as $K$ and $\ms{corr}(C) \in \calb_{K'} -
\calcu^{\cala}_{K'}$. Then, by virtue of condition~3, $K'$ is $\calp$-semi-compatible with $\ms{corr}(C)$,
from which it follows that $K$ is $\calp$-compatible with $C$. 

\item $K$ is an additional AEI and $C \in \cala$. By hypothesis, in $\cala$ there is an attachment between
an AEI $C'$ and $\ms{corr}(K)$, such that $C'$ is of the same type as $C$ and $C' \in \calb_{\ms{corr}(K)} -
\calcu^{\cala}_{\ms{corr}(K)}$. Then, by virtue of condition~1, $\ms{corr}(K)$ is $\calp$-compatible with
$C'$, from which it follows that $K$ is $\calp$-compatible with~$C$.

\item Both $K$ and $C$ are additional AEIs. By hypothesis, in $\cala$ there are two attached AEIs
$\ms{corr}(K)$ and $\ms{corr}(C)$ such that, by virtue of condition~1 applied to $\cala$, $\ms{corr}(K)$ is
$\calp$-compatible with $\ms{corr}(C)$. As a consequence, $K$ is $\calp$-compatible with $C$.

				\end{itemize}

\item If $\cala'$ is cyclic, then $\cala'$ satisfies condition~2. We first observe that $\cala'$ satisfies
condition~2.a because, by virtue of condition~4, the exogenous variation of $\kappa$ applied to $\cala'$
cannot generate a single cyclic union with empty frontier.

Now, suppose that $\calcu^{\cala'}_{K}$ is a cyclic union generated by the exogenous variation of $\kappa$.
By virtue of condition~4, we distinguish between the following two cases:

				\begin{itemize}

\item If $K$ is in $\cala$, then $\calcu^{\cala'}_{K} = \calcu^{\cala}_{K}$ and each $C_{i} \in
\calf_{\calcu^{\cala'}_{K}}$ belongs to $\calsf_{\calcu^{\cala}_{K}}$. Then, by virtue of condition~4 or by
virtue of condition~2.b applied to $\cala$, $C_i$ $\calp$-interoperates with the other AEIs of
$\calcu^{\cala'}_{K}$. Hence, $\calcu^{\cala'}_{K}$ satisfies condition~2.b. For the same reason, if
$\calcu^{\cala}_{K}$ satisfies condition~2.c, then so does $\calcu^{\cala'}_{K}$.

\item If $K$ is an additional AEI, then $\calcu^{\cala'}_{K}$ is strictly topologically equivalent to
$\calcu^{\cala}_{corr(K)} \in \calcu^{\cala}(\kappa)$. By means of an argument similar to the one applied
above, it follows that $\calcu^{\cala'}_{K}$ satisfies condition~2.b because so does
$\calcu^{\cala}_{corr(K)}$, and that if $\calcu^{\cala}_{corr(K)}$ satisfies condition~2.c, then so does
$\calcu^{\cala'}_{K}$.
\fullbox

				\end{itemize}

			\end{itemize}

		\end{proof}


%
%
\section*{Endogenous Variations}
\label{sec:endo_var}
%
%

We further extend the result to endogenous variations, which take place inside the topological frontier as
explained in~\cite{AB1,ABC1}. Following~\cite{ABC1}, we consider the notion of endo-coverability.


\medskip\noindent\textbf{Corollary 5.3 of~\cite{ABC1}}
Let $\cala$ be an architectural description and $\calp \in \Psi$ be a property for which the two conditions
of Thm.~5.1 of~\cite{ABC1} hold. Let $\cala'$ be an AT instance resulting from an endogenous variation
of $\cala$ for which the following additional conditions hold:

		\begin{enumerate}

\item[$\overline{\mathit{3}}.$] For each attachment in $\cala'$ from interaction $o$ of an AEI $C'_{1}$,
which belongs to an acyclic portion or to the intersection of some cycle with acyclic portions of the
abstract enriched flow graph of $\cala'$, to interaction $i$ of an AEI $C'_{2} \in \calb_{C'_{1}} -
\calcu^{\cala'}_{C'_{1}}$, there exists an attachment in $\cala$ from interaction $o$ of an AEI $C_{1}$ of
the same type as $C'_{1}$, with $C_{1}$ belonging to an acyclic portion or to the intersection of some cycle
with acyclic portions of the abstract enriched flow graph of $\cala$, to interaction $i$ of an AEI $C_{2}
\in \calb_{C_{1}} - \calcu^{\cala}_{C_{1}}$ of the same type as $C'_{2}$.

\item[$\overline{\mathit{4}}.$] No local interaction occurring in $\calp$ is involved in attachments
canceled by the endogenous variation.

\item[$\overline{\mathit{5}}.$] If $\cala$ or $\cala'$ is cyclic, then $\cala'$ is endo-coverable by
$\kappa$ and for each cyclic union~$\calcu^{\cala'}_{C}$ generated by the endogenous variation of $\kappa$:

			\begin{enumerate}

\item No local interaction of the AEIs of $\calcu^{\cala}_{C}$ that $\calp$-interoperate with the other AEIs
in $\calcu^{\cala}_{C}$ by virtue of condition~2 of Thm.~5.1 of~\cite{ABC1} is involved in attachments
canceled by the endogenous variation.

\item No possibly added AEI in $\calcu^{\cala'}_{C}$ belongs to the frontier of $\calcu^{\cala'}_{C}$.

\item If $C \in \cala$, then $\lsp \calcu^{\cala'}_{C} \rsp^{{\rm pc};
\#\calcu^{\cala'}_{C}}_{\calcu^{\cala'}_{C}} / \, H \: \wbis{\calp} \: \lsp \calcu^{\cala}_{C} \rsp^{{\rm
pc}; \#\calcu^{\cala}_{C}}_{\calcu^{\cala}_{C}}/ \, H$ where $H$ contains all local interactions of the
added/removed AEIs as well as those attached to them.

			\end{enumerate}

		\end{enumerate}

\noindent
Then $\lsp \cala' \rsp^{{\rm pc}; \#\cala'}_{\rm bbm}$ satisfies $\calp$ iff so does $\lsp C \rsp^{\rm pc;
wob}_{\cala}$ for some $C \in \cala$.

		\begin{proof}
We show that $\cala'$ satisfies the two conditions of Thm.~5.1 of~\cite{ABC1}, from which the result will
immediately follow thanks to condition~$\overline{\mathit{4}}$:

			\begin{itemize}

\item $\cala'$ satisfies condition~1.
Consider an AEI $K \in \cala'$ belonging to an acyclic portion or to the intersection of some cycle with
acyclic portions of the abstract enriched flow graph of $\cala'$, and an AEI $C \in \calb_{K} -
\calcu^{\cala'}_{K}$. We distinguish between the following two cases:

				\begin{itemize}

\item Both AEIs are in $\cala$. The only interesting case occurs whenever $K$ and $C$ are not attached in
$\cala$. In this case, by virtue of condition~$\overline{\mathit{3}}$, there exists an attachment in $\cala$
of the same kind between an AEI of the same type as $K$ and an AEI of the same type as $C$, from which the
result follows by virtue of condition~1 applied to $\cala$.

\item $K \in \cala$ and $C$ is an additional AEI, or $K$ is an additional AEI and $C \in \cala$, or both $K$
and $C$ are additional AEIs. It is sufficient to apply the same argument illustrated above.

				\end{itemize}

\item If $\cala'$ is cyclic, then $\cala'$ satisfies condition~2. Suppose that the endogenous variation of
$\kappa$ generates a single cyclic union $\calcu^{\cala'}_{C}$ with empty frontier. Then, by virtue of
condition~$\overline{\mathit{5}}.c$, $\lsp \calcu^{\cala'}_{C} \rsp^{{\rm pc};
\#\calcu^{\cala'}_{C}}_{\calcu^{\cala'}_{C}} / \, H \: \wbis{\calp} \: \lsp \calcu^{\cala}_{C} \rsp^{{\rm
pc}; \#\calcu^{\cala}_{C}}_{\calcu^{\cala}_{C}}/ \, H$ and by virtue of conditions~$\overline{\mathit{5}}.a$
and~$\overline{\mathit{5}}.b$ it turns out that conditions~2.a and~2.c are preserved. In particular, we now
show that if there exists $C_{i} \in \calcu^{\cala}_{C}$ that, by virtue of condition~2.a or~2.c applied to
$\cala$, $\calp$-interoperates with the other AEIs in $\calcu^{\cala}_{C}$, then $C_{i}$
$\calp$-interoperates with the other AEIs in $\calcu^{\cala'}_{C}$. By hypothesis:
\cws{0}{\hspace*{-0.8cm} \lsp \calcu^{\cala}_{C} \rsp^{{\rm tc};\#\calcu^{\cala}_{C};C_{i}}_{\cala} \, / \,
(\ms{Name} - \calv_{C_{i}; \cala}) \, / \, (H_{C_{i}, \calcu^{\cala}_{C}} \cup E_{C_{i},
\calcu^{\cala}_{C}}) \: \wbis{\calp} \: \lsp C_{i} \rsp^{\rm pc;wob}_{\cala}}
By condition~$\overline{\mathit{5}}.a$, $H \subseteq (\ms{Name} - \calv_{C_{i}; \cala})$. Hence, the
left-hand term of this equality is $\calp$-equivalent to: 
\cws{0}{\hspace*{-0.8cm} \lsp \calcu^{\cala}_{C} \rsp^{{\rm pc}; \#\calcu^{\cala}_{C}}_{\calcu^{\cala}_{C}}/
\, H  \, / \, (\ms{Name} - \calv_{C_{i}; \cala}) \, / \, (H_{C_{i}, \calcu^{\cala}_{C}} \cup E_{C_{i},
\calcu^{\cala}_{C}})}
and to:
\cws{0}{\hspace*{-0.8cm} \lsp \calcu^{\cala'}_{C} \rsp^{{\rm pc};
\#\calcu^{\cala'}_{C}}_{\calcu^{\cala'}_{C}} / \, H \, / \, (\ms{Name} - \calv_{C_{i}; \cala'}) \, / \,
(H_{C_{i}, \calcu^{\cala'}_{C}} \cup E_{C_{i}, \calcu^{\cala'}_{C}})}
which, for the same motivations, is $\calp$-equivalent to:
\cws{0}{\hspace*{-0.8cm} \lsp \calcu^{\cala'}_{C} \rsp^{{\rm tc};\#\calcu^{\cala'}_{C};C_{i}}_{\cala'} \, /
\, (\ms{Name} - \calv_{C_{i}; \cala'}) \, / \, (H_{C_{i}, \calcu^{\cala'}_{C}} \cup E_{C_{i},
\calcu^{\cala'}_{C}})}
from which the result follows.

Now, suppose that $\calcu^{\cala'}_{C}$ is a cyclic union with nonempty frontier generated by the endogenous
variation of $\kappa$. We distinguish between the following two cases:

				\begin{itemize}

\item If $\calcu^{\cala'}_{C}$ is equal to $\calcu^{\cala}_{C}$ (resp.\ strictly topologically equivalent to
a cyclic union $\caly \in \calcu^{\cala}(\kappa)$), then $\calcu^{\cala'}_{C}$ satisfies conditions~2.b
and~2.c because so does $\calcu^{\cala}_{C}$ (resp.\ $\caly$).

\item If $\calcu^{\cala'}_{C}$ includes some of the added AEIs or $\calcu^{\cala}_{C}$ includes some of the
removed AEIs, then it is sufficient to apply condition~$\overline{\mathit{5}}$ as shown above to derive that
conditions~2.b and~2.c are preserved.
\fullbox

				\end{itemize}

			\end{itemize}

		\end{proof}


%
%
\section*{Multiplicity Variations}
\label{sec:multi_var}
%
%

We finally extend the result to multiplicity variations, which take place at and-/or-interactions as
explained in~\cite{AB1,ABC1}.


\medskip\noindent\textbf{Corollary 5.4 of~\cite{ABC1}}
Let $\cala$ be an architectural description and $\calp \in \Psi$ be a property for which the two conditions
of Thm.~5.1 of~\cite{ABC1} hold. Let $\cala'$ be an AT instance resulting from a multiplicity variation of
$\cala$ for which the following additional conditions hold:

		\begin{enumerate}

\item[$\widetilde{\mathit{3}}.$] No local interaction occurring in $\calp$ is involved in attachments
canceled by the multiplicity variation.

\item[$\widetilde{\mathit{4}}.$] No local or-interaction involved in the multiplicity variation is attached
to a semi-synchronous uni-interaction or to an input asynchronous uni-interaction.

\item[$\widetilde{\mathit{5}}.$] Each local or-interaction involved in the multiplicity variation is enabled
infinitely often.

\item[$\widetilde{\mathit{6}}.$] If $\cala$ or $\cala'$ is cyclic, then $\calcu^{\cala}(\kappa) =
\calcu^{\cala'}(\kappa)$.

		\end{enumerate}

\noindent
Then $\lsp \cala' \rsp^{{\rm pc}; \#\cala'}_{\rm bbm}$ satisfies $\calp$ iff so does $\lsp C \rsp^{\rm pc;
wob}_{\cala}$ for some $C \in \cala$.

\pagebreak

		\begin{proof}
We show that $\cala'$ satisfies the two conditions of Thm.~5.1 of~\cite{ABC1}, from which the result will
immediately follow thanks to condition~$\widetilde{\mathit{3}}$:

			\begin{itemize}

\item $\cala'$ satisfies condition~1. Consider an AEI $K \in \cala'$ belonging to an acyclic portion or to
the intersection of some cycle with acyclic portions of the abstract enriched flow graph of $\cala'$, and an
AEI $C \in \calb_{K} - \calcu^{\cala'}_{K}$. We distinguish among the following five cases:

				\begin{itemize}

\item Both AEIs are in $\cala$ and are attached through interactions that are not subject to the
multiplicity variation. Then, $K$ is $\calp$-compatible with $C$ by virtue of condition~1 applied to
$\cala$.

\item $K$ is in $\cala$ and has an and-interaction (subject to the multiplicity variation) to which the AEI
$C$ is attached. If $C$ is in $\cala$, then $K$ is $\calp$-compatible with $C$ by virtue of condition~1
applied to $\cala$. If $C$ is an additional AEI, then $C$ is of the same type as an AEI $C'$ of $\cala$ that
is attached to $K$ through the same and-interaction. By virtue of condition~1, $K$ is $\calp$-compatible
with $C'$, from which we derive that $K$ is $\calp$-compatible with $C$.

\item $K$ is in $\cala$ and has an or-interaction (subject to the multiplicity variation) to which the AEI
$C$ is attached. First, assume that $C$ is in $\cala$. By virtue of condition~$\widetilde{\mathit{4}}$, both
in $\cala$ and in $\cala'$ the AEI $C$ does not raise any exception because of the attachments between $K$
and any other AEI attached to the or-interaction. Hence, $K$ is $\calp$-compatible with $C$ by virtue of
condition~1 applied to $\cala$. Second, assume that $C$ is an additional AEI. In this case, we can apply the
same argument, observing that $C$ is of the same type as an AEI $C'$ of $\cala$ that is attached to $K$
through the same or-interaction.

\item $C$ is in $\cala$ and has an and-interaction (subject to the multiplicity variation) to which $K$ is
attached. If $K$ is in $\cala$, then $K$ is $\calp$-compatible with $C$ by virtue of condition~1 applied to
$\cala$. If $K$ is an additional AEI, then it is of the same type as an AEI $K'$ of $\cala$ that is attached
to $C$ through the same and-interaction. By virtue of condition~1, $K'$ is $\calp$-compatible with $C$, from
which we derive that $K$ is $\calp$-compatible with $C$.

\item $C$ is in $\cala$ and has an or-interaction (subject to the multiplicity variation) to which the AEI
$K$ is attached. First, assume that $K$ is in $\cala$. By virtue of condition~$\widetilde{\mathit{4}}$, both
in $\cala$ and in $\cala'$ the AEI $K$ does not raise any exception because of the attachments between $C$
and any other AEI attached to the or-interaction. Moreover, by virtue of condition~$\widetilde{\mathit{5}}$,
$K$ eventually communicates with $C$ through the or-interaction of $C$. Hence, $K$ is $\calp$-compatible
with $C$ by virtue of condition~1 applied to $\cala$. Second, assume that $K$ is an additional AEI. In this
case, we can apply the same argument, observing that $K$ is of the same type as an AEI $K'$ of $\cala$ that
is attached to $C$ through the same or-interaction.

				\end{itemize}

\item $\cala'$ satisfies condition~2 because, by virtue of condition~$\widetilde{\mathit{6}}$, the set of
cyclic unions generated by $\kappa$ for~$\cala'$ is the same as the one generated by $\kappa$ for $\cala$.
\fullbox

			\end{itemize}

		\end{proof}


\newpage


\centerline{ORIGINAL PAPER}

\centerline{Handling Communications in Process Algebraic Architectural Description Languages:}
\centerline{Modeling, Verification, and Implementation}

\centerline{Marco Bernardo \quad Edoardo Bont\`a \quad Alessandro Aldini}

\begin{abstract}
Architectural description languages are a useful tool for modeling complex software systems at a high level
of abstraction. If based on formal methods, they can also serve for enabling the early verification of
various properties such as component coordination and for guiding the synthesis of code correct by
construction. This is the case with process algebraic architectural description languages, which are process
calculi enhanced with the main architectural concepts. However, the techniques with which those languages
have been equipped are mainly conceived to work with synchronous communications only. The objective of this
paper is threefold. On the modeling side, we show how to enhance the expressiveness of a typical process
algebraic architectural description language by including the capability of representing nonsynchronous
communications in such a way that the usability of the original language is preserved. On the verification
side, we show how to modify techniques for analyzing the absence of coordination mismatches like the
compatibility check for acyclic topologies and the interoperability check for cyclic topologies in such a
way that those checks are valid also for nonsynchronous communications. On the implementation side, we show
how to generate multithreaded object-oriented software in the presence of synchronous and nonsynchronous
communications in such a way that the properties proved at the architectural level are preserved at the code
level.
\end{abstract}

\section{Introduction}\label{intro}

\noindent
The growing complexity and the increasing size of modern software systems can be managed by adopting
notations for formal or semi-formal system modeling (model-driven approach). In this way, design documents
with a precise syntax can be produced and shared by all the people contributing to system development. In
order to avoid delays and cost increases due to the late discovery of errors in the development process,
another task that such notations should carry out is to enable the rigorous and hopefully automated analysis
of system properties and to guide the synthesis of code correct by construction. For instance, it is widely
recognized that property verification finds its own rightful place in the architectural design
phase~\cite{SG,BI}. The reason is that this phase precedes system implementation and provides support for
declarative/behavioral/topological system models that are complete at a high level of abstraction.

Many architectural description languages have been proposed. Some of them -- like, e.g.,
Wright~\cite{AG,ADG}, Darwin/FSP~\cite{MDEK,MK}, LEDA~\cite{CPT1}, PADL~\cite{AB}, and $\pi$-ADL~\cite{Oqu}
-- are based on process algebra~\cite{Mil,Hoa,BPS} due to its support to compositional modeling. It is worth
noting that process algebra is compositional, but not component-oriented. Thus, from the point of view of
process algebra, its architectural versions are a significant step forward in terms of usability. In fact,
they give special prominence to the main architectural concepts -- components, connectors, and styles --
while hiding the process algebraic technicalities to the software developer.

On the modeling side, this architectural upgrade has three important consequences. First, it permits to
describe the behavior of the components separately from the representation of the system topology, thus
overcoming the modeling difficulties deriving from the direct use of certain process algebraic operators
like, e.g., parallel composition. Second, it highlights the interactions among components and the
classification of their communications, thus allowing for static checks establishing system model
well-formedness. Third, it fosters the reuse of the specification of single components as well as of
complete systems, thus supporting the compositional and hierarchical modeling of entire system families.

On the verification side, process algebraic architectural description languages inherit all the techniques
applicable to process algebra, like model checking~\cite{CGP} and equivalence checking~\cite{CS}. In
addition, such languages are equipped with ad-hoc analysis techniques (see, e.g., \cite{AG,IWY,CPT2,AB})
mostly based on behavioral equivalences~\cite{Gla}, which are useful for detecting coordination mismatches
that may arise when assembling together components that are correct if taken in isolation. Moreover, they
can generate diagnostic information for pinpointing components responsible for mismatches.

The ad-hoc analysis techniques proposed in the literature deal only with synchronous communications. In that
setting, all ports of software components are blocking. A component waiting on a synchronous input port
cannot proceed until an output is sent by another component. Similarly, a component issuing an output via a
synchronous output port cannot proceed until another component is willing to receive. The limitation to
synchronous communications is not so restrictive for usual properties like deadlock freedom, which should
hold when no communication is blocking. In contrast, the validity of other properties related to activity
sequencing or message ordering may not be guaranteed in the presence of nonsynchronous communications.

In order to address the usual properties in a more general setting as well as the other properties mentioned
above, the first contribution of this paper is to show how to enhance the expressiveness of a typical
process algebraic architectural description language by including the capability of representing
nonsynchronous communications in such a way that the usability of the original language is preserved. More
specifically, we focus on PADL~\cite{AB} and we extend it by means of additional qualifiers useful to
distinguish among synchronous, semi-synchronous, and asynchronous ports.

Semi-synchronous ports are not blocking. A semi-synchronous port of a component succeeds if there is another
component ready to communicate with it, otherwise it raises an exception so as not to block the component to
which it belongs. For example, a semi-synchronous input port can be used to model accesses to a tuple space
via input or read probes~\cite{Gel}. A semi-synchronous output port can instead be used to model the
interplay between a graphical user interface and an underlying application, as the former must not block
whenever the latter cannot do certain tasks requested by the user.

Analogously, asynchronous ports are not blocking. Here the reason is that the beginning and the end of the
communications in which these ports are involved are completely decoupled. For instance, an asynchronous
output port can be used to model output operations on a tuple space. An asynchronous input port can instead
be used to model the periodical check for the presence of information received from an event notification
service~\cite{CRW}.

In the extended language, semi-synchronous ports can be easily handled with suitable semantic rules
generating exceptions whenever necessary, whereas asynchronous ports require the addition of implicit
repository-like components. In any case, the semantic treatment of nonsynchronous communications is
completely transparent to PADL users, as they only have to specify suitable synchronicity-related qualifiers
in their architectural descriptions. Therefore, the degree of usability of the original language is
unaffected.

The second contribution of this paper is to show how to modify techniques for analyzing the absence of
coordination mismatches -- like the compatibility check for acyclic topologies and the interoperability
check for cyclic topologies introduced in~\cite{AB} -- in such a way that those checks can still be applied
in the presence of nonsynchronous communications.

This is accomplished by viewing certain activities carried out through semi-synchronous and asynchronous
ports as internal activities when performing the above mentioned checks. The reason is that each such
activity has a specific outcome and takes place at a specific time instant when considered from the point of
view of the individual component executing that activity. However, in the overall architecture the same
activity can raise an exception (if the port is semi-synchronous and the other ports are not ready to
communicate with it) or can be delayed (if the port is asynchronous and the communication is buffered).
Thus, if we do not regard exceptions and all the activities carried out through asynchronous ports as
internal activities at verification time, the compatibility or interoperability check may fail even in the
absence of a real coordination mismatch.

The third contribution of this paper is to show how to generate multithreaded object-oriented software from
process algebraic architectural descriptions including various kinds of communications in such a way that
the properties proved at the architectural level are preserved at the code level. This last contribution is
related to one of the big issues in the software engineering field, i.e., guaranteeing that the
implementation of a software system complies with its architectural description~\cite{Gar}. Indeed, the
purpose of automatic code generation should be not only to speed up system implementation, but also to
ensure conformance by construction.

In order to bridge the gap between system modeling/verification and system implementation, we propose an
approach that automatically synthesizes multithreaded Java programs from PADL descriptions containing an
arbitrary combination of synchronous, semi-synchronous, and asynchronous ports. The choice of Java as target
language is due to the fact that Java offers a set of mechanisms for the well-structured management of
threads and their shared data, which should simplify the code generation task. Moreover, its object-oriented
nature -- and specifically its encapsulation capability -- makes Java an appropriate candidate for coping
with the high level of abstraction of process algebraic architectural descriptions during code generation.

The proposed approach is divided into two phases. In the first phase, we develop an architecture-driven
technique for thread coordination management. Similar to previous work (see, e.g., \cite{PR}), we advocate
the provision of a suitable software package that takes care of the details of thread coordination by means
of architecture-inspired units in a way that is completely transparent to the software developer. The
distinguishing feature of the first phase of our proposal is that also the employment of the package should
follow the same architecture-centric spirit. In other words, the use of the package units should be guided
by the architectural description of the system to be developed, as this description is a well suited tool
for achieving correct thread coordination in the case of concurrent object-oriented programs.

In the second phase, we handle the translation of the process algebraically specified behavior of individual
software components into threads. The separation of thread behavior generation from thread coordination
management turns out to be particularly appropriate in order to limit human intervention. In fact, while a
completely automated and architecture-driven technique can guarantee correct thread coordination, only a
partial translation based on stubs is possible for the generation of threads. In addition to the
considerations of~\cite{MK}, in which it is shown how a disciplined process algebraic modeling is beneficial
at subsequent stages, we also provide a set of guidelines for filling in stubs, which guarantee the
preservation at the code level of properties proved on the architectural description of the system under
construction.

This paper, which is a full and revised version of~\cite{BB3,BB1,BB2}, is organized as follows. After
recalling PADL in Sect.~\ref{padl}, in Sect.~\ref{nonsync} we extend its syntax with semi-synchronous and
asynchronous ports and we consequently revise its semantics. A running example based on a client-server
system is used throughout both sections. In Sect.~\ref{checks}, we modify the architectural compatibility
and interoperability checks in order to deal with nonsynchronous communications as well. The modified checks
are illustrated on the architectural description of an applet-based simulator for a cruise control system.
In Sect.~\ref{code}, we present the two-phase approach for synthesizing multithreaded Java software from
PADL descriptions including synchronous and nonsynchronous communications. The approach is exemplified by
means of the same cruise control system simulator as the previous section. Finally, in Sect.~\ref{concl} we
provide some concluding remarks and directions for future work.

\section{The Architectural Description Language PADL}\label{padl}

\noindent
PADL is a process algebraic architectural description language. In this section, after recalling some basic
notions of process algebra (Sect.~\ref{pa}), we present PADL syntax (Sect.~\ref{padlnot}) and semantics
(Sect.~\ref{padlsem}) by illustrating them through a client-server running example. For a complete
presentation and comparisons with related work, the interested reader is referred to~\cite{AB}.

\subsection{Process Algebra}\label{pa}

\noindent
Process calculi~\cite{Mil,Hoa,BPS} provide a set of operators by means of which the behavior of a system can
be described in an action-based, compositional way. Given a set $\ms{Name} = \ms{Name}_{\rm v} \cup \{ \tau
\}$ of action names including $\tau$ for denoting an invisible action, together with a set $\ms{Relab} = \{
\varphi : \ms{Name} \rightarrow{}{} \ms{Name} \mid \varphi^{-1}(\tau) = \{ \tau \} \}$ of relabeling
functions preserving action visibility, we consider a process calculus PA with the following process term
syntax:
\[\begin{array}{|rclclcl|}
\hline
P & \!\! ::= \!\! & \nil & & \textrm{inactive process} & & \\
& \!\! | \!\! & B & & \textrm{process constant} & & (B \eqdef P) \\[0.1cm]
& \!\! | \!\! & a \, . \, P & & \textrm{action prefix} & & (a \in \ms{Name}) \\[0.1cm]
& \!\! | \!\! & P + P & & \textrm{alternative composition} & & \\[0.1cm]
& \!\! | \!\! & P \pco{S} P & & \textrm{parallel composition} & & (S \subseteq \ms{Name}_{\rm v}) \\[0.1cm]
& \!\! | \!\! & P / H & & \textrm{hiding} & & (H \subseteq \ms{Name}_{\rm v}) \\[0.1cm]
& \!\! | \!\! & P[\varphi] & & \textrm{relabeling} & & (\varphi \in \ms{Relab}) \\
\hline
\end{array}\]

Operational semantic rules map every process term $P$ of PA to a state-transition graph $\lsp P \rsp$ called
labeled transition system. In this graph, each state corresponds to a process term derivable from $P$, the
initial state corresponds to $P$, and each transition is labeled with the corresponding action.

Observed that no rule is necessary for the inactive process $\nil$ -- as $\lsp \nil \rsp$ must be a
single-state graph with no transitions -- the operational semantic rules for dynamic operators (action
prefix and alternative composition) and process constants are the following:
\[\begin{array}{|c|}
\hline
a \, . \, P \arrow{a}{} P \hspace{1.0cm}
{\infr{B \eqdef P \hspace{0.5cm} P \arrow{a}{} P'}{B \arrow{a}{} P'}} \\[0.7cm]
{\infr{P_{1} \arrow{a}{} P'}{P_{1} + P_{2} \arrow{a}{} P'}} \hspace{1.0cm}
{\infr{P_{2} \arrow{a}{} P'}{P_{1} + P_{2} \arrow{a}{} P'}} \\[0.7cm]
\hline
\end{array}\]
Process $a \, . \, P$ can execute an action with name $a$ and then behaves as $P$. Process $P_{1} + P_{2}$
behaves as either $P_{1}$ or $P_{2}$ depending on which of them executes an action first (nondeterministic
choice). Constant $B$ behaves as the process term occurring in its possibly recursive defining equation.

The operational semantic rules for static operators (parallel composition, hiding, and relabeling) are the
following:
\[\begin{array}{|c|}
\hline
{\infr{P_{1} \arrow{a}{} P'_{1} \hspace{0.5cm} a \notin S}{P_{1} \pco{S} P_{2} \arrow{a}{} P'_{1} \pco{S}
P_{2}}} \hspace{1.0cm}
{\infr{P_{2} \arrow{a}{} P'_{2} \hspace{0.5cm} a \notin S}{P_{1} \pco{S} P_{2} \arrow{a}{} P_{1} \pco{S}
P'_{2}}} \\[0.7cm]
{\infr{P_{1} \arrow{a}{} P'_{1} \hspace{0.4cm} P_{2} \arrow{a}{} P'_{2} \hspace{0.4cm} a \in S}{P_{1}
\pco{S} P_{2} \arrow{a}{} P'_{1} \pco{S} P'_{2}}} \\[0.7cm]
{\infr{P \arrow{a}{} P' \hspace{0.5cm} a \in H}{P / H \arrow{\tau}{} P' / H}} \hspace{1.0cm}
{\infr{P \arrow{a}{} P' \hspace{0.5cm} a \notin H} {P / H \arrow{a}{} P' / H}} \\[0.7cm]
{\infr{P \arrow{a}{} P'}{P[\varphi] \arrow{\varphi(a)}{} P'[\varphi]}} \\
\hline
\end{array}\]
Process $P_{1} \pco{S} P_{2}$ behaves as $P_{1}$ in parallel with $P_{2}$ as long as actions are executed
whose name does not belong to $S$. In contrast, synchronizations are forced between any action executed by
$P_{1}$ and any action executed by $P_{2}$ that have the same name belonging to $S$. Process $P / H$ behaves
as $P$ with all executed actions occurring in $H$ made invisible. Process $P[\varphi]$ behaves as $P$ with
all executed actions relabeled via $\varphi$.

Process terms are compared and manipulated by means of behavioral equivalences~\cite{Gla}. Among the various
approaches, for PA we consider weak bisimilarity, according to which two process terms are equivalent if
they are able to mimic each other's visible behavior stepwise~\cite{Mil}.

Denoted by $\warrow{}{}$ the extension of $\arrow{}{}$ to action sequences, we say that a symmetric relation
$\calr$ is a weak bisimulation iff for all $(P_{1}, P_{2}) \in \calr$: \textit{(i)}~whenever $P_{1}
\arrow{a}{} P'_{1}$ for $a \in \ms{Name}_{\rm v}$, then $P_{2} \warrow{\tau^{*} a \tau^{*}}{} P'_{2}$ and
$(P'_{1}, P'_{2}) \in \calr$; \textit{(ii)}~whenever $P_{1} \arrow{\tau}{} P'_{1}$, then $P_{2}
\warrow{\tau^{*}}{} P'_{2}$ and $(P'_{1}, P'_{2}) \in \calr$. Weak bisimilarity $\wbis{\rm B}$, defined as
the union of all the weak bisimulations, is a congruence with respect to all the operators except for
alternative composition and has a modal characterization based on a weak variant of Hennessy-Milner logic.

\subsection{PADL Textual and Graphical Notations}\label{padlnot}

\noindent
A PADL description represents an architectural type, which is a family of software systems sharing certain
constraints on the observable behavior of their components as well as on their topology.

The textual description of an architectural type starts with the name and the formal parameters (initialized
with default values) of the architectural type. The textual description then comprises two sections, as
shown below:
{\small\[\begin{array}{|ll|}
\hline
\texttt{ARCHI\_TYPE} & \triangleleft \textit{name and initialized formal parameters} \triangleright
\\[0.2cm]
\hspace*{0.4cm} \texttt{ARCHI\_BEHAVIOR} & \\[-0.1cm]
\hspace*{1.5cm} \vdots & \hspace{0.4cm} \vdots \\
\hspace*{0.8cm} \texttt{ARCHI\_ELEM\_TYPE} & \triangleleft \textit{AET name and formal parameters}
\triangleright \\[0.05cm]
\hspace*{1.2cm} \texttt{BEHAVIOR} & \triangleleft \textit{sequence of PA defining equations built from} \\
& \tabspace{\triangleleft}
\textit{stop, action prefix, choice, and recursion} \triangleright \\
\hspace*{1.2cm} \texttt{INPUT\_INTERACTIONS} & \triangleleft \textit{input uni/and/or-interactions}
\triangleright \\
\hspace*{1.2cm} \texttt{OUTPUT\_INTERACTIONS} \hspace{0.1cm} & \triangleleft \textit{output
uni/and/or-interactions} \triangleright \\[-0.1cm]
\hspace*{1.5cm} \vdots & \hspace{0.4cm} \vdots \\[0.2cm]
\hspace*{0.4cm} \texttt{ARCHI\_TOPOLOGY} & \\[0.05cm]
\hspace*{0.8cm} \texttt{ARCHI\_ELEM\_INSTANCES} & \triangleleft \textit{AEI names and actual parameters}
\triangleright \\
\hspace*{0.8cm} \texttt{ARCHI\_INTERACTIONS} & \triangleleft \textit{architecture-level AEI interactions}
\triangleright \\
\hspace*{0.8cm} \texttt{ARCHI\_ATTACHMENTS} & \triangleleft \textit{attachments between AEI local
interactions} \triangleright \\[0.2cm]
\texttt{END} & \\
\hline
\end{array}\]}

The first section defines the behavior of the system family by means of types of software components and
connectors, which are collectively called architectural element types. The definition of an AET starts with
its name and its formal parameters and consists of the specification of its behavior and its interactions.

The behavior of an AET has to be provided in the form of a sequence of defining equations written in a
verbose variant of PA allowing only for the inactive process (rendered as \texttt{stop}), the value-passing
action prefix operator with a possible boolean guard condition, the alternative composition operator
(rendered as \texttt{choice}), and recursion (behavioral invocations).

The interactions are those actions occurring in the process algebraic specification of the behavior that act
as interfaces for the AET, while all the other actions are assumed to represent internal activities. Each
interaction has to be equipped with two qualifiers. The first one establishes whether the interaction is an
input or output interaction, whereas the second one describes the multiplicity of the communications in
which the interaction can be involved.

We distinguish among uni-interactions mainly involved in one-to-one communications (qualifier \texttt{UNI}),
and-interactions guiding inclusive one-to-many communications like multicasts (qualifier \texttt{AND}),
or-interactions guiding selective one-to-many communications like those between a server and its clients
(qualifier \texttt{OR}). It can also be established that an output or-interaction depends on an input
or-interaction, in order to guarantee that a selective one-to-many output is sent to the same element from
which the last selective many-to-one input was received (keyword \texttt{DEP}).

The second section of the PADL description defines the topology of the system family. It is composed of
three subsections. First, we have the declaration of the instances of the AETs -- called AEIs -- which
represent the actual system components and connectors, together with their actual parameters. Then, we have
the declaration of the architectural (as opposed to local) interactions, which are some of the interactions
of the AEIs that act as interfaces for the whole systems of the family. Finally, we have the declaration of
the architectural attachments among the local interactions of the AEIs, which make the AEIs communicate with
each other.

An attachment is admissible only if it goes from a local output interaction of an AEI to a local input
interaction of another AEI. Moreover, a local uni-interaction can be attached to only one local interaction,
whereas a local and-/or-interaction can be attached to (several) local uni-interactions only.

Besides the textual notation, PADL comes equipped with a graphical notation that is an extension of the flow
graph notation~\cite{Mil}. In an enriched flow graph, AEIs are depicted as boxes, local (resp.\
architectural) interactions are depicted as small black circles (resp.\ white squares) on the box border,
and attachments are depicted as directed edges between pairs each composed of a local output interaction and
a local input interaction. The small circle/square of an interaction is extended with a triangle (resp.\
bisected triangle) outside the AEI box if the interaction is an and-interaction (resp.\ or-interaction).
Or-dependences are depicted as dotted edges.

	\begin{example}\label{cssyn}

Suppose we need to model a scenario in which there is a server that can be contacted at any time by two
identically behaving clients. Assume that the server has no buffer for holding incoming requests and that,
after sending a request, a client cannot proceed until it receives a response from the server. The server
AET can be defined as follows:

		{\small\begin{verbatim}
 ARCHI_ELEM_TYPE Server_Type(void)
   BEHAVIOR
     Server(void; void) =
       receive_request . compute_response . send_response . Server()
   INPUT_INTERACTIONS  OR receive_request
   OUTPUT_INTERACTIONS OR send_response   DEP receive_request
		\end{verbatim}}

\noindent
where \texttt{compute\_response} is an internal action, while \texttt{send\_response} is declared to depend
on \texttt{receive\_request} in order to make sure that each response is sent back to the client that issued
the corresponding request. Since the behavior of the two clients is identical, a single client AET suffices:

		{\small\begin{verbatim}
 ARCHI_ELEM_TYPE Client_Type(void)
   BEHAVIOR
     Client(void; void) =
       process . send_request . receive_response . Client()
   INPUT_INTERACTIONS  UNI receive_response
   OUTPUT_INTERACTIONS UNI send_request
		\end{verbatim}}

\noindent
where \texttt{process} is an internal action. Finally, we declare the topology of the system as follows by
using the dot notation for the local interactions:

		{\small\begin{verbatim}
 ARCHI_TOPOLOGY
   ARCHI_ELEM_INSTANCES
     S   : Server_Type();
     C_1 : Client_Type();
     C_2 : Client_Type()
   ARCHI_INTERACTIONS
     void
   ARCHI_ATTACHMENTS
     FROM C_1.send_request TO S.receive_request;
     FROM C_2.send_request TO S.receive_request;
     FROM S.send_response  TO C_1.receive_response;
     FROM S.send_response  TO C_2.receive_response
		\end{verbatim}}

\noindent
The topology is better illustrated by the following enriched flow graph: \\[0.2cm]
\centerline{\hspace{0.3cm}\includegraphics{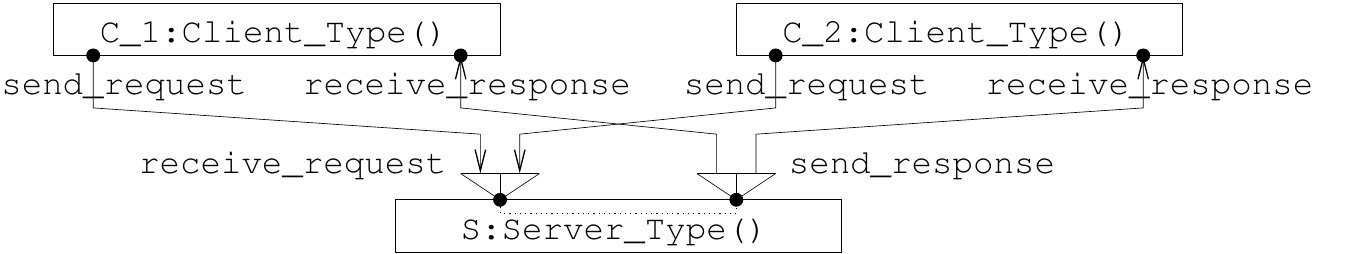}}
where the dotted edge linking the bisected triangles is the or-dependence.
\fullbox

	\end{example}

\subsection{The Semantics for PADL}\label{padlsem}

\noindent
The semantics for PADL is given by translation into PA. The first step of the translation focuses on the
semantics of each AEI. This is defined as the behavior of the corresponding AET with all the action
occurrences being preceded by the name of the AEI and the AET formal data parameters being substituted for
by the corresponding AEI actual data parameters.

Let $\calc$ be an AET with $m \in \natns_{\ge 0}$ formal parameters $\ms{fp}_{1}, \dots, \ms{fp}_{m}$ and
behavior given by a sequence $\cale$ of defining equations. Then the semantics of an AEI $C$ of type $\calc$
with actual parameters $\ms{ap}_{1}, \dots, \ms{ap}_{m}$ is defined as follows, where $C \, . \, \_$
introduces the dot notation on actions and $\{ \_ \hookrightarrow \_, \dots, \_ \hookrightarrow \_ \}$
denotes a syntactical substitution:
\[\begin{array}{|l|}
\hline
\lsp C \rsp \: = \: C . \cale \{ \ms{ap}_{1} \hookrightarrow \ms{fp}_{1}, \dots, \ms{ap}_{m} \hookrightarrow
\ms{fp}_{m} \} \\
\hline
\end{array}\]

If the AEI contains local or-interactions, each of them -- consistent with the fact that it guides a
selective one-to-many communication -- must be replaced by as many fresh local uni-interactions as there are
attachments involving the considered or-interaction. The fresh local uni-interactions are then attached to
the local uni-interactions of other AEIs to which the local or-interaction was originally attached, as shown
below:
\\[0.2cm]
\centerline{\includegraphics{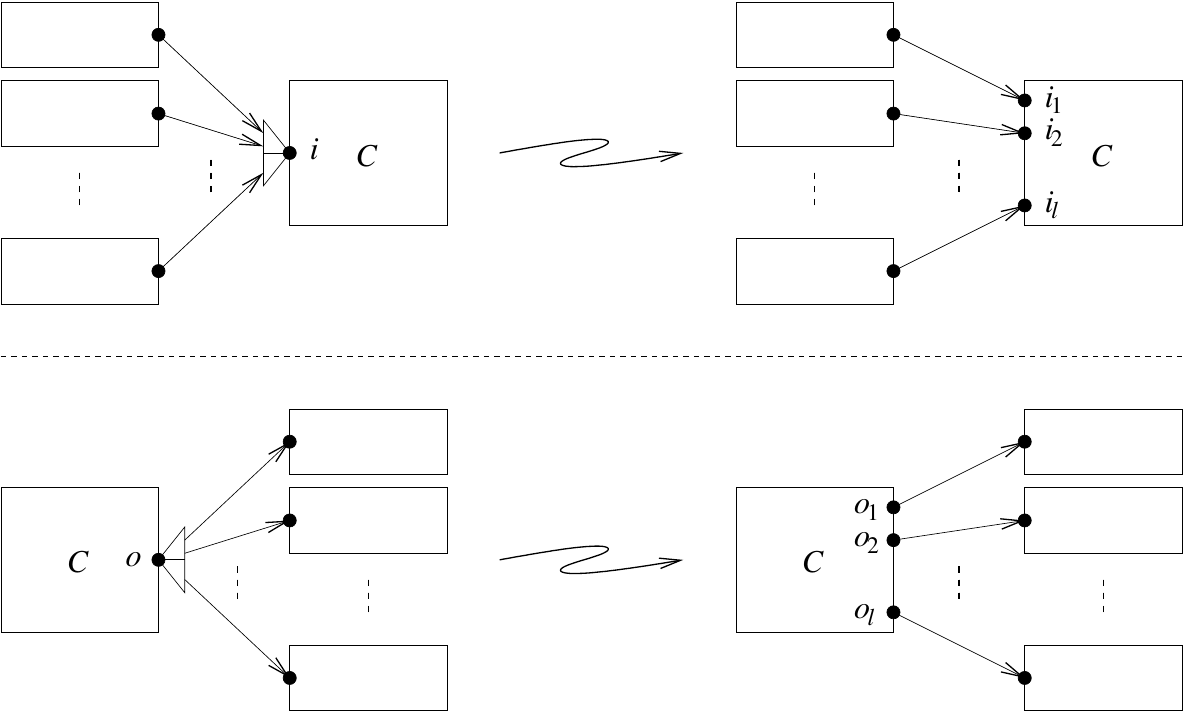}}
In that case, the semantics of $C$ is defined as follows:
\[\begin{array}{|l|}
\hline
\lsp C \rsp \: = \: \ms{or-rewrite}_{\emptyset}(C . \cale \{ \ms{ap}_{1} \hookrightarrow \ms{fp}_{1}, \dots,
\ms{ap}_{m} \hookrightarrow \ms{fp}_{m} \}) \\
\hline
\end{array}\]

In order to reflect the fact that an or-interaction can result in several alternative communications,
function $\ms{or-rewrite}$ inductively rewrites the body of any resulting defining equation by replacing
with fresh uni-interactions each occurrence of any local or-interaction whose number $\ms{attach-no}(\_)$ of
attachments is greater than~1. Or-dependences are dealt with by keeping track of the set $\ms{FI}$ of fresh
input uni-interactions currently in force that arise from input or-interactions on which some output
or-interaction depends. Here are the three basic clauses of the inductive definition of $\ms{or-rewrite}$:

	\begin{itemize}

\item If $a$ is either a local input or-interaction of $C$ on which no local output or-interaction depends,
or a local output or-interaction that does not depend on any local input or-interaction, and
$\ms{attach-no}(C.a) = l \ge 2$:
\[\begin{array}{|l|}
\hline
\ms{or-rewrite}_{\ms{FI}}(a \, . \, P) \: = \: \texttt{choice} \{ a_{1} \, . \,
\ms{or-rewrite}_{\ms{FI}}(P), \\[-0.2cm]
\tabspace{\ms{or-rewrite}_{\ms{FI}}(a \, . \, P) \: = \: \texttt{choice} \{}
\hspace*{1.0cm} \vdots \\[-0.2cm]
\tabspace{\ms{or-rewrite}_{\ms{FI}}(a \, . \, P) \: = \: \texttt{choice} \{}
a_{l} \, . \, \ms{or-rewrite}_{\ms{FI}}(P) \} \\
\hline
\end{array}\]

\item If $i$ is a local input or-interaction of $C$ on which a local output or-interaction depends and
$\ms{attach-no}(C.i) = l \ge 2$:
\[\begin{array}{|l|}
\hline
\ms{or-rewrite}_{\ms{FI}}(i \, . \, P) \: = \: \texttt{choice} \{ i_{1} \, . \, \ms{or-rewrite}_{\ms{FI} \,
\cup \, \{ i_{1} \} \, - \, \{ i_{j} \mid 1 \le j \le l \land j \neq 1 \}}(P), \\[-0.2cm]
\tabspace{\ms{or-rewrite}_{\ms{FI}}(i \, . \, P) \: = \: \texttt{choice} \{}
\hspace*{1.0cm} \vdots \\[-0.2cm]
\tabspace{\ms{or-rewrite}_{\ms{FI}}(i \, . \, P) \: = \: \texttt{choice} \{}
i_{l} \, . \, \ms{or-rewrite}_{\ms{FI} \, \cup \, \{ i_{l} \} \, - \, \{ i_{j} \mid 1 \le j \le l \land j
\neq l \}}(P) \} \\
\hline
\end{array}\]

\item If $o$ is a local output or-interaction of $C$ that depends on the local input or-interaction $i$ and
$\ms{attach-no}(C.i) = \ms{attach-no}(C.o) \ge 2$ and $i_{j} \in \ms{FI}$:
\[\begin{array}{|l|}
\hline
\ms{or-rewrite}_{\ms{FI}}(o \, . \, P) \: = \: o_{j} \, . \, \ms{or-rewrite}_{\ms{FI}}(P) \\
\hline
\end{array}\]

	\end{itemize}

	\begin{example}\label{cssem1}

Consider the client-server system described in Ex.~\ref{cssyn}. Then $\lsp \texttt{C\_1} \rsp$ and $\lsp
\texttt{C\_2} \rsp$ coincide with the defining equation for \texttt{Client}, where action names are preceded
by \texttt{C\_1} and \texttt{C\_2}, respectively. In contrast, $\lsp \texttt{S} \rsp$ is given by the
following defining equation obtained from the one for \texttt{Server} after rewriting the occurring
or-interactions:

		{\footnotesize\begin{verbatim}
RewSer(void; void) =
 choice
 {
  S.receive_request_1 . S.compute_response . S.send_response_1 . RewSer(),
  S.receive_request_2 . S.compute_response . S.send_response_2 . RewSer()
 }
		\end{verbatim}}

\vspace{-0.9cm}
\fullbox

	\end{example}

The second step of the translation defines the meaning of any set of AEIs $\{ C_{1}, \dots, C_{n} \}$ and
hence of an entire architectural description. Fixed an AEI $C_{j}$ in the set, let $\calli_{C_{j}}$ be the
set of local interactions of $C_{j}$ and $\calli_{C_{j}; C_{1}, \dots, C_{n}} \subseteq \calli_{C_{j}}$ be
the set of local interactions of $C_{j}$ attached to $\{ C_{1}, \dots, C_{n} \}$.

In order to make the process terms representing the semantics of these AEIs communicate in the presence of
attached interactions having different names -- in PA only actions with the same name can synchronize -- we
need a set $\cals(C_{1}, \dots, C_{n})$ of fresh action names, one for each pair of attached local
uni-interactions in $\{ C_{1}, \dots, C_{n} \}$ and for each set of local uni-interactions attached to the
same local and-interaction in $\{ C_{1}, \dots, C_{n} \}$. In order to ensure the uniqueness of the elements
of $\cals(C_{1}, \dots, C_{n})$, each of them is built by concatenating through symbol \# all the original
names in the corresponding maximal set of attached local interactions. Then, we need suitable injective
relabeling functions $\varphi_{C_{j}; C_{1}, \dots, C_{n}}$ mapping each set $\calli_{C_{j}; C_{1}, \dots,
C_{n}}$ to $\cals(C_{1}, \dots, C_{n})$ in such a way that:
\[\begin{array}{|l|}
\hline
\varphi_{C_{j}; C_{1}, \dots, C_{n}}(a_{1}) \: = \: \varphi_{C_{g}; C_{1}, \dots, C_{n}}(a_{2}) \\
\hline
\end{array}\]
if and only if $C_{j}.a_{1}$ and $C_{g}.a_{2}$ are attached to each other or to the same and-interaction.

The interacting semantics of $C_{j}$ with respect to $\{ C_{1}, \dots, C_{n} \}$ is defined as follows:
\[\begin{array}{|l|}
\hline
\lsp C_{j} \rsp_{C_{1}, \dots, C_{n}} \: = \: \lsp C_{j} \rsp [\varphi_{C_{j}; C_{1}, \dots, C_{n}}] \\
\hline
\end{array}\]
In general, the interacting semantics of $\{ C'_{1}, \dots, C'_{n'} \} \subseteq \{ C_{1}, \dots, C_{n} \}$
with respect to $\{ C_{1}, \dots, C_{n} \}$ is defined as follows (assuming parallel composition to be left
associative):
\[\begin{array}{|l|}
\hline
\lsp C'_{1}, \dots, C'_{n'} \rsp_{C_{1}, \dots, C_{n}} \: = \: \lsp C'_{1} \rsp_{C_{1}, \dots, C_{n}}
\pco{\cals(C'_{1}, C'_{2}; C_{1}, \dots, C_{n})} \\[0.1cm]
\hspace*{4.6cm} \lsp C'_{2} \rsp_{C_{1}, \dots, C_{n}} \pco{\cals(C'_{1}, C'_{3}; C_{1}, \dots, C_{n}) \cup
\cals(C'_{2}, C'_{3}; C_{1}, \dots, C_{n})} \; \dots \\[0.1cm]
\hspace*{4.9cm} \dots \; \pco{\mathop{\cup}\limits_{i = 1}^{n' - 1} \cals(C'_{i}, C'_{n'}; C_{1}, \dots,
C_{n})} \; \lsp C'_{n'} \rsp_{C_{1}, \dots, C_{n}} \\
\hline
\end{array}\]
where $\cals(C'_{j}, C'_{g}; C_{1}, \dots, C_{n}) = \cals(C'_{j}; C_{1}, \dots, C_{n}) \cap \cals(C'_{g};
C_{1}, \dots, C_{n})$ is the pairwise synchronization set of $C'_{j}$ and $C'_{g}$ with respect to $\{
C_{1}, \dots, C_{n} \}$, with $\cals(C'_{j}; C_{1}, \dots, C_{n}) = \varphi_{C'_{j}; C_{1}, \dots,
C_{n}}(\calli_{C'_{j}; C_{1}, \dots, C_{n}})$ being the synchronization set of $C'_{j}$ with respect to $\{
C_{1}, \dots, C_{n} \}$.

Finally, the semantics of an architectural type $\cala$ formed by the set of AEIs $\{ C_{1}, \dots, C_{n}
\}$ is defined as follows:
\[\begin{array}{|l|}
\hline
\lsp \cala \rsp \: = \: \lsp C_{1}, \dots, C_{n} \rsp_{C_{1}, \dots, C_{n}} \\
\hline
\end{array}\]

	\begin{example}\label{cssem2}

Consider again the client-server system of Ex.~\ref{cssyn}. Then the semantics of the entire description is
given by the following process term:
{\small\cws{0}{\begin{array}{l}
\lsp \texttt{S} \rsp [\texttt{receive\_request\_1} \mapsto
\texttt{C\_1.send\_request\#S.receive\_request\_1}, \\
\tabspace{\lsp \texttt{S} \rsp \;}
\texttt{send\_response\_1} \mapsto \texttt{S.send\_response\_1\#C\_1.receive\_response}, \\
\tabspace{\lsp \texttt{S} \rsp \;}
\texttt{receive\_request\_2} \mapsto \texttt{C\_2.send\_request\#S.receive\_request\_2}, \\
\tabspace{\lsp \texttt{S} \rsp \;}
\texttt{send\_response\_2} \mapsto \texttt{S.send\_response\_2\#C\_2.receive\_response}] \\
\hspace*{1.5cm} \pco{\{ {\tt C\_1.send\_request\#S.receive\_request\_1},} \\[-0.1cm]
\tabspace{\hspace*{1.5cm} \pco{\{}}
_{{\tt S.send\_response\_1\#C\_1.receive\_response} \}} \\[0.1cm]
\lsp \texttt{C\_1} \rsp [\texttt{send\_request} \mapsto \texttt{C\_1.send\_request\#S.receive\_request\_1},
\\
\tabspace{\lsp \texttt{C\_1} \rsp \;}
\texttt{receive\_response} \mapsto \texttt{S.send\_response\_1\#C\_1.receive\_response}] \\
\hspace*{1.5cm} \pco{\{ {\tt C\_2.send\_request\#S.receive\_request\_2},} \\[-0.1cm]
\tabspace{\hspace*{1.5cm} \pco{\{}}
_{{\tt S.send\_response\_2\#C\_2.receive\_response} \}} \\[0.1cm]
\lsp \texttt{C\_2} \rsp [\texttt{send\_request} \mapsto \texttt{C\_2.send\_request\#S.receive\_request\_2},
\\
\tabspace{\lsp \texttt{C\_2} \rsp \;}
\texttt{receive\_response} \mapsto \texttt{S.send\_response\_2\#C\_2.receive\_response}] \\
\end{array}}}
where $\lsp \texttt{S} \rsp$, $\lsp \texttt{C\_1} \rsp$, and $\lsp \texttt{C\_2} \rsp$ have been shown in
Ex.~\ref{cssem1}.
\fullbox

	\end{example}

\section{Semi-Synchronous and Asynchronous Interactions}\label{nonsync}

\noindent
The interactions occurring in a PADL description can be involved only in synchronous communications, hence
input and output interactions represent blocking operations. In order to increase the expressiveness of
PADL, within the interface of each AET we will provide support for distinguishing among synchronous,
semi-synchronous, and asynchronous interactions. The usability of the language will be preserved by means of
suitable synchronicity-related qualifiers that are made available to PADL users.

\begin{figure}[thb]

\centerline{\hspace{0.4cm}\includegraphics{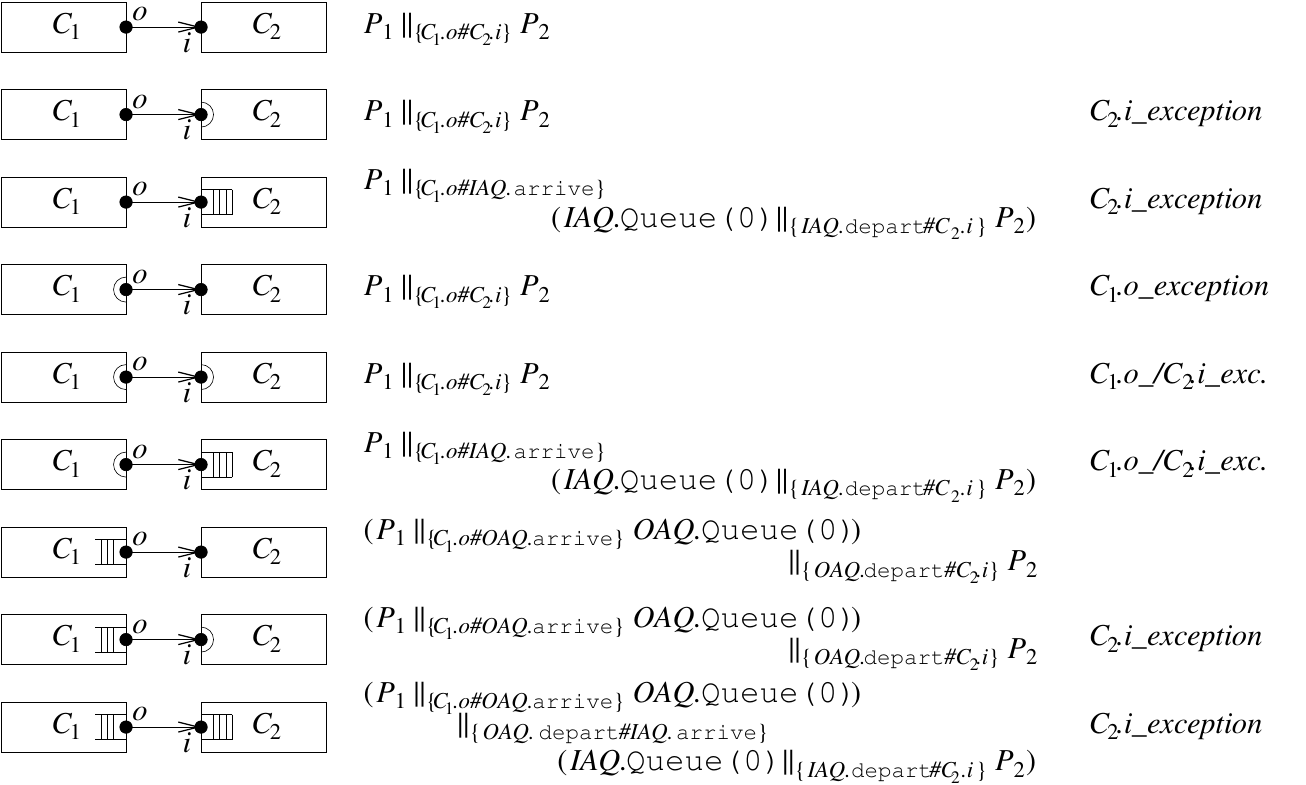}}
\caption{Forms of communications available in the extension of PADL}\label{comm}

	\end{figure}

In this section -- in which we use the same client-server running example as the previous section -- we
first enrich the textual and graphical notations in order to express nonsynchronous interactions
(Sect.~\ref{padlnotenriched}). Then, we discuss the treatment of semi-synchronous interactions through
additional semantic rules (Sect.~\ref{semisync}) and of asynchronous interactions through additional
implicit AEIs (Sect.~\ref{async}). Finally, we revise the semantics accordingly
(Sect.~\ref{padlsemenriched}). The nine resulting forms of communication are summarized by Fig.~\ref{comm},
with the first one being the only one originally available in PADL.

\subsection{Enriching PADL Textual and Graphical Notations}\label{padlnotenriched}

\noindent
In the textual notation of PADL, we introduce a third qualifier for each interaction, to be used in the
definition of the AETs. Such a qualifier establishes whether the interaction is synchronous (keyword
\texttt{SYNC}), semi-synchronous (keyword \texttt{SSYNC}), or asynchronous (keyword \texttt{ASYNC}).

While a synchronous interaction blocks the AEI executing it as long as the interactions attached to it are
not ready, this is not the case with nonsynchronous interactions. More precisely, a semi-synchronous
interaction raises an exception if it cannot take place immediately due to the (temporary or permanent)
unavailability of the interactions attached to it, so that the AEI executing it can proceed anyway.
Likewise, in the case of an asynchronous interaction, the beginning and the end of the communication are
decoupled, hence the AEI executing the interaction will never block.

A boolean variable $s.\texttt{success}$ is associated with each semi-synchronous interaction $s$. This
implicitly declared variable is automatically set at each execution of $s$ and is made available to PADL
users in order to catch exceptions. In this way, it is easy to model different behaviors to be undertaken
depending on the outcome of the execution of $s$.

In the graphical notation, a semi-synchronous interaction is depicted by extending the small circle/square
of the interaction with an arc inside the AEI box. An asynchronous interaction, instead, is depicted by
extending the small circle/square with a buffer inside the AEI box.

	\begin{example}\label{cssynmod}

Consider once more the client-server system of Ex.~\ref{cssyn}. Since the server has no buffer for
incoming requests, each client may want to send a request only if the server is not busy, so that the client
can keep working instead of passively waiting for the server to become available. This can be achieved by
transforming \texttt{send\_request} into a semi-synchronous interaction and by redefining the behavior of
\texttt{Client\_Type} as follows:

		\vspace{-0.2cm}
		{\small\begin{verbatim}
 ARCHI_ELEM_TYPE Client_Type(void)
   BEHAVIOR
     Client_Internal(void; void) =
       process . Client_Interacting();
     Client_Interacting(void; void) =
       send_request .
         choice
         {
           cond(send_request.success = true) ->
                       receive_response . Client_Internal(),
           cond(send_request.success = false) ->
                       keep_processing . Client_Interacting()
         }
   INPUT_INTERACTIONS  SYNC  UNI receive_response
   OUTPUT_INTERACTIONS SSYNC UNI send_request
		\end{verbatim}}

\noindent
On the other hand, the server should not make any assumption about the status of its clients, as these may
be much more complicated than the description above. In particular, when sending out a response to a client,
the server should not be blocked by the temporary or permanent unavailability of that client, as this would
decrease the quality of service. This can be achieved by using keyword \texttt{ASYNC} in the declaration of
output interaction \texttt{send\_response} within the definition of \texttt{Server\_Type}.
\fullbox

	\end{example}

\subsection{Semantics of Semi-Synchronous Interactions: Additional Rules}\label{semisync}

\noindent
A local semi-synchronous interaction $s$ executed by an AEI $C$ gives rise to a transition labeled with
$C.s$ within $\lsp C \rsp$, and hence to the setting of the related \texttt{success} variable to true.
However, in an interacting context, this transition has to be relabeled with $C.\ms{s\_exception}$ if $s$
cannot immediately participate in a communication. This is accomplished by means of additional semantic
rules encoding a context-sensitive variant of the relabeling operator.

Suppose that the local output interaction $o$ of an AEI $C_{1}$ is attached to the local input interaction
$i$ of an AEI $C_{2}$, where $C_{1}.o\#C_{2}.i$ is their fresh name. Let $P_{1}$ (resp.\ $P_{2}$) be the
process term representing the current state of $\lsp C_{1} \rsp_{C_{1}, C_{2}}$ (resp.\ $\lsp C_{2}
\rsp_{C_{1}, C_{2}}$) and $S = \cals(C_{1}, C_{2}; C_{1}, C_{2})$.

If $o$ is synchronous and $i$ is semi-synchronous -- which is the second form of communication depicted in
Fig.~\ref{comm} -- then the following additional semantic rule is necessary for handling exceptions:
\[\begin{array}{|c|}
\hline
{\infr{P_{1} \hspace{0.9cm} \not \hspace{-1.1cm} \larrow{C_{1}.o\#C_{2}.i}{} P'_{1} \hspace{0.8cm} P_{2}
\larrow{C_{1}.o\#C_{2}.i}{} P'_{2}}{P_{1} \pco{S} P_{2} \larrow{C_{2}.\ms{i\_exception}}{} P_{1} \pco{S}
P'_{2} \hspace{0.8cm} C_{2}.i.\texttt{success} = \texttt{false}}} \\
\hline
\end{array}\]

In the symmetric case in which $o$ is semi-synchronous and $i$ is synchronous -- which corresponds to the
fourth form of communication depicted in Fig.~\ref{comm} -- the following additional semantic rule is
necessary for handling exceptions:
\[\begin{array}{|c|}
\hline
{\infr{P_{1} \larrow{C_{1}.o\#C_{2}.i}{} P'_{1} \hspace{0.8cm} P_{2} \hspace{0.9cm} \not \hspace{-1.1cm}
\larrow{C_{1}.o\#C_{2}.i}{} P'_{2}}{P_{1} \pco{S} P_{2} \larrow{C_{1}.\ms{o\_exception}}{} P'_{1} \pco{S}
P_{2} \hspace{0.8cm} C_{1}.o.\texttt{success} = \texttt{false}}} \\
\hline
\end{array}\]

Finally, in the case in which both $o$ and $i$ are semi-synchronous -- which corresponds to the fifth form
of communication depicted in Fig.~\ref{comm} -- we have the previous two additional semantic rules together.

\subsection{Semantics of Asynchronous Interactions: Additional Implicit AEIs}\label{async}

\noindent
While semi-synchronous interactions are dealt with by means of suitable semantic rules accounting for
possible exceptions, asynchronous interactions need a different treatment because of the decoupling between
the beginning and the end of the communications in which those interactions are involved.

\begin{figure}[thb]

\centerline{\includegraphics{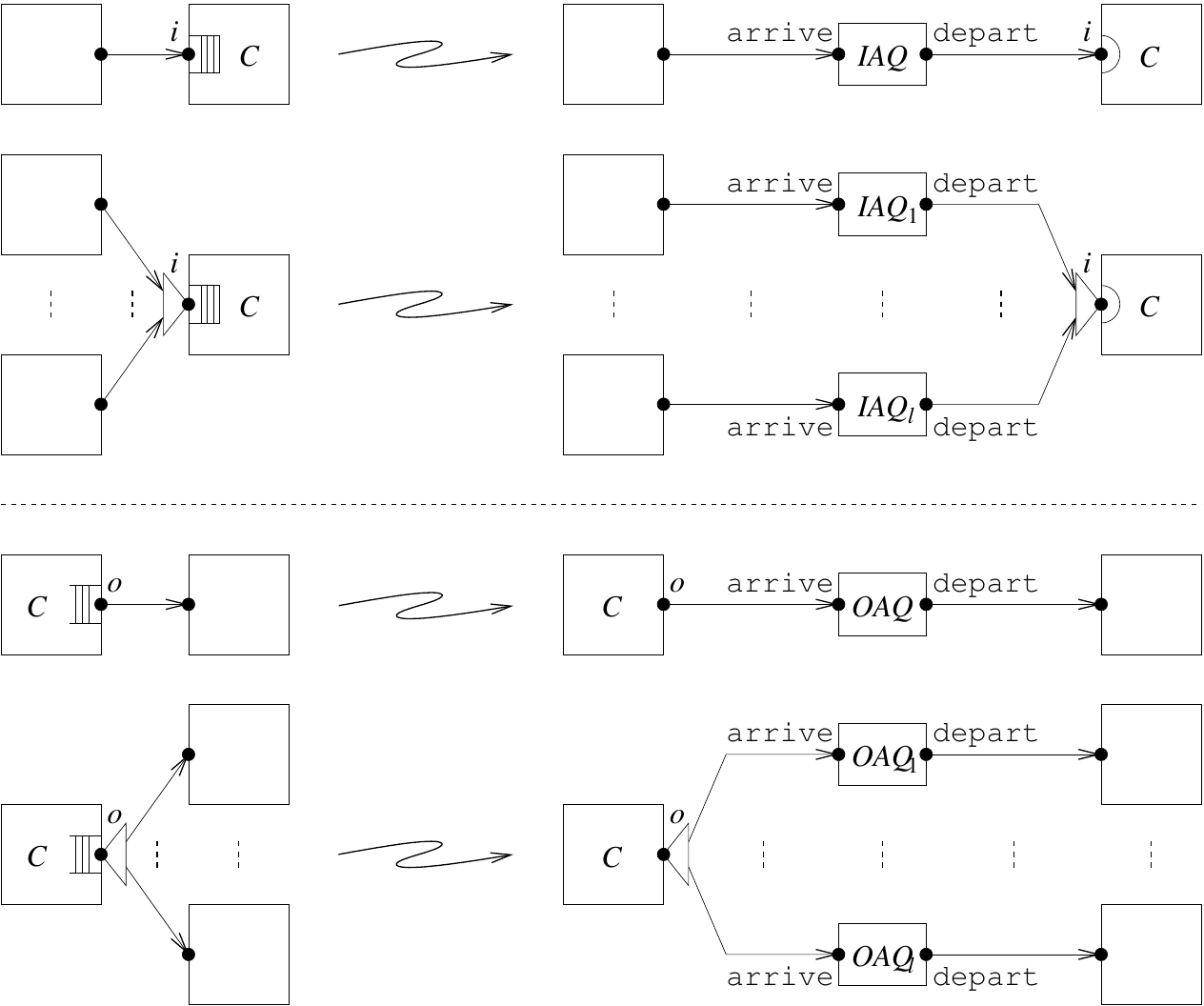}}
\caption{Topological management of local asynchronous uni-/and-interactions}\label{asyncrewrite}

		\end{figure}

After the or-rewriting process, for each local asynchronous uni-/and-interaction of an AEI $C$ we have to
introduce additional implicit AEIs that behave like unbounded buffers, thus realizing the third, sixth,
seventh, eighth, and ninth form of communication depicted in Fig.~\ref{comm}. As shown in
Fig.~\ref{asyncrewrite}, in the case of a local asynchronous and-interaction, it is necessary to introduce
as many additional implicit AEIs as there are attachments involving the
and-interaction.\footnote{In~\cite{BB3}, a single additional implicit AEI was introduced even in the case of
a local asynchronous and-interaction, thus determining an unnecessary synchronization among the AEIs
attached to the and-interaction.}

Each additional implicit input asynchronous queue (IAQ) and output asynchronous queue (OAQ) is of the
following type, where \texttt{arrive} is an always-enabled input synchronous uni-interaction while
\texttt{depart} is an output synchronous uni-interaction enabled only if the buffer is not empty:

	{\small\begin{verbatim}
 ARCHI_ELEM_TYPE Async_Queue_Type(void)
   BEHAVIOR
     Queue(int n := 0;
           void) =
       choice
       {
         cond(true)  -> arrive . Queue(n + 1),
         cond(n > 0) -> depart . Queue(n - 1)
       }
   INPUT_INTERACTIONS  SYNC UNI arrive
   OUTPUT_INTERACTIONS SYNC UNI depart
	\end{verbatim}}

In the case of a local input asynchronous uni-/and-interaction $i$ of $C$, each local output uni-interaction
originally attached to $i$ is implicitly re-attached to the \texttt{arrive} interaction of one of the
additional implicit IAQs. By contrast, the \texttt{depart} interaction of each additional implicit IAQ is
implicitly attached to $i$, which is implicitly converted into a semi-synchronous interaction. Note that $i$
becomes semi-synchronous because the communications between the \texttt{depart} interactions and $i$ must
not block $C$ whenever some buffers are empty.

In the case of a local output asynchronous uni-/and-interaction $o$ of $C$, this interaction is implicitly
converted into a synchronous interaction and re-attached to each \texttt{arrive} interaction of the
additional implicit OAQs. Note that $o$ is never blocked because all \texttt{arrive} interactions are always
enabled. By contrast, the \texttt{depart} interaction of each additional implicit OAQ is attached to one of
the input interactions originally attached to $o$.

\subsection{Revising PADL Semantics}\label{padlsemenriched}

\noindent
Due to the way nonsynchronous interactions have been managed, we only need to revise the definition of the
semantics of an AEI in isolation, while all the subsequent definitions given in Sect.~\ref{padlsem} are
unchanged. More precisely, we only have to take into account the possible presence of additional implicit
AEIs acting as unbounded buffers for local asynchronous interactions.

	\begin{definition}\label{revpadlsemdef}

Let $\calc$ be an AET, $\ms{fp}_{1}, \dots, \ms{fp}_{m}$ be its $m \in \natns_{\ge 0}$ formal parameters,
and $\cale$ be a sequence of defining equations giving its behavior. Let $C$ be an AEI of type $\calc$ with
actual parameters $\ms{ap}_{1}, \dots, \ms{ap}_{m}$ consistent by order and type with the formal parameters.
Suppose that $C$ has:

		\begin{itemize}

\item $h \in \natns_{\ge 0}$ local input asynchronous uni-interactions $i_{1}, \dots, i_{h}$ handled through
the related additional implicit AEIs $\ms{IAQ}_{1}, \dots, \ms{IAQ}_{h}$;

\item $h' \in \natns_{\ge 0}$ local input asynchronous and-interactions $i'_{1}, \dots, i'_{h'}$, where each
$i'_{j}$ is handled through the $\ms{attach-no}(C.i'_{j}) = \ms{il}_{j}$ related additional implicit AEIs
$\ms{IAQ}_{j, 1}, \dots, \ms{IAQ}_{j, \ms{il}_{j}}$;

\item $k \in \natns_{\ge 0}$ local output asynchronous uni-interactions $o_{1}, \dots, o_{k}$ handled
through the related additional implicit AEIs $\ms{OAQ}_{1}, \dots, \ms{OAQ}_{k}$;

\item $k' \in \natns_{\ge 0}$ local output asynchronous and-interactions $o'_{1}, \dots, o'_{k'}$, where
each $o'_{j}$ is handled through the $\ms{attach-no}(C.o'_{j}) = \ms{ol}_{j}$ related additional implicit
AEIs $\ms{OAQ}_{j, 1}, \dots, \ms{OAQ}_{j, \ms{ol}_{j}}$.

		\end{itemize}

\noindent
Then the semantics of $C$ is the result of a cascade of function applications:
\[\begin{array}{|c|}
\hline
\lsp C \rsp \: = \: \ms{o-and}^{k'}_{\ms{ol}_{k'}}( ...
\ms{o-and}^{1}_{\ms{ol}_{1}}(\ms{o-uni}_{k}(\ms{i-and}^{h'}_{\ms{il}_{h'}}( ...
\ms{i-and}^{1}_{\ms{il}_{1}}(\ms{i-uni}_{h}(C)) ... ))) ... ) \\
\hline
\end{array}\]
where, denoted by $f(C)$ the current result, we define:
\[\begin{array}{|rcl|}
\hline
\ms{i-uni}_{0}(C) & \!\! = \!\! & \ms{or-rewrite}_{\emptyset}(C.\cale \{ \ms{ap}_{1} \hookrightarrow
\ms{fp}_{1}, \dots, \ms{ap}_{m} \hookrightarrow \ms{fp}_{m} \}) \, [\varphi_{C, {\rm async}}] \\
\ms{i-uni}_{j}(C) & \!\! = \!\! & \ms{IAQ}_{j}.\texttt{Queue(0)} \, [\varphi_{C, {\rm async}}]
\hspace{3.8cm} 1 \le j \le h \\
& & \hspace{0.5cm} \pco{\{ \ms{IAQ}_{j}.{\tt depart}\#C.i_{j} \}} \, (\ms{i-uni}_{j - 1}(C)) \\
\hline
\ms{i-and}^{j}_{1}(f(C)) & \!\! = \!\! & \ms{IAQ}_{j, 1}.\texttt{Queue(0)} \, [\varphi_{C, {\rm async}}]
\hspace{3.5cm} 1 \le j \le h' \\
& & \hspace{0.5cm} \pco{\{ \ms{IAQ}_{j, 1}.{\tt depart}\# \dots \#\ms{IAQ}_{j, \ms{il}_{j}}.{\tt
depart}\#C.i'_{j} \}} \, (f(C)) \\
\ms{i-and}^{j}_{j'}(f(C)) & \!\! = \!\! & \ms{IAQ}_{j, j'}.\texttt{Queue(0)} \, [\varphi_{C, {\rm async}}]
\hspace{3.2cm} 2 \le j' \le \ms{il}_{j} \\
& & \hspace{0.5cm} \pco{\{ \ms{IAQ}_{j, 1}.{\tt depart}\# \dots \#\ms{IAQ}_{j, \ms{il}_{j}}.{\tt
depart}\#C.i'_{j} \}} \, (\ms{i-and}^{j}_{j' - 1}(f(C))) \\
\hline
\ms{o-uni}_{0}(f(C)) & \!\! = \!\! & f(C) \\
\ms{o-uni}_{j}(f(C)) & \!\! = \!\! & (\ms{o-uni}_{j - 1}(f(C))) \pco{\{ C.o_{j}\#\ms{OAQ}_{j}.{\tt arrive}
\}} \\
& & \hspace{0.5cm} \ms{OAQ}_{j}.\texttt{Queue(0)} \, [\varphi_{C, {\rm async}}] \hspace{3.1cm} 1 \le j \le k
\\
\hline
\ms{o-and}^{j}_{1}(f(C)) & \!\! = \!\! & (f(C)) \pco{\{ C.o'_{j}\#\ms{OAQ}_{j, 1}.{\tt arrive}\# \dots
\#\ms{OAQ}_{j, \ms{ol}_{j}}.{\tt arrive} \}} \\
& & \hspace{0.5cm} \ms{OAQ}_{j, 1}.\texttt{Queue(0)} \, [\varphi_{C, {\rm async}}] \hspace{2.8cm} 1 \le j
\le k' \\
\ms{o-and}^{j}_{j'}(f(C)) & \!\! = \!\! & (\ms{o-and}^{j}_{j' - 1}(f(C))) \pco{\{
C.o'_{j}\#\ms{OAQ}_{j, 1}.{\tt arrive}\# \dots \#\ms{OAQ}_{j, \ms{ol}_{j}}.{\tt arrive} \}} \\
& & \hspace{0.5cm} \ms{OAQ}_{j, j'}.\texttt{Queue(0)} \, [\varphi_{C, {\rm async}}] \hspace{2.5cm} 2 \le j'
\le \ms{ol}_{j} \\
\hline
\end{array}\]
with relabeling function $\varphi_{C, {\rm async}}$ transforming the originally asynchronous local
interactions of $C$ and the local interactions of the additional implicit AEIs attached to them into the
respective fresh names occurring in the synchronization sets above.
\fullbox

	\end{definition}

	\begin{example}\label{cssemmod}

Consider the variant of the client-server system described in Ex.~\ref{cssynmod}. With respect to
Ex.~\ref{cssem1}, $\lsp S \rsp$ changes as follows due to the presence of its two local output asynchronous
uni-interactions:
{\small\cws{0}{\begin{array}{l}
\texttt{RewSer} [\texttt{S.send\_response\_1} \mapsto \texttt{S.send\_response\_1\#OAQ\_1.arrive},
\\[-0.1cm]
\tabspace{\texttt{RewSer} \;}
\texttt{S.send\_response\_2} \mapsto \texttt{S.send\_response\_2\#OAQ\_2.arrive}] \\
\hspace*{0.4cm} \pco{\{ {\tt S.send\_response\_1\#OAQ\_1.arrive} \}} \\[0.1cm]
\hspace*{0.8cm} \texttt{OAQ\_1.Queue(0)} [\texttt{OAQ\_1.arrive} \mapsto
\texttt{S.send\_response\_1\#OAQ\_1.arrive}] \\
\hspace*{1.2cm} \pco{\{ {\tt S.send\_response\_2\#OAQ\_2.arrive} \}} \\[0.1cm]
\hspace*{1.6cm} \texttt{OAQ\_2.Queue(0)} [\texttt{OAQ\_2.arrive} \mapsto
\texttt{S.send\_response\_2\#OAQ\_2.arrive}] \\
\end{array}}}
As a consequence, with respect to Ex.~\ref{cssem2}, the semantics of the whole description changes as
follows, where the roles of \texttt{S.send\_response\_1} and \texttt{S.send\_response\_2} are now played by
\texttt{AOQ\_1.depart} and \texttt{AOQ\_2.depart}, respectively:
{\small\cws{8}{\begin{array}{l}
\lsp \texttt{S} \rsp [\texttt{receive\_request\_1} \mapsto
\texttt{C\_1.send\_request\#S.receive\_request\_1}, \\
\tabspace{\lsp \texttt{S} \rsp \;}
\texttt{OAQ\_1.depart} \mapsto \texttt{OAQ\_1.depart\#C\_1.receive\_response}, \\
\tabspace{\lsp \texttt{S} \rsp \;}
\texttt{receive\_request\_2} \mapsto \texttt{C\_2.send\_request\#S.receive\_request\_2}, \\
\tabspace{\lsp \texttt{S} \rsp \;}
\texttt{OAQ\_2.depart} \mapsto \texttt{OAQ\_2.depart\#C\_2.receive\_response}] \\
\hspace*{1.5cm} \pco{\{ {\tt C\_1.send\_request\#S.receive\_request\_1},} \\[-0.1cm]
\tabspace{\hspace*{1.5cm} \pco{\{}}
_{{\tt OAQ\_1.depart\#C\_1.receive\_response} \}} \\[0.1cm]
\lsp \texttt{C\_1} \rsp [\texttt{send\_request} \mapsto \texttt{C\_1.send\_request\#S.receive\_request\_1},
\\
\tabspace{\lsp \texttt{C\_1} \rsp \;}
\texttt{receive\_response} \mapsto \texttt{OAQ\_1.depart\#C\_1.receive\_response}] \\
\hspace*{1.5cm} \pco{\{ {\tt C\_2.send\_request\#S.receive\_request\_2},} \\[-0.1cm]
\tabspace{\hspace*{1.5cm} \pco{\{}}
_{{\tt OAQ\_2.depart\#C\_2.receive\_response} \}} \\[0.1cm]
\lsp \texttt{C\_2} \rsp [\texttt{send\_request} \mapsto \texttt{C\_2.send\_request\#S.receive\_request\_2},
\\
\tabspace{\lsp \texttt{C\_2} \rsp \;}
\texttt{receive\_response} \mapsto \texttt{OAQ\_2.depart\#C\_2.receive\_response}] \\
\end{array}}}
\fullbox

	\end{example}

\section{Modifying Architectural Checks}\label{checks}

\noindent
The objective of the architectural checks for PADL developed in~\cite{AB} is to infer certain architectural
properties like correct component coordination from the properties of the individual AEIs. The idea is to
verify the absence of coordination mismatches resulting in property violations through a topological
reduction process based on equivalence checking. Given an architectural description, the starting point is
constituted by an abstract variant of its enriched flow graph, where vertices correspond to AEIs and two
vertices are linked by an edge if and only if attachments have been declared among their interactions. From
a topological viewpoint, the resulting graph is a combination of possibly intersecting stars (see
Sect.~\ref{compadapt}) and cycles (see Sect.~\ref{interopadapt}), which are thus viewed as basic topological
formats.

The strategy then consists of applying specific checks locally to all stars and cycles occurring in the
abstract graph. Each check verifies whether the star/cycle contains an AEI behaviorally equivalent to the
whole star/cycle, in which case the star/cycle can be replaced by that AEI. The process successfully
terminates when the whole graph has been reduced to a single behaviorally equivalent AEI, as at that point
it is sufficient to verify whether that AEI satisfies the given property or not. In case of failure, the
mentioned checks provide diagnostic information useful to pinpoint components responsible for possible
property violations within a single star/cycle.

In order to be applicable, the strategy requires the existence of a behavioral equivalence that possesses
the following characteristics. Firstly, the equivalence must preserve the property of interest -- i.e., it
cannot relate two models such that one of them enjoys the property whereas the other one does not -- which
is fundamental for enabling the topological reduction process. Secondly, it must be a congruence with
respect to static operators, thus allowing the topological reduction process to be applied to single
portions of the topology of an architectural description -- which is more efficient than considering the
entire topology at once -- without affecting the possible validity of the property. Thirdly, it must be able
to abstract from internal actions, as an architectural property is typically expressed in terms of the
possibility/necessity of executing local interactions in a certain order. Fourthly, it must have a modal
logic characterization, which is necessary for producing diagnostic information in case of failure of the
topological reduction process.

In this section, after introducing some further notation (Sect.~\ref{closedsemadapt}), we show how to modify
the compatibility check for stars (Sect.~\ref{compadapt}) and the interoperability check for cycles
(Sect.~\ref{interopadapt}) in such a way that both checks can still be applied in the presence of
nonsynchronous interactions. Then, the two revised checks are exemplified on the architectural description
of an applet-based simulator for a cruise control system (Sect.~\ref{cruise}). Although these checks are
conceived for an entire class of properties characterizable through behavioral equivalences that meet the
four constraints mentioned above, for the sake of simplicity here the considered property is deadlock
freedom and the behavioral equivalence chosen among those preserving deadlock freedom is weak bisimilarity
$\wbis{\rm B}$ introduced in Sect.~\ref{pa}.

\subsection{Revising Closed Interacting Semantics}\label{closedsemadapt}

\noindent
Before applying the architectural checks to a star/cycle given by the set of AEIs $\{ C_{1}, \dots, C_{n}
\}$, for each AEI $C_{j}$ in the set we have to hide all of its internal actions and architectural
interactions as well as all of its local interactions that are not attached to $\{ C_{1}, \dots, C_{n} \}$.
The reason is that these actions cannot result in mismatches within the star/cycle, but may hamper the
topological reduction process if left visible. Following the terminology of~\cite{AB}, we thus have to
consider closed variants of the interacting semantics of the AEIs in the set, in which the mentioned actions
are made unobservable.

In the framework of PADL enriched with nonsynchronous interactions, for each AEI $C_{j}$ in the set we also
have to hide all of its additional implicit AEIs that are not attached to $\{ C_{1}, \dots, C_{n} \}$, as
those additional implicit AEIs are necessary only in the presence of the AEIs not in $\{ C_{1}, \dots, C_{n}
\}$ to which they are attached. Therefore, the only actions that remain observable are those in
$\calli_{C_{j}; C_{1}, \dots, C_{n}}$ and those in $\caloali_{C_{j}}$. The latter set contains the
originally asynchronous local interactions of $C_{j}$ together with the local interactions of the related
additional implicit AEIs to which they have been re-attached, including the exceptions that may be raised by
the local input semi-synchronous interactions in the set corresponding to local input asynchronous
interactions. We point out that $\caloali_{C_{j}}$ is disjoint from $\calli_{C_{j}}$, as it essentially
comprises the action names forming the composite names occurring in the synchronization sets of the
semantics of an AEI in isolation provided in Def.~\ref{revpadlsemdef}.

In order to set the visibility of each action of $C_{j}$ according to the needs of the topological reduction
process, we introduce a partially closed variant of the interacting semantics of an AEI defined at the end
of Sect.~\ref{padlsem}, in which we hide all the actions not in $\calli_{C_{j}; C_{1}, \dots, C_{n}} \cup
\caloali_{C_{j}}$. Since in many cases we also have to hide all the actions in $\caloali_{C_{j}}$, we
introduce a totally closed variant too. Thus, unlike~\cite{AB}, we have two closed variants of the
interacting semantics of an AEI. Both variants are parameterized with respect to a set of AEIs $\{ C''_{1},
\dots, C''_{n''} \}$, $n'' \in \natns$, determining the additional implicit AEIs to be included.

	\begin{definition}

Let $\calc$ be an AET, $\ms{fp}_{1}, \dots, \ms{fp}_{m}$ be its $m \in \natns_{\ge 0}$ formal parameters,
and $\cale$ be a sequence of defining equations giving its behavior. Let $C_{j} \in \{ C_{1}, \dots, C_{n}
\}$ be an AEI of type $\calc$ with actual parameters $\ms{ap}_{1}, \dots, \ms{ap}_{m}$ consistent by order
and type with the formal parameters. The interacting semantics of $C_{j}$ with respect to $\{ C_{1}, \dots,
C_{n} \}$ without buffers for its originally asynchronous local interactions is defined as follows:
\[\begin{array}{|rcl|}
\hline
\lsp C_{j} \rsp^{\rm wob}_{C_{1}, \dots, C_{n}} & \!\! = \!\! & \ms{or-rewrite}_{\emptyset}(C_{j}.\cale \{
\ms{ap}_{1} \hookrightarrow \ms{fp}_{1}, \dots, \ms{ap}_{m} \hookrightarrow \ms{fp}_{m} \}) \\
& & \hspace{0.5cm} [\varphi_{C_{j}, {\rm async}}] \, [\varphi_{C_{j}; C_{1}, \dots, C_{n}}] \\
\hline
\end{array}\]
We denote by $\lsp C_{j} \rsp^{\#C''_{1}, \dots, C''_{n''}}_{C_{1}, \dots, C_{n}}$ the variant of $\lsp
C_{j} \rsp^{\rm wob}_{C_{1}, \dots, C_{n}}$ including the buffers for the originally asynchronous local
interactions of $C_{j}$ attached to $\{ C''_{1}, \dots, C''_{n''} \}$.
\fullbox

	\end{definition}

	\begin{definition}

Let $C_{j} \in \{ C_{1}, \dots, C_{n} \}$. The partially closed interacting semantics of $C_{j}$ with
respect to $\{ C_{1}, \dots, C_{n} \}$ including its buffers attached to $\{ C''_{1}, \dots, C''_{n''} \}$
is defined as follows:
\[\begin{array}{|c|}
\hline
\lsp C_{j} \rsp^{{\rm pc}; \#C''_{1}, \dots, C''_{n''}}_{C_{1}, \dots, C_{n}} \: = \: \lsp C_{j}
\rsp^{\#C''_{1}, \dots, C''_{n''}}_{C_{1}, \dots, C_{n}} \, / \, (\ms{Name} - \calv_{C_{j}; C_{1}, \dots,
C_{n}}) \\
\hline
\end{array}\]
with $\calv_{C_{j}; C_{1}, \dots, C_{n}} = \varphi_{C_{j}; C_{1}, \dots, C_{n}}(\calli_{C_{j}; C_{1}, \dots,
C_{n}}) \cup \varphi_{C_{j}, {\rm async}}(\caloali_{C_{j}})$ and we write $\lsp C_{j} \rsp^{\rm pc;
wob}_{C_{1}, \dots, C_{n}}$ if $n'' = 0$.
\fullbox

	\end{definition}

	\begin{definition}

Let $\{ C'_{1}, \dots, C'_{n'} \} \subseteq \{ C_{1}, \dots, C_{n} \}$. The partially closed interacting
semantics of $\{ C'_{1}, \dots, C'_{n'} \}$ with respect to $\{ C_{1}, \dots, C_{n} \}$ including their
buffers attached to $\{ C''_{1}, \dots, C''_{n''} \}$ is defined as follows:
\[\begin{array}{|l|}
\hline
\lsp C'_{1}, \dots, C'_{n'} \rsp^{{\rm pc}; \#C''_{1}, \dots, C''_{n''}}_{C_{1}, \dots, C_{n}} \: =
\\[0.1cm]
\hspace*{3.6cm} \lsp C'_{1} \rsp^{{\rm pc}; \#C''_{1}, \dots, C''_{n''}}_{C_{1}, \dots, C_{n}}
\pco{\cals(C'_{1}, C'_{2}; C_{1}, \dots, C_{n})} \\[0.2cm]
\hspace*{4.1cm} \lsp C'_{2} \rsp^{{\rm pc}; \#C''_{1}, \dots, C''_{n''}}_{C_{1}, \dots, C_{n}}
\pco{\cals(C'_{1}, C'_{3}; C_{1}, \dots, C_{n}) \cup \cals(C'_{2}, C'_{3}; C_{1}, \dots, C_{n})} \; \dots
\\[0.2cm]
\hspace*{4.6cm} \dots \; \pco{\mathop{\cup}\limits_{i = 1}^{n' - 1} \cals(C'_{i}, C'_{n'}; C_{1}, \dots,
C_{n})} \; \lsp C'_{n'} \rsp^{{\rm pc}; \#C''_{1}, \dots, C''_{n''}}_{C_{1}, \dots, C_{n}} \\
\hline
\end{array}\]
where the synchronization sets are built as at the end of Sect.~\ref{padlsem}.
\fullbox

	\end{definition}

	\begin{definition}

Let $C_{j} \in \{ C_{1}, \dots, C_{n} \}$. The totally closed interacting semantics of $C_{j}$ with respect
to $\{ C_{1}, \dots, C_{n} \}$ including its buffers attached to $\{ C''_{1}, \dots, C''_{n''} \}$ is
defined as follows:
\[\begin{array}{|c|}
\hline
\lsp C_{j} \rsp^{{\rm tc}; \#C''_{1}, \dots, C''_{n''}}_{C_{1}, \dots, C_{n}} \: = \: \lsp C_{j} \rsp^{{\rm
pc}; \#C''_{1}, \dots, C''_{n''}}_{C_{1}, \dots, C_{n}} \, / \, \varphi_{C_{j}, {\rm
async}}(\caloali_{C_{j}}) \\
\hline
\end{array}\]
and we write $\lsp C_{j} \rsp^{\rm tc; wob}_{C_{1}, \dots, C_{n}}$ if $n'' = 0$.
\fullbox

	\end{definition}

	\begin{definition}

Let $\{ C'_{1}, \dots, C'_{n'} \} \subseteq \{ C_{1}, \dots, C_{n} \}$. The totally closed interacting
semantics of $\{ C'_{1}, \dots, C'_{n'} \}$ with respect to $\{ C_{1}, \dots, C_{n} \}$ including their
buffers attached to $\{ C''_{1}, \dots, C''_{n''} \}$ is defined as follows:
\[\begin{array}{|l|}
\hline
\lsp C'_{1}, \dots, C'_{n'} \rsp^{{\rm tc}; \#C''_{1}, \dots, C''_{n''}}_{C_{1}, \dots, C_{n}} \: = \\
\hspace*{3.6cm} \lsp C'_{1} \rsp^{{\rm tc}; \#C''_{1}, \dots, C''_{n''}}_{C_{1}, \dots, C_{n}}
\pco{\cals(C'_{1}, C'_{2}; C_{1}, \dots, C_{n})} \\[0.2cm]
\hspace*{4.1cm} \lsp C'_{2} \rsp^{{\rm tc}; \#C''_{1}, \dots, C''_{n''}}_{C_{1}, \dots, C_{n}}
\pco{\cals(C'_{1}, C'_{3}; C_{1}, \dots, C_{n}) \cup \cals(C'_{2}, C'_{3}; C_{1}, \dots, C_{n})} \; \dots
\\[0.2cm]
\hspace*{4.6cm} \dots \; \pco{\mathop{\cup}\limits_{i = 1}^{n' - 1} \cals(C'_{i}, C'_{n'}; C_{1}, \dots,
C_{n})} \; \lsp C'_{n'} \rsp^{{\rm tc}; \#C''_{1}, \dots, C''_{n''}}_{C_{1}, \dots, C_{n}} \\
\hline
\end{array}\]
where the synchronization sets are built as at the end of Sect.~\ref{padlsem}. The variant totally closed up
to $\{ C'''_{1}, \dots, C'''_{n'''} \} \subset \{ C'_{1}, \dots, C'_{n'} \}$, i.e., in which $\lsp C'''_{j}
\rsp^{{\rm pc}; \#C''_{1}, \dots, C''_{n''}}_{C_{1}, \dots, C_{n}}$ is considered in place of $\lsp C'''_{j}
\rsp^{{\rm tc}; \#C''_{1}, \dots, C''_{n''}}_{C_{1}, \dots, C_{n}}$, is denoted by $\lsp C'_{1}, \dots,
C'_{n'} \rsp^{{\rm tc}; \#C''_{1}, \dots, C''_{n''}; C'''_{1}, \dots, C'''_{n'''}}_{C_{1}, \dots, C_{n}}$.
\fullbox

	\end{definition}

\subsection{Adapting Architectural Compatibility for Stars}\label{compadapt}

\noindent
A star is a portion of the abstract enriched flow graph of an architectural description, which is not part
of a cyclic subgraph. It is formed by a central AEI $K$ and a border $\calb_{K} = \{ C_{1}, \dots, C_{n} \}$
including all the AEIs attached to~$K$. As explained in~\cite{AB}, the validity of an architectural property
over a star can be investigated by analyzing the interplay between the central AEI $K$ and each of the AEIs
in the border, as there cannot be attachments among AEIs in the border. In order to achieve a correct
coordination between $K$ and $C_{j} \in \calb_{K}$, the actual observable behavior of $C_{j}$ should
coincide with the observable behavior expected by $K$. In other words, the observable behavior of $K$ should
not be altered by the insertion of $C_{j}$ into the border of the star.

In order to cope with the presence of nonsynchronous interactions, we modify the architectural compatibility
check for stars as follows.

	\begin{definition}\label{compdef}

Given an architectural description $\cala$, let $K$ be the central AEI of a star of $\cala$, $\calb_{K} = \{
C_{1}, \dots, C_{n} \}$ be the border of the star, $C_{j}$ be an AEI in $\calb_{K}$, $H_{K, C_{j}}$ be the
set of interactions of additional implicit AEIs of $K$ that are attached to interactions of $C_{j}$, and
$E_{K, C_{j}}$ be the set of exceptions that may be raised by semi-synchronous interactions involved in
attachments between $K$ and~$C_{j}$. We say that $K$ is compatible with $C_{j}$ iff:
\cws{11}{(\lsp K \rsp^{{\rm pc}; \#C_{j}}_{\cala} \pco{\cals(K, C_{j}; \cala)} \, \lsp C_{j} \rsp^{{\rm
tc}; \#K}_{K, \calb_{K}}) \, / \, (H_{K, C_{j}} \cup E_{K, C_{j}}) \: \wbis{\rm B} \: \lsp K \rsp^{\rm pc;
wob}_{\cala}}
\fullbox

	\end{definition}

All possible originally asynchronous local interactions of $C_{j}$ and all of its interactions possibly
attached to AEIs outside the star have been hidden by taking the totally closed interacting semantics of
$C_{j}$ with respect to the AEIs inside the star. We also observe that $H_{K, C_{j}} \cup E_{K, C_{j}} =
\emptyset$ whenever there are no local nonsynchronous interactions involved in attachments inside the star,
in which case all partially closed interacting semantics between $\lsp K \rsp^{{\rm pc}; \#\cala}_{\cala}$
and $\lsp K \rsp^{\rm pc; wob}_{\cala}$ coincide with $\lsp K \rsp^{\rm tc; wob}_{\cala}$.

We now extend the compatibility theorem of~\cite{AB} to nonsynchronous interactions. This provides a
sufficient condition for reducing the deadlock verification of the entire star to the deadlock verification
of its central AEI.

	\begin{theorem}\label{compthm}

Let $\cala$, $K$, $\calb_{K}$, $C_{j}$, $H_{K, C_{j}}$, and $E_{K, C_{j}}$ be the same as
Def.~\ref{compdef}. Whenever $K$ is compatible with every $C_{j} \in \calb_{K}$, then:
\cws{0}{\lsp K, \calb_{K} \rsp^{{\rm tc}; \#K, \calb_{K}; K}_{K, \calb_{K}} \, / \, \mathop{\cup}\limits_{j
= 1}^{n} (H_{K, C_{j}} \cup E_{K, C_{j}}) \: \wbis{\rm B} \: \lsp K \rsp^{\rm pc; wob}_{\cala}}
hence $\lsp K, \calb_{K} \rsp^{{\rm tc}; \#K, \calb_{K}; K}_{K, \calb_{K}} \, / \, \mathop{\cup}\limits_{j =
1}^{n} (H_{K, C_{j}} \cup E_{K, C_{j}})$ is deadlock free iff so is $\lsp K \rsp^{\rm pc; wob}_{\cala}$.

		\begin{proof}
Since there cannot be attachments between interactions of the AEIs of $\calb_{K}$, it turns out that $\lsp
K, \calb_{K} \rsp^{{\rm tc}; \#K, \calb_{K}; K}_{K, \calb_{K}} \, / \, \mathop{\cup}\limits_{j = 1}^{n}
(H_{K, C_{j}} \cup E_{K, C_{j}})$ is given by:
\cws{4}{\begin{array}{l}
(\lsp K \rsp^{{\rm pc}; \#\calb_{K}}_{\cala} \pco{\cals(K, C_{1}; \cala)} \, \lsp C_{1} \rsp^{{\rm tc};
\#K}_{K, \calb_{K}} \pco{\cals(K, C_{2}; \cala)} \, \lsp C_{2} \rsp^{{\rm tc}; \#K}_{K, \calb_{K}} \\[0.2cm]
\hspace*{3.5cm} \pco{\cals(K, C_{3}; \cala)} \, \dots \pco{\cals(K, C_{n}; \cala)} \, \lsp C_{n} \rsp^{{\rm
tc}; \#K}_{K, \calb_{K}}) / \, \mathop{\cup}\limits_{j = 1}^{n} (H_{K, C_{j}} \cup E_{K, C_{j}}) \\
\end{array}}

Since each local asynchronous and-interaction is dealt with by introducing as many additional implicit AEIs
as there are attachments involving the and-interaction, $H_{K, C_{j}} \cap H_{K, C_{g}} = \emptyset$ for all
$j \neq g$. Hence, every hiding set $H_{K, C_{j}}$ can be distributed in such a way that it is applied
as soon as possible. Similarly, every hiding set $E_{K, C_{j}}$ can be anticipated too, because $E_{K,
C_{j}}$ is independent from $E_{K, C_{g}}$ for all $j \neq g$ due to the fact that any exception is raised
locally at a single AEI. As a consequence, $\lsp K, \calb_{K} \rsp^{{\rm tc}; \#K, \calb_{K}; K}_{K,
\calb_{K}} \, / \, \mathop{\cup}\limits_{j = 1}^{n} (H_{K, C_{j}} \cup E_{K, C_{j}})$ can be rewritten in
the following way:
\cws{4}{\begin{array}{l}
( \dots ((\lsp K \rsp^{{\rm pc}; \#\calb_{K}}_{\cala} \pco{\cals(K, C_{1}; \cala)} \, \lsp C_{1} \rsp^{{\rm
tc}; \#K}_{K, \calb_{K}}) \, / \, (H_{K, C_{1}} \cup E_{K, C_{1}}) \\[0.2cm]
\hspace*{1.5cm} \pco{\cals(K, C_{2}; \cala)} \, \lsp C_{2} \rsp^{{\rm tc}; \#K}_{K, \calb_{K}}) \, / \,
(H_{K, C_{2}} \cup E_{K, C_{2}}) \\[0.2cm]
\hspace*{3.0cm} \pco{\cals(K, C_{3}; \cala)} \, \dots \pco{\cals(K, C_{n}; \cala)} \, \lsp C_{n} \rsp^{{\rm
tc}; \#K}_{K, \calb_{K}}) \, / \, (H_{K, C_{n}} \cup E_{K, C_{n}}) \\
\end{array}}

Denoted by $\ms{IAQ}_{K, C_{j}}$ (resp.\ $\ms{OAQ}_{K, C_{j}}$) the parallel composition of the behaviors of
the input (resp.\ output) asynchronous queues of $K$ attached to~$C_{j}$ -- whose local interactions are
relabeled according to $\varphi_{K, {\rm async}}$ and $\varphi_{K; \cala}$ -- and by $\caloali^{\rm
input}_{K, C_{j}}$ (resp.\ $\caloali^{\rm output}_{K, C_{j}}$) their local interactions attached to $K$, it
turns out that $\lsp K \rsp^{{\rm pc}; \#\calb_{K}}_{\cala}$ is given by:
\cws{4}{\begin{array}{l}
\ms{IAQ}_{K, C_{n}} \pco{\varphi_{K, {\rm async}}(\caloali^{\rm input}_{K, C_{n}})} \\
\hspace*{1.0cm} ( \cdots \\
\hspace*{2.0cm} (\ms{IAQ}_{K, C_{2}} \pco{\varphi_{K, {\rm async}}(\caloali^{\rm input}_{K, C_{2}})} \\
\hspace*{3.0cm} (\ms{IAQ}_{K, C_{1}} \pco{\varphi_{K, {\rm async}}(\caloali^{\rm input}_{K, C_{1}})} \\
\hspace*{4.0cm} \lsp K \rsp^{\rm pc; wob}_{\cala} \\
\hspace*{3.0cm} \pco{\varphi_{K, {\rm async}}(\caloali^{\rm output}_{K, C_{1}})} \ms{OAQ}_{K, C_{1}}) \\
\hspace*{2.0cm} \pco{\varphi_{K, {\rm async}}(\caloali^{\rm output}_{K, C_{2}})} \ms{OAQ}_{K, C_{2}}) \\
\hspace*{1.0cm} \cdots ) \\
\pco{\varphi_{K, {\rm async}}(\caloali^{\rm output}_{K, C_{n}})} \ms{OAQ}_{K, C_{n}} \\
\end{array}}

Since from the point of view of $C_{j}$ and $C_{g}$ the asynchronous queues of $K$ attached to $C_{j}$ are
independent from the asynchronous queues of $K$ attached to $C_{g}$ for all $j \neq g$, we have that $\lsp
K, \calb_{K} \rsp^{{\rm tc}; \#K, \calb_{K}; K}_{K, \calb_{K}} \, / \, \mathop{\cup}\limits_{j = 1}^{n}
(H_{K, C_{j}} \cup E_{K, C_{j}})$ can be rewritten in the following way:
\cws{4}{\begin{array}{l}
(\ms{IAQ}_{K, C_{n}} \pco{\varphi_{K, {\rm async}}(\caloali^{\rm input}_{K, C_{n}})} \\
\hspace*{0.5cm} ( \cdots \\
\hspace*{1.0cm} (\ms{IAQ}_{K, C_{2}} \pco{\varphi_{K, {\rm async}}(\caloali^{\rm input}_{K, C_{2}})} \\
\hspace*{1.5cm} (\ms{IAQ}_{K, C_{1}} \pco{\varphi_{K, {\rm async}}(\caloali^{\rm input}_{K, C_{1}})} \\
\hspace*{2.0cm} \lsp K \rsp^{\rm pc; wob}_{\cala} \\
\hspace*{1.5cm} \pco{\varphi_{K, {\rm async}}(\caloali^{\rm output}_{K, C_{1}})} \ms{OAQ}_{K, C_{1}}
\pco{\cals(K, C_{1}; \cala)} \, \lsp C_{1} \rsp^{{\rm tc}; \#K}_{K, \calb_{K}}) \, / \, (H_{K, C_{1}} \cup
E_{K, C_{1}}) \\
\hspace*{1.0cm} \pco{\varphi_{K, {\rm async}}(\caloali^{\rm output}_{K, C_{2}})} \ms{OAQ}_{K, C_{2}}
\pco{\cals(K, C_{2}; \cala)} \, \lsp C_{2} \rsp^{{\rm tc}; \#K}_{K, \calb_{K}}) \, / \, (H_{K, C_{2}} \cup
E_{K, C_{2}}) \\
\hspace*{0.5cm} \cdots ) \\
\pco{\varphi_{K, {\rm async}}(\caloali^{\rm output}_{K, C_{n}})} \ms{OAQ}_{K, C_{n}} \pco{\cals(K, C_{n};
\cala)} \, \lsp C_{n} \rsp^{{\rm tc}; \#K}_{K, \calb_{K}}) \, / \, (H_{K, C_{n}} \cup E_{K, C_{n}}) \\
\end{array}}

Since $\ms{IAQ}_{K, C_{1}} \pco{\varphi_{K, {\rm async}}(\caloali^{\rm input}_{K, C_{1}})} \, \lsp K
\rsp^{\rm pc; wob}_{\cala} \pco{\varphi_{K, {\rm async}}(\caloali^{\rm output}_{K, C_{1}})} \ms{OAQ}_{K,
C_{1}}$ is precisely $\lsp K \rsp^{{\rm pc}; \#C_{1}}_{\cala}$ and \linebreak $(\lsp K \rsp^{{\rm pc};
\#C_{1}}_{\cala} \pco{\cals(K, C_{1}; \cala)} \, \lsp C_{1} \rsp^{{\rm tc}; \#K}_{K, \calb_{K}}) \, / \,
(H_{K, C_{1}} \cup E_{K, C_{1}}) \wbis{\rm B} \lsp K \rsp^{\rm pc; wob}_{\cala}$ due to the compatibility of
$K$ with $C_{1}$, by virtue of the congruence property of $\wbis{\rm B}$ with respect to static operators --
and also with respect to the context-sensitive variant of the relabeling operator governing exceptions as it
encodes an injective relabeling function over actions in dot notation, which thus preserves action
qualifiers -- it turns out that $\lsp K, \calb_{K} \rsp^{{\rm tc}; \#K, \calb_{K}; K}_{K, \calb_{K}} \, / \,
\mathop{\cup}\limits_{j = 1}^{n} (H_{K, C_{j}} \cup E_{K, C_{j}})$ is weakly bisimilar to:
\cws{4}{\begin{array}{l}
(\ms{IAQ}_{K, C_{n}} \pco{\varphi_{K, {\rm async}}(\caloali^{\rm input}_{K, C_{n}})} \\
\hspace*{0.5cm} ( \cdots \\
\hspace*{1.0cm} (\ms{IAQ}_{K, C_{2}} \pco{\varphi_{K, {\rm async}}(\caloali^{\rm input}_{K, C_{2}})} \\
\hspace*{2.0cm} \lsp K \rsp^{\rm pc; wob}_{\cala} \\
\hspace*{1.0cm} \pco{\varphi_{K, {\rm async}}(\caloali^{\rm output}_{K, C_{2}})} \ms{OAQ}_{K, C_{2}}
\pco{\cals(K, C_{2}; \cala)} \, \lsp C_{2} \rsp^{{\rm tc}; \#K}_{K, \calb_{K}}) \, / \, (H_{K, C_{2}} \cup
E_{K, C_{2}}) \\
\hspace*{0.5cm} \cdots ) \\
\pco{\varphi_{K, {\rm async}}(\caloali^{\rm output}_{K, C_{n}})} \ms{OAQ}_{K, C_{n}} \pco{\cals(K, C_{n};
\cala)} \, \lsp C_{n} \rsp^{{\rm tc}; \#K}_{K, \calb_{K}}) \, / \, (H_{K, C_{n}} \cup E_{K, C_{n}}) \\
\end{array}}

By reasoning in the same way for each of the other AEIs $C_{2}, \dots, C_{n}$ of $\calb_{K}$, we end up with
\linebreak $\lsp K, \calb_{K} \rsp^{{\rm tc}; \#K, \calb_{K}; K}_{K, \calb_{K}} \, / \,
\mathop{\cup}\limits_{j = 1}^{n} (H_{K, C_{j}} \cup E_{K, C_{j}}) \wbis{\rm B} \lsp K \rsp^{\rm pc;
wob}_{\cala}$. The second part of the result then follows from the fact that $\wbis{\rm B}$ preserves
deadlock freedom.

(An induction on the size of $\calb_{K}$ is hampered by the variability of the set of AEIs with respect to
which the interacting semantics are defined.)
\fullbox

		\end{proof}

	\end{theorem}

\subsection{Adapting Architectural Interoperability for Cycles}\label{interopadapt}

\noindent
A cycle is a closed simple path in the abstract enriched flow graph of an architectural description, which
traverses a set $\caly = \{ C_{1}, \dots, C_{n} \}$ of $n \ge 3$ AEIs. As explained in~\cite{AB}, the
validity of an architectural property over a cycle cannot be investigated by analyzing the interplay between
pairs of AEIs, because of the possible presence of arbitrary interferences among the various AEIs in the
cycle. In order to achieve a correct coordination between any $C_{j}$ and the rest of the cycle, the actual
observable behavior of $C_{j}$ should coincide with the observable behavior expected by the rest of the
cycle. In other words, the observable behavior of $C_{j}$ should not be altered by the insertion of $C_{j}$
itself into the cycle.

In order to cope with the presence of nonsynchronous interactions, we modify the architectural
interoperability check for cycles as follows.

	\begin{definition}\label{interopdef}

Given an architectural description $\cala$, let $\caly = \{ C_{1}, \dots, C_{n} \}$ be the set of AEIs
traversed by a cycle of $\cala$, $C_{j}$ be an AEI in the cycle, $H_{C_{j}, \caly}$ be the set of
interactions of additional implicit AEIs of $C_{j}$ that are attached to $\caly$, and $E_{C_{j}, \caly}$ be
the set of exceptions that may be raised by semi-synchronous interactions involved in attachments between
$C_{j}$ and $\caly$. We say that $C_{j}$ interoperates with the other AEIs in the cycle iff:
\cws{11}{\lsp \caly \rsp^{{\rm tc}; \#\caly; C_{j}}_{\cala} \, / \, (\ms{Name} - \calv_{C_{j}; \cala}) \, /
\, (H_{C_{j}, \caly} \cup E_{C_{j}, \caly}) \: \wbis{\rm B} \: \lsp C_{j} \rsp^{\rm pc; wob}_{\cala}}
\fullbox

	\end{definition}

All possible originally asynchronous local interactions of the other AEIs in the cycle and all of their
interactions that are not attached to $C_{j}$ have been hidden by taking the totally closed interacting
semantics of those AEIs and by leaving visible only the actions in $\calv_{C_{j}; \cala}$. We also observe
that, whenever $C_{j}$ has no local nonsynchronous interactions and is not attached to semi-synchronous
interactions of other AEIs in the cycle, then $H_{C_{j}, \caly} \cup E_{C_{j}, \caly} = \emptyset$ and both
$\lsp C_{j} \rsp^{{\rm pc}; \#\caly}_{\cala}$ and $\lsp C_{j} \rsp^{\rm pc; wob}_{\cala}$ coincide with
$\lsp C_{j} \rsp^{\rm tc; wob}_{\cala}$.

We now extend the interoperability theorem of~\cite{AB} to nonsynchronous interactions. This provides a
sufficient condition for reducing the deadlock verification of the entire cycle to the deadlock verification
of one of its AEIs.

	\begin{theorem}\label{interopthm}

Let $\cala$, $\caly$, $C_{j}$, $H_{C_{j}, \caly}$, and $E_{C_{j}, \caly}$ be the same as
Def.~\ref{interopdef}. Whenever $C_{j}$ interoperates with the other AEIs in the cycle, then $\lsp \caly
\rsp^{{\rm tc}; \#\caly; C_{j}}_{\cala} \, / (\ms{Name} - \calv_{C_{j}; \cala}) \, / \, (H_{C_{j}, \caly}
\cup E_{C_{j}, \caly})$ is deadlock free iff so is $\lsp C_{j} \rsp^{\rm pc; wob}_{\cala}$.

		\begin{proof}
A straightforward consequence of Def.~\ref{interopdef} and of the fact that $\wbis{\rm B}$ preserves
deadlock freedom.
\fullbox

		\end{proof}

	\end{theorem}

\subsection{Case Study: An Applet-Based Simulator for a Cruise Control System}\label{cruise}

\noindent
In this section, we discuss the application of the modified architectural checks by revisiting the cruise
control system considered in~\cite{MK,BCD}.

Once the engine has been turned on, this system is governed by the two standard pedals of the automobile --
accelerator and brake -- and by three additional buttons -- on, off, and resume. When on is pressed, the
cruise control system records the current speed and maintains the automobile at that speed. When the
accelerator, the brake, or off is pressed, the cruise control system disengages but retains the speed
setting. If resume is pressed later on, then the system is able to accelerate or decelerate the automobile
to the previously recorded speed.

The cruise control system is formed by four software components: a sensor, a speed controller, a speed
detector, and a speed actuator. The sensor detects the driver commands and forwards them to the speed
controller, which in turn triggers the speed actuator. The speed detector periodically measures the number
of wheel revolutions per time unit. The speed actuator adjusts the throttle on the basis of the triggers
received from the controller and of the speed measured by the detector.

Let us describe the cruise control system with PADL. The sensor AET is defined as follows:

	{\footnotesize\begin{verbatim}
 ARCHI_ELEM_TYPE Sensor_Type(void)
   BEHAVIOR
     Sensor_Off(void; void) =
       detected_engine_on . turn_engine_on . Sensor_On();
     Sensor_On(void; void) =
       choice
       {
         detected_accelerator . press_accelerator . Sensor_On(),
         detected_brake . press_brake . Sensor_On(),
         detected_on . press_on . Sensor_On(),
         detected_off . press_off . Sensor_On(),
         detected_resume . press_resume . Sensor_On(),
         detected_engine_off . turn_engine_off . Sensor_Off()
       }
   INPUT_INTERACTIONS  SYNC UNI detected_engine_on; detected_engine_off;
                                detected_accelerator; detected_brake;
                                detected_on; detected_off; detected_resume
   OUTPUT_INTERACTIONS SYNC UNI press_accelerator; press_brake;
                                press_on; press_off; press_resume
                            AND turn_engine_on; turn_engine_off
	\end{verbatim}}

The speed controller triggers the speed actuator on the basis of the commands forwarded by the sensor. It
can be inactive (when the engine is off), active (when the engine is on), cruising (after pressing the on
button in the active state or the resume button in the suspended state), or suspended (after pressing any
pedal or button different from on/resume in the cruising state):

	{\footnotesize\begin{verbatim}
 ARCHI_ELEM_TYPE Controller_Type(void)
   BEHAVIOR
     Inactive(void; void) =
       turned_engine_on . Active();
     Active(void; void) =
       choice
       {
         pressed_accelerator . Active(),
         pressed_brake . Active(),
         pressed_on . trigger_record . Cruising(),
         pressed_off . Active(),
         pressed_resume . Active(),
         turned_engine_off . Inactive()
       };
     Cruising(void; void) =
       choice
       {
         pressed_accelerator . trigger_disable . Suspended(),
         pressed_brake . trigger_disable . Suspended(),
         pressed_on . Cruising(),
         pressed_off . trigger_disable . Suspended(),
         pressed_resume . Cruising(),
         turned_engine_off . trigger_disable . Inactive()
       };
     Suspended(void; void) =
       choice
       {
         pressed_accelerator . Suspended(),
         pressed_brake . Suspended(),
         pressed_on . trigger_record . Cruising(),
         pressed_off . Suspended(),
         pressed_resume . trigger_resume . Cruising(),
         turned_engine_off . Inactive()
       }
   INPUT_INTERACTIONS  SYNC UNI turned_engine_on; turned_engine_off;
                                pressed_accelerator; pressed_brake;
                                pressed_on; pressed_off; pressed_resume
   OUTPUT_INTERACTIONS SYNC UNI trigger_record; trigger_resume;
                                trigger_disable
	\end{verbatim}}

The speed detector periodically communicates the number of wheel revolutions per time unit to the speed
actuator:

	{\footnotesize\begin{verbatim}
 ARCHI_ELEM_TYPE Detector_Type(void)
   BEHAVIOR
     Detector_Off(void; void) =
       turned_engine_on . Detector_On();
     Detector_On(void; void) =
       choice
       {
         measure_speed . signal_speed . Detector_On(),
         turned_engine_off . Detector_Off()
       }
   INPUT_INTERACTIONS  SYNC UNI turned_engine_on; turned_engine_off
   OUTPUT_INTERACTIONS SYNC UNI signal_speed
	\end{verbatim}}

The speed actuator adjusts the throttle on the basis of the triggers received from the controller and of the
speed measured by the detector. It can be disabled (until the on/resume button is pressed) or enabled (until
any pedal or button different from on/resume is pressed):

	{\footnotesize\begin{verbatim}
 ARCHI_ELEM_TYPE Actuator_Type(void)
   BEHAVIOR
     Disabled(void; void) =
       choice
       {
         signaled_speed . Disabled(),
         triggered_record . record_speed . Enabled(),
         triggered_resume . resume_speed . Enabled()
       };
     Enabled(void; void) =
       choice
       {
         signaled_speed . adjust_throttle . Enabled(),
         triggered_disable . disable_control . Disabled()
       }
   INPUT_INTERACTIONS  SYNC UNI triggered_record; triggered_resume;
                                triggered_disable; signaled_speed
   OUTPUT_INTERACTIONS void
	\end{verbatim}}

Suppose we want to design an applet-based simulator for the cruise control system. The applet must contain a
panel with seven software buttons -- corresponding to turning the engine on/off, the two pedals, and the
three hardware buttons -- and a text area showing the sequence of buttons that have been pressed. When
pressing one of the seven software buttons, the corresponding operation either succeeds or fails. In the
first case, the panel can interact with the sensor and the text area is updated accordingly. In the second
case -- think, e.g., of pressing the accelerator button when the engine is off -- the panel cannot interact
with the sensor, rather it emits a beep.

In order not to block the simulator when the pressure of a software button fails, we need to model several
operations of the panel through semi-synchronous interactions, as shown below:

	{\footnotesize\begin{verbatim}
 ARCHI_ELEM_TYPE Panel_Type(void)
   BEHAVIOR
     Unallocated(void; void) =
       init_applet . start_applet . Active();
     Active(void; void) =
       choice
       {
         signal_engine_on . Checking(signal_engine_on.success),
         signal_accelerator . Checking(signal_accelerator.success),
         signal_brake . Checking(signal_brake.success),
         signal_on . Checking(signal_on.success),
         signal_off . Checking(signal_off.success),
         signal_resume . Checking(signal_resume.success),
         signal_engine_off . Checking(signal_engine_off.success),
         stop_applet . Inactive()
       };
     Checking(boolean success; void) =
       choice
       {
         cond(success = true)  -> update . Active(),
         cond(success = false) -> beep . Active(),
       };
     Inactive(void; void) =
       choice
       {
         start_applet . Active(),
         destroy_applet . Unallocated()
       }
   INPUT_INTERACTIONS  SYNC  UNI init_applet; start_applet;
                                 stop_applet; destroy_applet
   OUTPUT_INTERACTIONS SSYNC UNI signal_engine_on; signal_engine_off;
                                 signal_accelerator; signal_brake;
                                 signal_on; signal_off; signal_resume
	\end{verbatim}}

\noindent
where all the input interactions are related to user commands for starting/stopping the simulator.

	\begin{figure}[p]

\centerline{\includegraphics{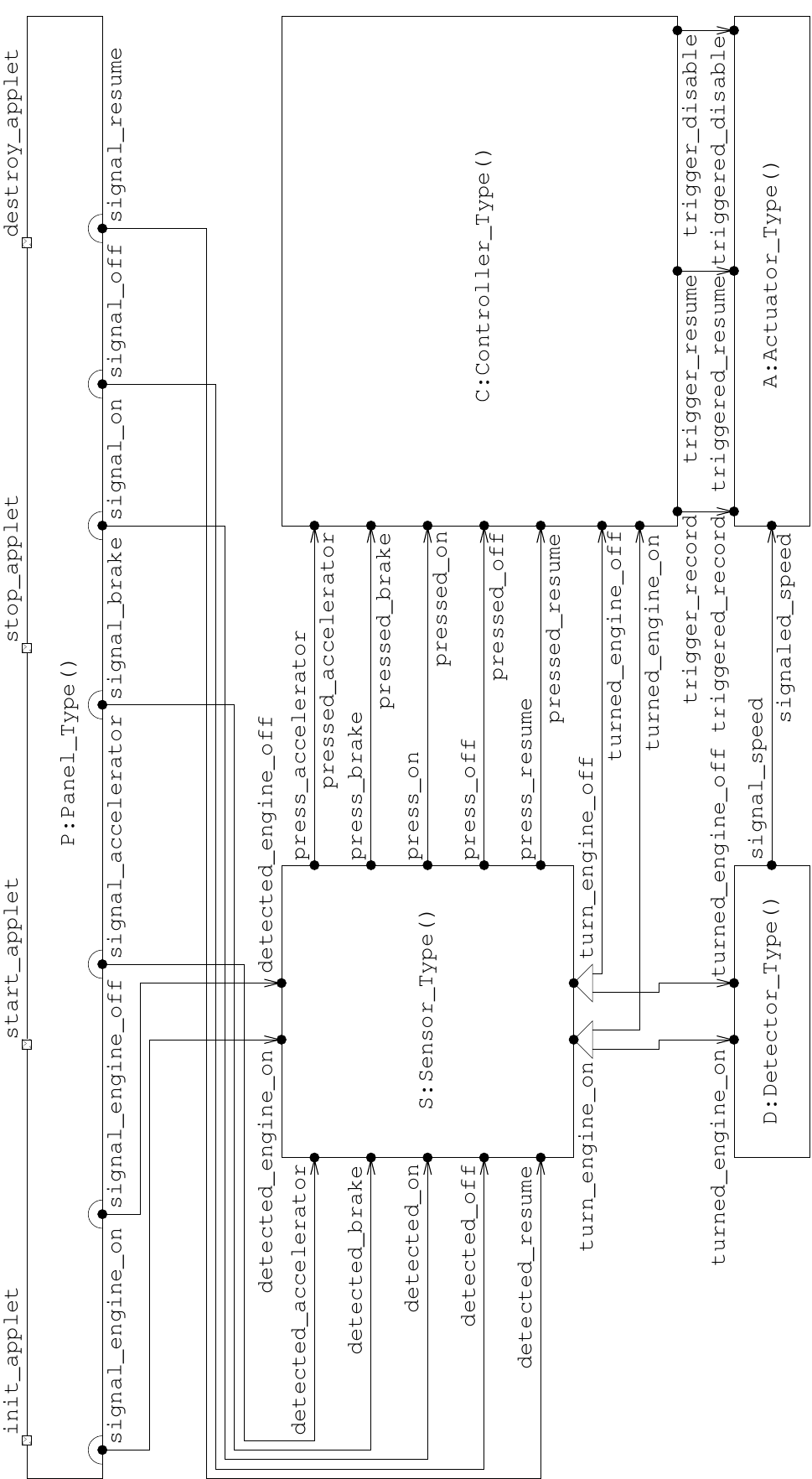}}
\caption{Enriched flow graph of the cruise control system simulator}\label{ccssfg}

	\end{figure}

The topology of the overall system is described through the enriched flow graph of Fig.~\ref{ccssfg}. As can
be noted, it is an intersection of a cycle and a star, where the cycle traverses AEIs \texttt{S},
\texttt{C}, \texttt{D}, and \texttt{A}.

Suppose we wish to verify whether the applet-based simulator is deadlock free. First of all, we observe that
all the AEIs are deadlock free. Then, we consider the cycle and we apply to it the interoperability check.
Since there are no nonsynchronous interactions within the cycle, applying the modified interoperability
check to the cycle boils down to applying the original check. The outcome is thus known from~\cite{BCD}:
\texttt{S} interoperates with \texttt{C}, \texttt{D}, and \texttt{A}, hence the cycle can be reduced to
$\lsp \texttt{S} \rsp^{\rm pc; wob}_{\tt P, S, C, D, A}$ (Thm.~\ref{interopthm}).

Now, it is easy to see that \texttt{S} is compatible with the only AEI in the only acyclic portion of the
topology -- \texttt{P} -- because all the exceptions that the semi-synchronous interactions of \texttt{P}
may raise are hidden when applying the check. Therefore, we can conclude that the entire architectural
description can be reduced to $\lsp \texttt{S} \rsp^{\rm pc; wob}_{\tt P, S, C, D, A}$, and hence it is
deadlock free because so is $\lsp \texttt{S} \rsp^{\rm pc; wob}_{\tt P, S, C, D, A}$ (Thm.~\ref{compthm}).

\section{Code Generation}\label{code}

\noindent
The architectural description of a software system should be used not only for modeling and verification
purposes, but also for guiding subsequent stages of the software development process. In particular, it
should serve as a basis for synthesizing the software system itself. As an attempt to bridge the gap between
system modeling/verification and system implementation, we propose an approach that automatically generates
multithreaded Java software from PADL descriptions containing an arbitrary combination of synchronous,
semi-synchronous, and asynchronous interactions.

Similar to~\cite{MK}, the choice of Java as target language is made for two reasons. First, Java supports
multithreading and offers a set of mechanisms for the well-structured management of threads and their shared
data, which should simplify the code generation task given the concurrency inherent in process algebraic
architectural descriptions. Second, its object-oriented nature -- and specifically its encapsulation
capability -- makes Java an appropriate candidate for coping with the high level of abstraction typical of
process algebraic architectural description languages. In fact, as can be expected, the translation of PADL
descriptions into Java software cannot be complete, and hence will require the intervention of the software
developer in specific positions of the generated code, e.g., for inserting the Java statements corresponding
to internal actions.

Another good reason for selecting Java is that it is equipped with software model-checking tools -- like,
e.g., Java PathFinder~\cite{VHBPL} -- that complement the analysis conducted on PADL descriptions by making
it possible the verification of property preservation at the code level. In fact, although property
preservation is guaranteed under certain constraints as we will see, an inappropriate intervention of the
software developer on the generated code may lead to the violation of properties proved at the architectural
level.

In order to compare our approach with related work for automatically generating code from architectural
descriptions, it is worth recalling two families distinguished on the basis of the distance between the
formalism used for describing software architectures (in our case, PADL) and the implementation language in
which code is generated (in our case, Java). The first family, characterized by an exogenous transformation,
is the long-distance one. In this family, the formalism is kept well separated from the implementation
language and descriptions are entirely translated into code. To this family belong architectural description
languages endowed with code generation facilities like Aesop~\cite{GAO}, C2SADEL~\cite{MRT}, and
Darwin~\cite{MDEK}. The second family, characterized by a semi-endogenous transformation, is the
short-distance one. In this family, the formalism is embedded in the implementation code in form of special
comments, as in SyncGen~\cite{DDHM}, or in form of special keywords and statements, as in
ArchJava~\cite{ACN}. In this case, only special symbols are translated into implementation code, while the
rest is left unchanged.

While the latter transformation can offer a flexible support to software developers -- as classical
programming techniques and patterns can be applied when designing systems -- the level of abstraction of the
underlying architectural formalism is usually low. In the case of process algebraic architectural
description languages, the transformation typically adopted for generating code is the former. The reason is
that such languages are specifically conceived for abstracting high-level properties of entire software
systems, hence they are distant from implementation languages.

One of the ideas at the basis of our approach is the provision of a library of software components -- a Java
package called \texttt{Sync} -- for adding architectural capabilities to the target programming language, in
order to shorten the distance from the architectural description language from which the code will be
generated. Hence, our approach can be viewed as a semi-exogenous transformation, as opposed to the
semi-endogenous one. Based on the classification given in~\cite{CH}, we can say that if the target model is
considered as a text, a semi-endogenous transformation can be simply realized as a model-to-text,
template-based transformation. Unfortunately, this transformation cannot be applied with the same simplicity
to our semi-exogenous approach, as the distance from model to text is still long. However, thanks to the
presence of a package like \texttt{Sync}, a model-to-text transformation can be implemented very easily
based on the pattern Visitor~\cite{GHJV}.

In this section, we present the two phases of our approach by illustrating them on the same architectural
description of an applet-based simulator for a cruise control system as the previous section. In the first
phase, we develop an architecture-driven technique for thread coordination management based on package
\texttt{Sync} (Sect.~\ref{phase1}). In the second phase, we handle the translation of the
process-algebraically-specified behavior of individual software components into threads
(Sect.~\ref{phase2}). Then, we focus on the preservation at the code level of the properties proved at the
architectural level (Sect.~\ref{mismfree}).

\subsection{First Phase: Thread Coordination Management}\label{phase1}

\noindent
The first phase of our approach to the generation of multithreaded Java code from PADL descriptions deals
with the thread coordination management. This is accomplished by developing a Java package called
\texttt{Sync}, which automatically takes care of the details of thread synchronization. Both the
implementation of the package and the use of its units for coordinating threads are guided by
architecture-level abstractions.

In the following, we present a reference thread communication model and then we illustrate the structure and
the usage of \texttt{Sync}.

\subsubsection{Thread Communication Model}\label{commmodel}

\noindent
The thread communication model adopted by package \texttt{Sync}, which fully complies with the interaction
qualifiers of PADL, relies on two roles (sender and receiver) and encompasses two different dimensions.

The first dimension -- thread communication synchronicity -- comprises nine values arising from all possible
combinations of a synchronous, semi-synchronous, or asynchronous activity on the sender side with a
synchronous, semi-synchronous, or asynchronous activity on the receiver side. Similar to PADL, in a
synchronous-to-synchronous communication both threads wait for the other to become ready. In the
semi-synchronous case, the thread checks whether the other is ready and, if not, raises an exception without
blocking. In the asynchronous case, the thread simply sends/receives a signal or message through a buffer
and then proceeds independently of the status of the other thread (an exception is raised at the
asynchronous receiving side if no signal or message is available in the buffer).

The second dimension -- thread communication multiplicity -- comprises the following five values:
uni-to-uni, and-to-uni, uni-to-and, or-to-uni, uni-to-or. As with PADL, in a uni-to-uni communication only
two threads are involved (unicast). In an and-to-uni/uni-to-and communication, a thread simultaneously
communicates with several other threads (multicast). Finally, in an or-to-uni/uni-to-or communication a
thread communicates with only one thread selected out of a set of other threads (server-clients).

\subsubsection{Structure of Package {\tt Sync}}

\begin{figure}[p]

\centerline{\includegraphics{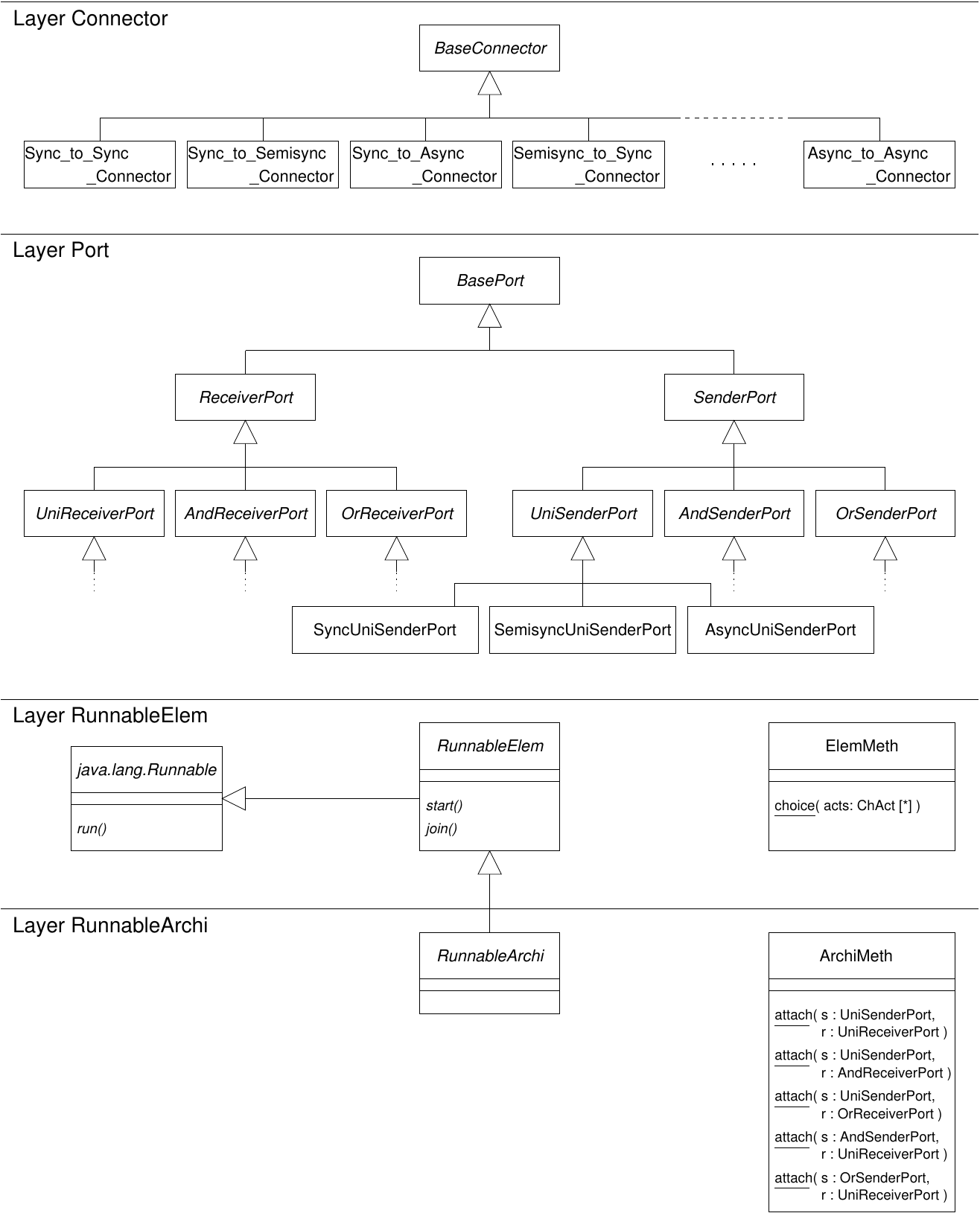}}
\caption{Class diagrams of the conceptual layers of \texttt{Sync}}\label{packsync}

        \end{figure}

\noindent
The Java package \texttt{Sync}, which adheres to the above mentioned communication model, is structured into
four conceptual layers: \texttt{Connector}, \texttt{Port}, \texttt{RunnableElem}, and
\texttt{RunnableArchi}. Each of them corresponds to a different architectural abstraction and comprises a
set of components realized through Java classes and interfaces, some of which are visible by the software
developer. The related class diagrams are shown in Fig.~\ref{packsync}.

The first layer, called \texttt{Connector}, is a set of invisible Java classes and interfaces that are used
within \texttt{Sync} to perform synchronizations and data transfers between two threads. As sketched in
Fig.~\ref{packsync}, there are nine classes realizing -- consistently with the adopted thread communication
model -- the nine combinations of communication synchronicities.

This layer is inspired by the traditional producer-consumer model, where the producer and the consumer
represent the sender and receiver threads, respectively. This is implemented by equipping every
\texttt{Connector} class with a buffer shared only by the two related threads. Each such class also has two
interface methods -- \texttt{send()} and \texttt{receive()} -- for accessing the buffer in a mutually
exclusive way and two interface methods -- \texttt{obsSnd()} and \texttt{obsRcv()} -- for observing the
status of the threads using a \texttt{Connector} object in the case of nonasynchronous and-/or-interactions.

The second layer, called \texttt{Port}, is a set of Java classes and interfaces that realize the abstraction
corresponding to a set of statements through which a thread interacts with possibly many other threads. As
sketched in Fig.~\ref{packsync}, there are eighteen classes combining -- through \texttt{Connector} objects
-- the three cases of communication multiplicity with the three cases of communication synchronicity both on
the receiver side and on the sender side.

This layer gives rise to suitable objects, each of which will be attached either to a single
\texttt{Connector} object -- if representing a uni-interaction -- or to several \texttt{Connector} objects
-- if representing an and-/or-interaction. Every \texttt{Port} object must be instantiated by specifying its
owner thread in the \texttt{Port} constructor. Each of the nine \texttt{Port} classes related to the sender
side is equipped with a \texttt{send()} method, whereas each of the nine \texttt{Port} classes related to
the receiver side is equipped with a \texttt{receive()} method. Each of the two methods makes use of the
homonymous method of the associated \texttt{Connector} objects and raises an
\texttt{UnattachedPortException} whenever the \texttt{Port} object is not connected to any other
\texttt{Port} object. This happens if the interaction associated with the \texttt{Port} object is an
architectural interaction not attached to any other interaction. A different exception,
\texttt{SemisyncPortNotReadyException}, is raised if the associated interaction is semi-synchronous and the
connected \texttt{Port} objects are not ready to communicate. Both exceptions are defined as derived classes
of a base class called \texttt{SyncException} defined within \texttt{Sync}.

The third layer, called \texttt{RunnableElem}, is an interface derived from the standard \texttt{Runnable}
interface that realizes the abstraction corresponding to a thread in a concurrent Java program. This layer
provides support for the generation of Java threads from the process-algebraically-specified behavior of
AETs in a PADL description. In particular, as shown in Fig.~\ref{packsync}, it is also equipped with a class
called \texttt{ElemMeth}, which defines a static method called \texttt{choice()} used for the translation of
alternative compositions.

The fourth layer, called \texttt{RunnableArchi}, is a \texttt{RunnableElem}-derived interface that realizes
the abstraction corresponding to a concurrent Java program. The compatibility of \texttt{RunnableArchi} with
\texttt{RunnableElem} provides support for hierarchical software development. As shown in
Fig.~\ref{packsync}, in order to avoid the direct handling of \texttt{Connector} objects, this layer is also
equipped with a class called \texttt{ArchiMeth}, which defines a family of five static methods called
\texttt{attach()}. Each of them receives two parameters, which must be a sender \texttt{Port} object and a
receiver \texttt{Port} object, and connects them by creating a \texttt{Connector} object only if they refer
to two different owner threads according to one of the five combinations of communication multiplicities
admitted by the adopted thread communication model.

\subsubsection{Usage of Package {\tt Sync}}

\begin{figure}[thb]

\centerline{\includegraphics{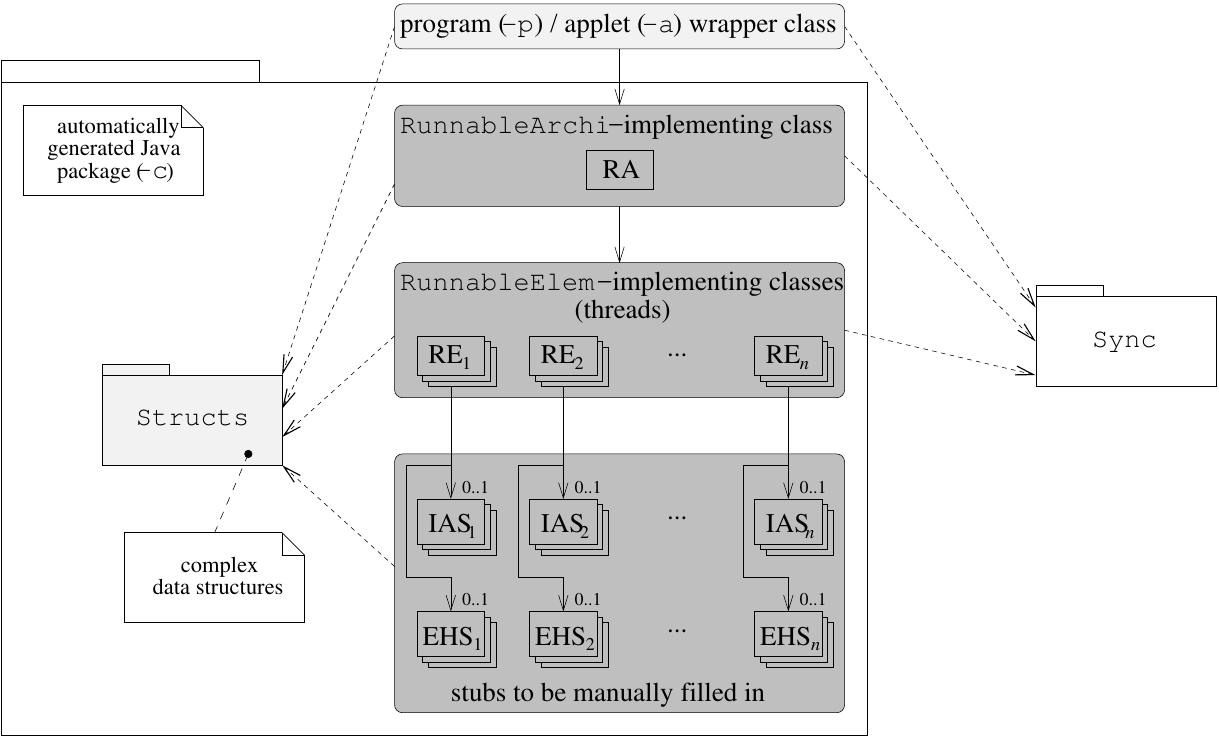}}
\caption{Structure of the \texttt{Sync}-based generated code}\label{filedep}

	\end{figure}

\noindent
From a code generation viewpoint, package \texttt{Sync} is a repository of architectural abstractions
ensuring the correct and transparent handling of synchronizations and data exchanges among the threads of
the software system being designed. The structure of the generated code is shown in Fig.~\ref{filedep}.

Starting from a PADL description of the system, it is necessary to create a \texttt{RunnableArchi} object
for the overall system, together with as many \texttt{RunnableElem} objects as there are AEIs in the PADL
description (see the upper part of Fig.~\ref{filedep}). Then, it is necessary to create as many
\texttt{Port} objects as there are interactions in the AEIs synthesized as \texttt{RunnableElem} objects.
Afterwards, it is necessary to create all the \texttt{Connector} objects that are needed to make the AEIs
synthesized as \texttt{RunnableElem} objects communicate -- according to the attachments in the PADL
description -- through their interactions synthesized as \texttt{Port} objects.

The lower part of Fig.~\ref{filedep} shows stub classes to be manually filled in. As we will see in the
second phase, these stubs are generated for managing the internal actions of the AETs (IAS) and for handling
exceptions that may be raised by the interactions of the instances of those AETs (EHS) -- i.e.,
\texttt{UnattachedPortException} and \texttt{SemisyncPortNotReadyException}.

Fig.~\ref{filedep} also shows a package called \texttt{Structs}. The reason is that most of the data types
provided by PADL -- such as boolean, (bounded) integer, real, list, and array -- can be trivially translated
into built-in Java data types. However, the record data type and the generic object data type provided by
PADL need ad-hoc classes and interfaces made available through \texttt{Structs}.

Each of the classes generated by PADL2Java for a given PADL description is stored into a distinct
\texttt{.java} file. All these files are contained in a single package with the same name as the PADL
description source file. Further \texttt{.java} files may be generated depending on the specified
translation option.

Fig.~\ref{filedep} shows the various options that are available at code generation time: \texttt{-c},
\texttt{-p}, and \texttt{-a}. The default option \texttt{-c} stands for the generation of the package
depicted in the figure, with no additional classes. Option \texttt{-p} stands for the synthesis of a full
Java program. This can be achieved through the generation of a further public class that contains only
method \texttt{main()}, which acts as a wrapper for the \texttt{RunnableArchi}-implementing class. Option
\texttt{-a} stands for the synthesis of a Java applet. This can be achieved through the generation of a
further public class derived from the standard \texttt{JApplet} class, which gives rise to a wrapper for the
\texttt{RunnableArchi}-implementing class.

Both for \texttt{-p} and for \texttt{-a}, the wrapper class creates an instance of the
\texttt{RunnableArchi}-implementing class and declares a \texttt{SyncUniSenderPort} or
\texttt{SyncUniReceiverPort} object for each architectural interaction declared in the
\texttt{RunnableArchi}-implementing class. In this way, the wrapper is allowed to dispatch suitable commands
to the right \texttt{RunnableElem} objects whenever necessary. Moreover, in the \texttt{-a} case the wrapper
class defines a number of stubs related to the initialization, activation, deactivation, and destruction of
the applet, which will be filled in by the software developer if needed.

	\begin{example}\label{ccsapplet}

Consider the cruise control system simulator of Sect.~\ref{cruise} and suppose we want to synthesize it as
multithreaded Java software from its PADL description. Since it is an applet-based simulator, we need to
generate an applet wrapper. The following \texttt{.java} file contains the \texttt{JApplet}-derived class
synthesized for the wrapper, where \texttt{Cruise\_Control\_System\_package} is the package for the cruise
control system:

		{\footnotesize\begin{verbatim}
 import Sync.*;
 import Cruise_Control_System_package.Cruise_Control_System;
 import javax.swing.JApplet;

 public class MainAppletClass extends JApplet {
  
   //---------- DECLARING ARCHITECTURE -----------//
   Cruise_Control_System archiInstance = null;

   //---------- DECLARING APPLET PORTS -----------//
   SyncUniSenderPort to_P_init_applet;
   SyncUniSenderPort to_P_start_applet;
   SyncUniSenderPort to_P_stop_applet;
   SyncUniSenderPort to_P_destroy_applet;

   //---------- DEFINING APPLET MEMBERS ----------//
   public void init() {
     if (archiInstance == null) {
       archiInstance = new Cruise_Control_System();
       // to_P_init_applet = new SyncUniSenderPort(this);
       // to_P_start_applet = new SyncUniSenderPort(this);
       // to_P_stop_applet = new SyncUniSenderPort(this);
       // to_P_destroy_applet = new SyncUniSenderPort(this);
       // try {
       //   ArchiMeth.attach(to_P_init_applet,
       //                    archiInstance.P_init_applet);
       //   ArchiMeth.attach(to_P_start_applet,
       //                    archiInstance.P_start_applet);
       //   ArchiMeth.attach(to_P_stop_applet,
       //                    archiInstance.P_stop_applet);
       //   ArchiMeth.attach(to_P_destroy_applet,
       //                    archiInstance.P_destroy_applet);
       // } catch(BadAttachmentException e) {}
       archiInstance.start();
       // FILL IN THE FINAL PART OF THE METHOD BODY IF NEEDED
   }

   public void start() {
     // FILL IN THE METHOD BODY IF NEEDED
   }

   public void stop() {
     // FILL IN THE METHOD BODY IF NEEDED
   }

   public void destroy() {
     // FILL IN THE INITIAL PART OF THE METHOD BODY IF NEEDED
     archiInstance = null;
   }
 }
		\end{verbatim}}

As can be noted, for each of the four architectural input interaction of AEI \texttt{P}, we have the
declaration of a sender port whose name coincides with the name of the corresponding architectural input
interaction prefixed by \texttt{to\_} (the prefix is \texttt{from\_} in the case of an architectural output
interaction). Each such sender port will manage the commands given by the user of the applet-based simulator
when initializing, starting, stopping, or destroying the applet-based simulator itself depending on how the
software developer fills in the body of the methods associated with those operations.

As far as method \texttt{init()} is concerned, the initial part of its body includes some commented
statements for instantiating the applet ports and attaching them to the corresponding architectural
interactions. Those statements are commented because, in general, it is not necessarily the case that all
the architectural interactions must be attached to the corresponding ports in the wrapper. For instance,
some of them should be attached to other software units. It is the responsibility of the software developer
to uncomment the statements that are appropriate for the specific system.
\fullbox

	\end{example}

\subsubsection{Structure of the {\tt RunnableArchi}-Implementing Class}

\noindent
When using package \texttt{Sync} in the translation process, the Java class synthesized to implement the
\texttt{RunnableArchi} interface for the overall system has the same name as the architectural type and is
structured as follows:
\[\begin{array}{|l|}
\hline
\texttt{public class} \triangleleft \! \textit{architectural type name} \! \triangleright \texttt{implements
RunnableArchi \{} \\
\hspace*{0.4cm} \triangleleft \textit{Declaring Runnable Elements} \triangleright \\
\hspace*{0.4cm} \triangleleft \textit{Declaring Architectural Interactions} \triangleright \\
\hspace*{0.4cm} \triangleleft \textit{Defining Constructor} \triangleright \\
\hspace*{0.4cm} \triangleleft \textit{Building Architecture} \triangleright: \\
\hspace*{0.8cm} \triangleleft \textit{Instantiating Runnable Elements} \triangleright \\
\hspace*{0.8cm} \triangleleft \textit{Assigning Architectural Interactions} \triangleright \\
\hspace*{0.8cm} \triangleleft \textit{Attaching Local Interactions} \triangleright \\
\hspace*{0.4cm} \triangleleft \textit{Running Architecture} \triangleright \\
\texttt{\}} \\
\hline
\end{array}\]

The first section, \textit{Declaring Runnable Elements}, declares an object of a \linebreak
\texttt{RunnableElem}-implementing class -- without instantiating it -- for each AEI declared in the PADL
description.

The second section, \textit{Declaring Architectural Interactions}, declares a public \texttt{Port} object
for each architectural interaction declared in the PADL description. Such objects are public because the
architectural interactions are the interfaces of the whole system, hence support must be provided for them
to be used for the hierarchical or compositional modeling of complex systems as well as within wrapper
classes.

The third section, \textit{Defining Constructor}, defines the class constructor together with its
parameters, which coincide with the parameters of the architectural type.

The fourth section, \textit{Building Architecture}, defines a method called \linebreak
\texttt{buildArchiTopology()} that constructs the architecture topology in a way similar to the
architectural topology section of the PADL description. First, it instantiates the previously declared
\texttt{RunnableElem} objects. Second, it assigns the previously declared public \texttt{Port} objects
through the corresponding \texttt{Port} objects of the newly instantiated \texttt{RunnableElem} objects.
Third, it invokes the method \texttt{attach()} defined within \texttt{Sync} in order to connect the
\texttt{Port} objects of the newly instantiated \texttt{RunnableElem} objects according to the attachments
declared in the PADL description.

Finally, the fifth section, \textit{Running Architecture}, declares the thread associated with the class
itself. Moreover, it defines the public methods \texttt{start()}, which starts the previously declared
thread, \texttt{join()}, which allows other threads to wait for the end of the execution of the previously
declared thread, and \texttt{run()}, which starts all the previously instantiated \texttt{RunnableElem}
objects and then waits for their termination.

	\begin{example}\label{ccsra}

We now continue Ex.~\ref{ccsapplet}. The following \texttt{.java} file contains the \texttt{RunnableArchi}
interface for the cruise control system:

		{\footnotesize\begin{verbatim}
 package Cruise_Control_System_package;

 import Sync.*;

 public class Cruise_Control_System implements RunnableArchi {

   //-------- DECLARING RUNNABLE ELEMENTS --------//
   Panel_Type P;
   Sensor_Type S;
   Controller_Type C;
   Detector_Type D;
   Actuator_Type A;

   //--- DECLARING ARCHITECTURAL INTERACTIONS ----//
   public SyncUniReceiverPort P_init_applet;
   public SyncUniReceiverPort P_start_applet;
   public SyncUniReceiverPort P_stop_applet;
   public SyncUniReceiverPort P_destroy_applet;

   //----------- DEFINING CONSTRUCTOR ------------//
   Cruise_Control_System() {
     buildArchiTopology();
   }

   //----------- BUILDING ARCHITECTURE -----------//
   void buildArchiTopology() {

     // INSTANTIATING RUNNABLE ELEMENTS:
     P = new Panel_Type();
     S = new Sensor_Type();
     C = new Controller_Type();
     D = new Detector_Type();
     A = new Actuator_Type();
    
     // ASSIGNING ARCHITECTURAL INTERACTIONS:
     this.P_init_applet = P.init_applet;
     this.P_start_applet = P.start_applet;
     this.P_stop_applet = P.stop_applet;
     this.P_destroy_applet = P.destroy_applet;

     // ATTACHING LOCAL INTERACTIONS:
     try {
       ArchiMeth.attach(P.signal_engine_on,
                        S.detected_engine_on);
       ArchiMeth.attach(P.signal_engine_off,
                        S.detected_engine_off);
       ArchiMeth.attach(P.signal_accelerator,
                        S.detected_accelerator);
       ArchiMeth.attach(P.signal_brake,
                        S.detected_brake);
       ArchiMeth.attach(P.signal_on,
                        S.detected_on);
       ArchiMeth.attach(P.signal_off,
                        S.detected_off);
       ArchiMeth.attach(P.signal_resume,
                        S.detected_resume);
       ArchiMeth.attach(S.turn_engine_on,
                        C.turned_engine_on);
       ArchiMeth.attach(S.turn_engine_on,
                        D.turned_engine_on);
       ArchiMeth.attach(S.turn_engine_off,
                        C.turned_engine_off);
       ArchiMeth.attach(S.turn_engine_off,
                        D.turned_engine_off);
       ArchiMeth.attach(S.press_accelerator,
                        C.pressed_accelerator);
       ArchiMeth.attach(S.press_brake,
                        C.pressed_brake);
       ArchiMeth.attach(S.press_on,
                        C.pressed_on);
       ArchiMeth.attach(S.press_off,
                        C.pressed_off);
       ArchiMeth.attach(S.press_resume,
                        C.pressed_resume);
       ArchiMeth.attach(C.trigger_record,
                        A.triggered_record);
       ArchiMeth.attach(C.trigger_resume,
                        A.triggered_resume);
       ArchiMeth.attach(C.trigger_disable,
                        A.triggered_disable);
       ArchiMeth.attach(D.signal_speed,
                        A.signaled_speed);
     } catch(BadAttachmentException e) {}
   }

   //----------- RUNNING ARCHITECTURE ------------//
   Thread th_Cruise_Control_System = null;
 
   public void start() {
     (th_Cruise_Control_System = new Thread(this)).start();
   }

   public void join() throws InterruptedException {
     th_Cruise_Control_System.join();
   }

   public void run() {
     P.start();
     S.start();
     C.start();
     D.start();
     A.start();
     try {
       P.join();
       S.join();
       C.join();
       D.join();
       A.join();
     } catch(InterruptedException e) {}
   }
 }
		\end{verbatim}}

\noindent
Note that it faithfully implements the topology of Fig.~\ref{ccssfg}.
\fullbox

	\end{example}

\subsubsection{Structure of {\tt RunnableElem}-Implementing Classes}\label{runnableelem}

\noindent
Similar to the case of \texttt{RunnableArchi}, each of the Java classes synthesized to implement the
\texttt{RunnableElem} interface corresponding to an AET has the same name as the AET and is structured as
follows:
\[\begin{array}{|l|}
\hline
\texttt{class} \triangleleft \! \textit{architectural element type name} \! \triangleright
\texttt{implements RunnableElem \{} \\
\hspace*{0.4cm} \triangleleft \textit{Declaring Behavioral Equations Interfaces} \triangleright \\
\hspace*{0.4cm} \triangleleft \textit{Instantiating Interactions} \triangleright \\
\hspace*{0.4cm} \triangleleft \textit{Declaring Stubs} \triangleright \\
\hspace*{0.4cm} \triangleleft \textit{Defining Constructor} \triangleright \\
\hspace*{0.4cm} \triangleleft \textit{Defining Behavior} \triangleright \\
\hspace*{0.4cm} \triangleleft \textit{Running Element} \triangleright \\
\texttt{\}} \\
\hline
\end{array}\]

The first section, \textit{Declaring Behavioral Equations Interfaces}, defines an interface called
\linebreak \texttt{BehavioralEquationInterface} and declares an equation object of such an interface for
each behavioral equation occurring in the AET definition.

The second section, \textit{Instantiating Interactions}, instantiates various \texttt{Port} objects of
various types (see the second layer of Fig.~\ref{packsync}) on the basis of the interactions occurring in
the AET definition and their qualifiers.

The third section, \textit{Declaring Stubs}, declares two stub objects to be manually filled in later on. As
already mentioned, one stub object is for the translation of the internal actions occurring in the AET
definition, while the other stub is for handling exceptions related to the interactions occurring in the AET
definition.

The fourth section, \textit{Defining Constructor}, defines the class constructor together with its
parameters, which coincide with the parameters of the AET. The constructor declares the parameters as
nonpublic members in order to store their values and make them available throughout the thread execution.
Then, the constructor invokes the method defined in the next section.

The fifth section, \textit{Defining Behavior}, creates instances of anonymous classes implementing
\linebreak \texttt{BehavioralEquationInterface} and assigns them to the previously mentioned equation
objects. Each anonymous class translates a different behavioral equation occurring in the AET. The only
method declared by the interface, \texttt{behavEqCall()}, is defined here for each equation object and will
be discussed in the second phase.

Finally, the sixth section, \textit{Running Element}, declares the thread associated with the class itself.
Moreover, it defines the public methods \texttt{start()}, which starts the previously declared thread,
\texttt{join()}, which allows other threads to wait for the end of the execution of the previously declared
thread, and \texttt{run()}, which instantiates the two stub classes and executes the various equation
objects.

	\begin{example}\label{ccsre}

We now continue Ex.~\ref{ccsapplet} and Ex.~\ref{ccsra}. The following \texttt{.java} file contains the
\texttt{RunnableElem} interface for \texttt{Panel\_Type}:

		{\footnotesize\begin{verbatim}
 package Cruise_Control_System_package;

 import Sync.*;

 class Panel_Type implements RunnableElem {

   //- DECLARING BEHAVIORAL EQUATIONS INTERFACES -//
   interface BehavioralEquationInterface { void behavEqCall(); }
   BehavioralEquationInterface Unallocated,
                               Active,
                               Checking,
                               Inactive;
   BehavioralEquationInterface nextBehavEq;
   Object[] actualPars;

   //-------- INSTANTIATING INTERACTIONS ---------//
   SyncUniReceiverPort init_applet = new SyncUniReceiverPort(this);
   SyncUniReceiverPort start_applet = new SyncUniReceiverPort(this);
   SyncUniReceiverPort stop_applet = new SyncUniReceiverPort(this);
   SyncUniReceiverPort destroy_applet = new SyncUniReceiverPort(this);
   SemisyncUniSenderPort signal_engine_on = new SemisyncUniSenderPort(this);
   SemisyncUniSenderPort signal_engine_off = new SemisyncUniSenderPort(this);
   SemisyncUniSenderPort signal_accelerator = new SemisyncUniSenderPort(this);
   SemisyncUniSenderPort signal_brake = new SemisyncUniSenderPort(this);
   SemisyncUniSenderPort signal_on = new SemisyncUniSenderPort(this);
   SemisyncUniSenderPort signal_off = new SemisyncUniSenderPort(this);
   SemisyncUniSenderPort signal_resume = new SemisyncUniSenderPort(this);

   //-------------- DECLARING STUBS --------------//
   IAS_Panel_Type internal_Panel_Type;
   EHS_Panel_Type exception_Panel_Type;

   //----------- DEFINING CONSTRUCTOR ------------//
   Panel_Type() {
     defineBehavEquations();
   }

   //------------- DEFINING BEHAVIOR -------------//
   void defineBehavEquations() {
     ...
   }

   //-------------- RUNNING ELEMENT --------------//
   Thread th_Panel_Type = null;
  
   public void start() {
     (th_Panel_Type = new Thread(this)).start();
   }

   public void join() throws InterruptedException {
     th_Panel_Type.join();
   }
  
   public void run() {
     ...
   }
 }
		\end{verbatim}}

As can be noted, the definitions of methods \texttt{defineBehavEquations()} and \texttt{run()} have been
omitted because they are strictly related to the synthesis of the thread behavior. Therefore, they will be
shown after presenting the second phase of our approach.
\fullbox

	\end{example}

\subsection{Second Phase: Thread Behavior Generation}\label{phase2}

\noindent
The second phase of our approach to the generation of multithreaded Java code from PADL descriptions deals
with the translation of the process algebraic specification of the behavior of the AETs into thread classes.
Due to the different level of abstraction of an architectural description language and of a programming
language, only a partial translation based on stubs is possible, with the preservation of architectural
properties depending on the way in which the stubs are filled in by the software developer.

In the following, we introduce a reference thread generation model and we show how to synthesize thread
method \texttt{run()} and how to translate process algebraic operators consistent with the adopted model in
order to synthesize method \texttt{defineBehavEquations()}. We also show how to synthesize stubs for
handling exceptions and internal actions.

\subsubsection{Thread Generation Model}\label{genermodel}

\noindent
A finite state machine model is adopted to guide the generation of a thread from the description of an AET
made out of a sequence of process algebraic defining equations. Large conditional statements or table-based
approaches do not guarantee high efficiency when many conditions or associations have to be checked at run
time for each invocation of a behavioral equation. Therefore, we generate code on the basis of the
behavioral pattern State~\cite{GHJV}.

As shown in Sect.~\ref{runnableelem}, each AET is translated into a class implementing the interface
\texttt{RunnableElem}. Each state of the behavior of the AET is defined as an inner class implementing the
interface \linebreak \texttt{BehavioralEquationInterface} and defining the method \texttt{behavEqCall()}.
The idea is that every behavioral equation is translated into an inner state class having the same name as
the equation. The code for an inner state class is generated by proceeding by induction on the syntactical
structure -- \texttt{stop}, behavioral invocation, action prefix, and \texttt{choice} -- of the process
algebraic term occurring on the right-hand side of the corresponding behavioral equation.

The class associated with an AET also defines the member \texttt{nextBehavEq}, a reference to an object
implementing \texttt{BehavioralEquationInterface}, and the member \texttt{actualPars}, a reference to an
array of objects. These references, which are shared and visible by all the inner state classes, are in
charge of defining the state transitions, i.e., the next behavioral equation to be executed and the actual
parameters to be passed.

We conclude by observing that a different treatment is needed for action prefixes depending on whether the
related actions are interactions or internal actions. An interaction is involved in communications, hence it
is automatically managed via package \texttt{Sync}. However, as we have seen in Sect.~\ref{commmodel}, the
execution of an interaction may result in two kinds of exception, whose handling is left to the software
developer. Our thread generation model thus includes the possibility of associating an exception handling
stub (EHS) with each interaction, to be filled in by the software developer. By contrast, an internal action
is not involved at all in communications, as it takes place inside an AET. In this case, the architectural
description does not provide any information about how to translate the action into a sequence of Java
statements. As a consequence, our thread generation model also includes the possibility of associating an
internal action stub (IAS) with each internal action, to be filled in by the software developer.

\subsubsection{Synthesizing Thread Method {\tt run()}}

\noindent
The class corresponding to an AET comprises the constructor and method \texttt{run()}. While the former
instantiates the inner state classes by calling method \texttt{defineBehavEquations()}, as shown below the
latter first of all instantiates the EHSs and the IASs declared as members of the class:
\[\begin{array}{|l|}
\hline
\texttt{public void run() \{} \\
\hspace*{0.4cm} \triangleleft \textit{EHS instantiation} \triangleright \\
\hspace*{0.4cm} \triangleleft \textit{IAS instantiation} \triangleright \\
\hspace*{0.4cm} \texttt{nextBehavEq =} \, \triangleleft \! \textit{first behavioral equation} \triangleright
\! \texttt{;} \\
\hspace*{0.4cm} \texttt{actualPars =} \, \triangleleft \! \textit{actual parameters of the first behavioral
equation} \triangleright \!
\texttt{;} \\ 
\hspace*{0.4cm} \texttt{while (nextBehavEq != null)} \\
\hspace*{0.8cm} \texttt{nextBehavEq.behavEqCall();} \\
\texttt{\}} \\
\hline
\end{array}\]
Method \texttt{run()} then assigns the state class instance representing the first behavioral equation to
\texttt{nextBehavEq} and the related parameters to \texttt{actualPars}. A \texttt{while} statement carries
out the execution of the behavioral equations starting from the first one, by repeatedly invoking method
\texttt{behavEqCall()} on the state class instance referenced by \texttt{nextBehavEq}. When
\texttt{nextBehavEq} is set to \texttt{null}, the execution of method \texttt{run()} terminates.

	\begin{example}\label{ccsrun}

We complete the second omitted part of Ex.~\ref{ccsre} by showing the definition of method \texttt{run()}
for the class associated with \texttt{Panel\_Type}:

		{\footnotesize\begin{verbatim}
 public void run() {
   internal_Panel_Type = new IAS_Panel_Type();
   exception_Panel_Type = new EHS_Panel_Type();
   nextBehavEq = Unallocated;
   actualPars = null;
   while (nextBehavEq != null)
     nextBehavEq.behavEqCall();
 }
		\end{verbatim}}
\fullbox

	\end{example}

\subsubsection{Translating {\tt stop}}

\noindent
Process term \texttt{stop} represents the situation in which no further action can be executed. It is
therefore translated by assigning \texttt{null} both to \texttt{nextBehavEq} and to \texttt{actualPars}. As
a consequence, when encountering \texttt{stop}, method \texttt{run()} terminates its execution.

\subsubsection{Translating Behavioral Invocations}

\noindent
The behavioral invocation $B(\underline{e})$ represents a process term that behaves as the behavioral
equation whose identifier is $B$, when passing the possibly empty sequence of actual parameters
$\underline{e}$. A behavioral invocation, which can occur only within the scope of an action prefix
operator, is not translated into a method call, as this may result in the generation of inefficient code in
case of recursion. Instead, a behavioral invocation is translated into an assignment to \texttt{nextBehavEq}
of an instance of the inner state class that corresponds to the next behavioral equation, followed by an
assigment to \texttt{actualPars} of the actual parameters needed by the next behavioral equation.

\subsubsection{Translating Action Prefixes}

\noindent
The action prefix operator is used to represent a process term that can execute an action and then behaves
as described by another process term. As already mentioned, the translation of the action depends on whether
it is an interaction or an internal action.

In the first case, the action is translated into an invocation of method \texttt{send()} -- if it is an
output interaction -- or method \texttt{receive()} -- if it is an input interaction -- of the corresponding
\texttt{Port} object. If the interaction is semi-synchronous or architectural, its translation must then be
completed by filling in the related method in an EHS.

In the second case, the action translation is completely left to the software developer, as internal actions
cannot be treated automatically at all. A method for each of them is placed in a distinct IAS, which has to
be filled in by the software developer with the corresponding Java statements. As a consequence, every
occurrence of an internal action is translated into an invocation of the related method in an IAS.

	\begin{example}\label{ccsstub}

We now continue Ex.~\ref{ccsapplet}, Ex.~\ref{ccsra}, and Ex.~\ref{ccsre}. The following \texttt{.java} file
contains the exception handling stubs for \texttt{Panel\_Type}:

		{\footnotesize\begin{verbatim}
 package Cruise_Control_System_package;

 class EHS_Panel_Type {

   EHS_Panel_Type() {
     // FILL IN THE CONSTRUCTOR BODY IF NEEDED
   }

   void signal_engine_on_ssyncEH() {
     // FILL IN THE METHOD BODY IF NEEDED
   }

   void signal_engine_off_ssyncEH() {
     // FILL IN THE METHOD BODY IF NEEDED
   }

   void signal_accelerator_ssyncEH() {
     // FILL IN THE METHOD BODY IF NEEDED
   }

   void signal_brake_ssyncEH() {
     // FILL IN THE METHOD BODY IF NEEDED
   }

   void signal_on_ssyncEH() {
     // FILL IN THE METHOD BODY IF NEEDED
   }

   void signal_off_ssyncEH() {
     // FILL IN THE METHOD BODY IF NEEDED
   }

   void signal_resume_ssyncEH() {
     // FILL IN THE METHOD BODY IF NEEDED
   }

   void init_applet_archiEH() {
     // FILL IN THE METHOD BODY IF NEEDED
   }

   void start_applet_archiEH() {
     // FILL IN THE METHOD BODY IF NEEDED
   }

   void stop_applet_archiEH() {
     // FILL IN THE METHOD BODY IF NEEDED
   }

   void destroy_applet_archiEH() {
     // FILL IN THE METHOD BODY IF NEEDED
   }
 }
		\end{verbatim}}

\noindent
We have a method for each semi-synchronous interaction of \texttt{Panel\_Type}, together with a method for
each architectural interaction of \texttt{P}. Note that each method of the first (resp.\ second) group has
the same name as the corresponding interaction augmented with suffix \texttt{\_ssyncEH} (resp.\
\texttt{\_archiEH}).

The following \texttt{.java} file instead contains the internal action stubs for \texttt{Panel\_Type}:

		{\footnotesize\begin{verbatim}
 package Cruise_Control_System_package;

 class IAS_Panel_Type {

   IAS_Panel_Type() {
     // FILL IN THE CONSTRUCTOR BODY IF NEEDED
   }

   void update() {
     // FILL IN THE METHOD BODY
   }

   void beep() {
     // FILL IN THE METHOD BODY
   }
 }
		\end{verbatim}}

\noindent
In this case, we have a method for each of the two internal actions of \texttt{Panel\_Type}, which must
necessarily be filled in by the software developer.
\fullbox

	\end{example}

\subsubsection{Translating {\tt choice}}\label{choice}

\noindent
The choice operator expresses a selection among a certain number of alternative behaviors described through
process terms. A choice-based process term is translated into a \texttt{switch-case} statement, whose
condition is given by an invocation of the static method \texttt{choice()} defined in the class
\texttt{ElemMeth} of package Sync.

There are two cases that must be addressed in order to translate the choice operator. The first one is the
case where every process term involved in the choice starts with an action prefix operator. In this case,
the method \texttt{choice()} is directly employed, which accepts as input an array of objects of class
\texttt{ChAct}, each of which contains a boolean guard expressing the possible constraint under which the
corresponding starting action is enabled (default value \texttt{true}). Should one of the starting actions
be an interaction, an additional piece of information is contained in the corresponding object, which is a
reference to the \texttt{Port} object associated with the interaction. Method \texttt{choice()} returns the
index (within the array) of the starting action selected for execution.

A starting action is enabled (and hence can be selected for execution) if its guard evaluates to
\texttt{true} and -- in the case of a synchronous interaction -- the corresponding \texttt{Port} object is
ready to communicate. If all the starting actions with guard evaluating to \texttt{true} are synchronous
interactions, method \texttt{choice()} waits -- and the thread that contains it passivates -- until one of
the associated \texttt{Port} objects is ready to communicate. If all the guards of the starting actions
evaluate to \texttt{false}, method \texttt{choice()} returns a negative value. Normally, at most one
starting action is enabled, as the guards associated with alternative actions are usually in conflict with
each other. Should this not be the case, i.e., if several starting actions are enabled, a probabilistic
mechanism is applied to select one of those actions. This is useful when generating code from quantitative
variants of PADL like \aemilia~\cite{BBS}, in which the probability or the duration of actions can be
expressed.

Based on the index returned by \texttt{choice()}, the \texttt{switch-case} statement invokes the method
associated with the execution of the selected starting action. This method is \texttt{send()} or
\texttt{receive()} in the case of an interaction, whereas for an internal action it is the corresponding
method in the related IAS. The invocation of this method is followed in turn by the translation of the
process term prefixed by the selected action. In the default clause, which comes into play when a negative
value is returned by \texttt{choice()}, process term \texttt{stop} is invoked by assigning \texttt{null}
both to \texttt{nextBehavEq} and to \texttt{actualPars}.

The second case is the one in which some of the process terms involved in the choice do not start with an
action prefix operator. If one of these process terms is \texttt{stop}, then nothing has to be added for it
in the \texttt{ChAct} array and the \texttt{switch-case} statement, because it is selected by default
whenever the other involved process terms cannot be selected. If instead one of these process terms is a
nested choice, then a flattening of the nested choice takes place during the translation.

	\begin{example}\label{ccsbeh}

We complete the first omitted part of Ex.~\ref{ccsre} by showing the definition of method \linebreak
\texttt{defineBehavEquations()} for the class associated with \texttt{Panel\_Type}:

		{\footnotesize\begin{verbatim}
 void defineBehavEquations() {

   Unallocated = new BehavioralEquationInterface() {
     public void behavEqCall() {
       _Unallocated();
     }
     private void _Unallocated() {
       try {
         init_applet.receive();
       } catch(UnattachedPortException e) {
         exception_Panel_Type.init_applet_archiEH();
       } 
       try {
         start_applet.receive();
       } catch(UnattachedPortException e) {
         exception_Panel_Type.start_applet_archiEH();
       } 
       nextBehavEq = Active;
       actualPars  = null;
     }
   };

   Active = new BehavioralEquationInterface() {
     public void behavEqCall() {
       _Active();
     }
     private void _Active() {
       switch (
         ElemMeth.choice(
           new ChAct[] {
             new ChAct(true, signal_engine_on),
             new ChAct(true, signal_accelerator),
             new ChAct(true, signal_brake),
             new ChAct(true, signal_on),
             new ChAct(true, signal_off),
             new ChAct(true, signal_resume),
             new ChAct(true, signal_engine_off),
             new ChAct(true, stop_applet)
           }
         )
       )
       {
         case 0:
           try {
             signal_engine_on.send();
           } catch(NotReadyPortException e) {
             exception_Panel_Type.signal_engine_on_ssyncEH();
           }
           nextBehavEq = Checking;
           actualPars  = new Object[] {signal_engine_on.success()};
           break;
         case 1:
           try {
             signal_accelerator.send();
           } catch(NotReadyPortException e) {
             exception_Panel_Type.signal_accelerator_ssyncEH();
           }
           nextBehavEq = Checking;
           actualPars  = new Object[] {signal_accelerator.success()};
           break;
         case 2:
           try {
             signal_brake.send();
           } catch(NotReadyPortException e) {
             exception_Panel_Type.signal_brake_ssyncEH();
           }
           nextBehavEq = Checking;
           actualPars  = new Object[] {signal_brake.success()};
           break;
         case 3:
           try {
             signal_on.send();
           } catch(NotReadyPortException e) {
             exception_Panel_Type.signal_on_ssyncEH();
           }
           nextBehavEq = Checking;
           actualPars  = new Object[] {signal_on.success()};
           break;
         case 4:
           try {
             signal_off.send();
           } catch(NotReadyPortException e) {
             exception_Panel_Type.signal_off_ssyncEH();
           }
           nextBehavEq = Checking;
           actualPars  = new Object[] {signal_off.success()};
           break;
         case 5:
           try {
             signal_resume.send();
           } catch(NotReadyPortException e) {
             exception_Panel_Type.signal_resume_ssyncEH();
           }
           nextBehavEq = Checking;
           actualPars  = new Object[] {signal_resume.success()};
           break;
         case 6:
           try {
             signal_engine_off.send();
           } catch(NotReadyPortException e) {
             exception_Panel_Type.signal_engine_off_ssyncEH();
           }
           nextBehavEq = Checking;
           actualPars  = new Object[] {signal_engine_off.success()};
           break;
         case 7:
           try {
             stop_applet.receive();
           } catch(UnattachedPortException e) {
             exception_Panel_Type.stop_applet_archiEH();
           } 
           nextBehavEq = Inactive;
           actualPars  = null;
           break;
         default:
           nextBehavEq = null;
           actualPars  = null;
       }
     }
   };

   Checking = new BehavioralEquationInterface() {
     public void behavEqCall() {
       _Checking((Boolean)actualPars[0]);
     }
     private void _Checking(boolean success) {
       switch (
         ElemMeth.choice(
           new ChAct[] {
             new ChAct(success == true, null),
             new ChAct(success == false, null)
           }
         )
       )
       {
         case 0:
           internal_Panel_Type.update();
           nextBehavEq = Active;
           actualPars  = null;
           break;
         case 1:
           internal_Panel_Type.beep();
           nextBehavEq = Active;
           actualPars  = null;
           break;
         default:
           nextBehavEq = null;
           actualPars  = null;
       }
     }
   };

   Inactive = new BehavioralEquationInterface() {
     public void behavEqCall() {
       _Inactive();
     }
     private void _Inactive() {
       switch (
         ElemMeth.choice(
           new ChAct[] {
             new ChAct(true, start_applet),
             new ChAct(true, destroy_applet)
           }
         )
       )
       {
         case 0:
           try {
             start_applet.receive();
           } catch(UnattachedPortException e) {
             exception_Panel_Type.start_applet_archiEH();
           } 
           nextBehavEq = Active;
           actualPars  = null;
           break;
         case 1:
           try {
             destroy_applet.receive();
           } catch(UnattachedPortException e) {
             exception_Panel_Type.destroy_applet_archiEH();
           } 
           nextBehavEq = Unallocated;
           actualPars  = null;
           break;
         default:
           nextBehavEq = null;
           actualPars  = null;
       }
     }
   };
 }
		\end{verbatim}}

\noindent
As can be noted, we have a \texttt{BehavioralEquationInterface} for each of the four behavioral equations of
\texttt{Panel\_Type}.
\fullbox

	\end{example}

\subsection{Preservation of Architectural Properties}\label{mismfree}

\noindent
We conclude by discussing the issue of guaranteeing that the properties proved at the architectural level --
through the techniques illustrated in Sect.~\ref{checks} -- are preserved at the code level. Since we have
taken an approach based on automatic code generation, property preservation should be achieved by
construction. In other words, the translation from PADL to Java illustrated before should have been defined
in a way that ensures property preservation. We now investigate this by separately considering the code
generated for thread coordination, the code generated for translating behavioral equations, and the code
provided for filling in stubs.

The code generated for coordinating threads cannot infringe the preservation of architectural properties, up
to the methods for handling the exceptions that semi-synchronous or architectural interactions may raise. In
fact, the code for thread coordination is completely generated in an automatic way by means of package
\texttt{Sync}. As far as the system topology is concerned, this is built in the
\texttt{RunnableArchi}-implementing class in the same way as prescribed by the second section of the PADL
description. Moreover, both PADL and \texttt{Sync} adhere to the same communication model. On the PADL side,
each interaction is given three qualifiers: input vs.\ output, synchronous vs.\ semi-synchronous vs.\
asynchronous, uni vs.\ and vs.\ or. Each interaction is then translated into an invocation of method
\texttt{receive()} or \texttt{send()} defined in the corresponding \texttt{Port} object, depending on
whether it is an input or an output interaction, respectively. Additionally, the kind of this \texttt{Port}
object -- synchronous vs.\ semi-synchronous vs.\ asynchronous, uni vs.\ and vs.\ or -- is the same as that
of the interaction.

Each behavioral equation occurring in an AET definition is translated into an inner state class of the
corresponding \texttt{RunnableElem}-implementing class. The translation proceeds by induction on the
syntactical structure of the process term occurring on the right-hand side of the behavioral equation. The
way in which the translation is carried out, together with the way in which the thread execution flow
proceeds according to the order established by the invocations of the behavioral equations, ensures the
preservation of the AET behavior, up to the methods for translating internal actions.

As a consequence, the preservation of architectural properties critically depends on the way in which the
software developer manually fills in EHSs and IASs. Here we shall consider only IASs, as EHSs can be treated
similarly.

	\begin{figure}[thb]

\centerline{\includegraphics{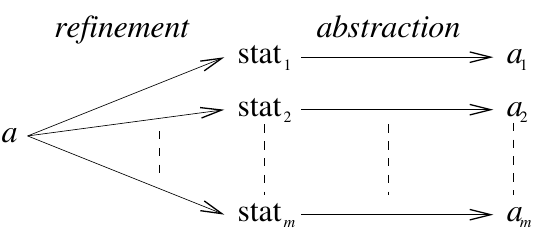}}
\caption{Refinement of an internal action and related statement abstraction}\label{actionsref}

        \end{figure}

In order to be able to reason about architectural property preservation, we have to compare the internal
actions and the corresponding sequences of Java statements on the same process algebraic ground. As shown in
Fig.~\ref{actionsref}, each of the Java statements into which an internal action is refined during the
translation process can be abstractly viewed as a fresh action. The following theorem provides a sufficient
condition for ensuring the preservation of an architectural property. For the sake of simplicity, as
mentioned at the beginning of Sect.~\ref{checks}, we concentrate on deadlock freedom. Below, we denote by
$\obis{\rm B}$ the observational equivalence of~\cite{Mil}, which is a refinement of $\wbis{\rm B}$ that
turns out to be a congruence with respect to alternative composition too.

	\begin{theorem}

Let $T$ be the process algebraic description of the behavior of a thread and $a$ be an internal action
occurring in $T$. Let $a_{1}, a_{2}, \dots, a_{m}$ be the fresh actions abstracting the statements into
which $a$ is translated and $T'$ be the process algebraic description of the behavior of the thread obtained
from $T$ by replacing every occurrence of $a \, . \_$ with $a_{1} \, . \, a_{2} \, . \, \dots \, . \, a_{m}
\, . \_$. Let $H$ be the set of internal actions occurring in $T$ or $T'$. Whenever $T$ is deadlock free and
$a \, . \, \texttt{stop} / H \obis{\rm B} a_{1} \, . \, a_{2} \, . \, \dots \, . \, a_{m} \, . \,
\texttt{stop} / H$, then $T'$ is deadlock free as well.

		\begin{proof}
Since $\obis{\rm B}$ is a congruence with respect to all the process algebraic operators, from $a \, . \,
\texttt{stop} / H \obis{\rm B} a_{1} \, . \, a_{2} \, . \, \dots \, . \, a_{m} \, . \, \texttt{stop} / H$ it
follows that $T / H \obis{\rm B} T' / H$, hence $T / H \wbis{\rm B} T' / H$ because $\wbis{\rm B}$ is
coarser than $\obis{\rm B}$. Since $T$ is deadlock free, $\wbis{\rm B}$ preserves deadlock freedom, and
deadlock freedom is expressed in terms of interactions (which cannot belong to $H$), it follows that $T'$ is
deadlock free as well.
\fullbox
		\end{proof}

	\end{theorem}

Note that, in the theorem above, it is not necessarily the case that all of the actions $a_{1}, a_{2},
\dots, a_{m}$ associated with the Java statements provided by the software developer belong to $H$. As an
example, one of such actions may correspond to an invocation of \texttt{send()} or \texttt{receive()} or of
a method such as \texttt{behavEqCall()} belonging to a state class. Fortunately, in practice both cases are
prevented from occurring by the fact that the \texttt{Port} objects -- which contain methods \texttt{send()}
and \texttt{receive()} -- and the \texttt{RunnableElem}-implementing class instances -- which contain the
state classes -- are not visible within the stubs.

In order to preserve architectural properties, we provide some guidelines that the software developer should
follow when filling in the stubs for handling exceptions and internal actions. The purpose of these
guidelines is to achieve as much as possible the benefits of code generation, i.e., speeding up system
implementation in a way that conforms by construction to the architectural model of the system:

	\begin{itemize}

\item No synchronized or thread control methods -- like \texttt{wait()} and \texttt{notify()} -- should be
invoked within the stubs, as their invocations would be internal to the threads but at the same time could
affect the way threads communicate with each other.

\item No further thread should be created within the stubs, as this would have an observable impact on the
system topology and on thread coordination.

\item There should be no variables/objects that are visible from several stub classes. This means that all
the data shared by several threads should be exchanged only through suitable units of package \texttt{Sync}.

\item The stub method associated with the first internal action following an invocation of \texttt{send()}
or \texttt{receive()} should copy every object passed in that invocation, and all the stub methods
associated with the subsequent internal actions should work on those copies of the objects. This would avoid
interferences among threads stemming from the fact that \texttt{send()} always keeps a reference to the
passed objects -- so that it can be defined within \texttt{Sync} in a way that supports arbitrarily many
parameters of arbitrary types -- and such objects may be modified by the stub method associated with some
internal action.

\item All the exceptions that can be raised when executing a stub method should be caught -- or prevented
from being raised -- inside the method, rather than propagating to the \texttt{RunnableElem}-implementing
class.

\item Nonterminating statements should not occur within stub methods.

	\end{itemize}

We conclude by observing that we have found stubs more appropriate than abstract classes because the former
allow the generated code to be compiled, including invocations of stub methods as they are concrete. In
order to effectively remind the software developer to fill in the stubs, it would suffice to automatically
introduce a statement in the definition of each stub method, which prints out an explicative message
whenever an empty stub method is invoked at run time.

\section{Conclusion}\label{concl}

\noindent
In this paper, we have extended a typical process algebraic architectural description language by including
semi-synchronous interactions -- handled by means of suitable semantic rules -- and asynchronous
interactions -- managed by adding implicit buffer-like components. Besides enhancing the expressiveness of
the language without compromising its usability, we have shown that the architectural compatibility check
and the architectural interoperability check can be easily adapted to cope with the presence of
nonsynchronous interactions.  Moreover, we have illustrated an architecture-inspired approach to the
generation of multithreaded object-oriented code from process algebraic architectural descriptions
containing an arbitrary combination of synchronous and nonsynchronous interactions in such a way that
properties proved at the architectural level are preserved at the code level.

On the modeling and verification side, our work constitutes -- as far as we know -- the first systematic
attempt to deal with semi-synchronous and asynchronous communications in process algebraic architectural
description languages. Although we have focused on PADL (the interested reader is referred to~\cite{AB} for
a comparison with similar notations), we believe that our ideas can be applied to the other process
algebraic architectural description languages appeared in the literature with minor modifications.

Concerning future work, we observe that, in the case of asynchronous interactions, the semantic model
underlying a process algebraic architectural description may have infinitely many states due to the
additional implicit components behaving like unbounded buffers. As shown in~\cite{BZ}, it is hard to reason
about component-based systems under an asynchronous semantics, because many properties such as deadlock
freedom are undecidable. In order for the modified architectural checks to be effectively applicable in this
case, one option is to allow users to limit the size of buffers statically. Another option is to derive
sufficient conditions under which the state space is guaranteed to be finite, which requires further
investigation.

On the code generation side, there are several related work. First of all we mention ArchJava~\cite{ACN}.
This is an extension of Java aiming at the unification of software architecture with implementation, in
order to ensure that the implementation conforms to the architectural specification with respect to
communication integrity. According to this property, each component in the implementation may only
communicate directly with the components to which it is connected in the architecture.

Our approach differs from ArchJava in several ways. First, it does not extend Java, but generates Java code
from process algebraic architectural descriptions. In our approach the developer is then required to fill in
some stubs to complete the code for the behavior of the threads, thus giving a certain degree of
flexibility. The price to be paid is that the guidelines may be violated, whereas a similar situation is not
possible in ArchJava. Second, our approach adopts a richer communication model, implemented and
transparently made available through package \texttt{Sync}. This guarantees a property even stronger than
communication integrity: implementation threads directly communicate only with the threads they are
connected to in the architectural description in the way prescribed by the architectural description itself
with respect to communication synchronicity (synchronous, semi-synchronous, asynchronous) and communication
multiplicity (uni, and, or). Third, since it keeps the architectural description language separated from the
implementation language, our approach provides a higher-level support than ArchJava for the preservation of
behavioral properties. On the other hand, the strong integration between architecture and implementation
endows ArchJava with useful dynamic capabilities, like run-time creation of components -- although
communication integrity places restrictions on the way in which their instances can be used -- and
connections among them.

We then mention C2SADEL~\cite{MRT}. This is an architectural description language tied to the C2 style,
which combines the usual architectural concepts with type theory. Type checking is used to analyze the
architectural descriptions for consistency by unifying corresponding operations required and provided by
different components. Moreover, Java code can be automatically generated from C2SADEL descriptions. Since
type checking is a static analysis technique, while the architectural features on which we focus are
behavioral and concerned with a rich communication model, we believe that our approach can guarantee the
preservation of more complex properties than C2SADEL.

In the future, we plan to investigate the combination of our approach with those discussed before. In
particular, we would like to investigate the applicability of our approach to C2SADEL, in order to take
advantage of both type checking and behavioral analysis from the architectural level to the code level.
Similarly, we would like to experiment our approach with ArchJava -- by generating ArchJava code instead of
Java code -- in order to exploit the complementary strengths of the two approaches.

On the application side, we recall that the performance-oriented version \aemilia~\cite{BBS} of PADL is the
input language of TwoTowers~\cite{Ber}, an open-source software tool for the functional verification,
security analysis, and performance evaluation of software architectures. TwoTowers is being extended in
order to include the capability of expressing semi-synchronous and asynchronous interactions, as well as to
investigate the absence of architectural mismatches through the modified compatibility and interoperability
checks. Moreover, we have recently implemented another tool called PADL2Java, which translates PADL
descriptions into multithreaded Java code according to the approach presented in this paper. PADL2Java will
soon be integrated in TwoTowers, in order to construct an architecture-centric toolset encompassing
modeling, verification, and implementation of software architectures. This will allow us to investigate the
effectiveness and the scalability of our techniques when tackling larger case studies.

With regard to our toolset, we would like to integrate it with software model checking tools, like Java
PathFinder~\cite{VHBPL}, and to define specific rules for static analysis tools, like TPTP~\cite{GM}. The
reason is that the preservation at the code level of the properties proved at the architectural level is
guaranteed only if (the underlying platform is correct and) the software developer follows the guidelines
provided in Sect.~\ref{mismfree} when filling in IAS and EHS stubs. Having a software model checker such as
PathFinder -- possibly driven by techniques like, e.g., those developed in~\cite{PPK,GPC,CCGJV} -- available
within TwoTowers would permit the verification of the overall system after the possible intervention of the
software developer, whereas customized static analysis tools such as TPTP may be exploited for guiding the
previously mentioned intervention.


\end{document}